\documentclass[twocolumn]{aastex631}

\usepackage{xspace}
\usepackage{natbib}
\shorttitle{The Lyman $\alpha$ Reference Sample}
\shortauthors{Melinder et al.}
\graphicspath{{./}{figures/}}

\newcommand{\lya}{Ly$\alpha$\xspace}
\newcommand{\ha}{H$\alpha$\xspace}
\newcommand{\hb}{H$\beta$\xspace}

\newcommand{\oi}{[O\,{\sc i}]\xspace}
\newcommand{\oii}{[O\,{\sc ii}]\xspace}
\newcommand{\oiii}{[O\,{\sc iii}]\xspace}
\newcommand{\nii}{[N\,{\sc ii}]\xspace}
\xspace
\newcommand{\hi}{H\,{\sc i}\xspace}
\newcommand{\hii}{H\,{\sc ii}\xspace}

\newcommand{\msun}{M$_{\odot}$\xspace}
\newcommand{\msunyr}{M$_{\odot}$~yr$^{-1}$\xspace}

\newcommand{\lfuv}{$L_{\rm FUV}$\xspace}
\newcommand{\lsun}{L$_{\odot}$\xspace}

\newcommand{\fesc}{$f_{\mathrm{esc}}$\xspace}

\newcommand{\z}{$z$\xspace}

\newcommand\ebvn{E(B-V)$_{\mathrm{neb}}$\xspace}

\newcommand\llya{$L_{\mathrm{Ly\alpha}}$\xspace}
\newcommand\fesclya{$f_{\mathrm{esc}}$\xspace}
\newcommand\wlya{$W_{\mathrm{Ly\alpha}}$\xspace}

\newcommand\wha{$W_{\mathrm{H\alpha}}$\xspace}
\newcommand\lyaha{Ly$\alpha/$H$\alpha$\xspace}
\newcommand\hahb{H$\alpha/$H$\beta$\xspace}
\newcommand\oiiihb{[O\,{\sc iii}]/H$\beta$\xspace}
\newcommand{\ergc}{{erg/s/cm$^2$/\AA}\xspace}
\newcommand{\ergl}{{erg/s/cm$^2$}\xspace}
\newcommand{\kms}{km s$^{-1}$\xspace}

\begin{document}

\title{The Lyman $\alpha$ Reference Sample XIV: \\
Lyman $\alpha$ imaging of 45 low redshift star-forming galaxies and inferences on global emission.}

\correspondingauthor{Jens Melinder}
\email{jens@astro.su.se}

\author[0000-0003-0470-8754]{Jens Melinder}
\affil{Stockholm University, Department of Astronomy and Oskar Klein Centre for Cosmoparticle Physics, AlbaNova University Centre, SE-10691,
Stockholm, Sweden}

\author{G{\"o}ran {\"O}stlin}
\affil{Stockholm University, Department of Astronomy and Oskar Klein Centre for Cosmoparticle Physics, AlbaNova University Centre, SE-10691,
Stockholm, Sweden}

\author[0000-0001-8587-218X]{Matthew Hayes}
\affil{Stockholm University, Department of Astronomy and Oskar Klein Centre for Cosmoparticle Physics, AlbaNova University Centre, SE-10691,
Stockholm, Sweden}

\author{Armin Rasekh}
\affil{Stockholm University, Department of Astronomy and Oskar Klein Centre for Cosmoparticle Physics, AlbaNova University Centre, SE-10691,
Stockholm, Sweden}

\author[0000-0002-8823-9723]{J. Miguel Mas-Hesse}
\affil{Centro de Astrobiologia (CAB), CSIC--INTA, ESAC, 28692 Villanueva de la Ca\~{n}ada, Spain}

\author[0000-0002-1821-7019]{John M. Cannon}
\affil{Department of Physics \& Astronomy, Macalester College, 1600 Grand Avenue, Saint Paul, MN 55105, USA}

\author[0000-0002-1821-7019]{Daniel Kunth}
\affil{Institut d’Astrophysique, Paris, 98 bis Boulevard Arago, F-75014 Paris, France}

\author[0000-0003-4207-0245]{Peter Laursen}
\affil{Cosmic Dawn Center (DAWN)}
\affil{Niels Bohr Institute, University of Copenhagen, Jagtvej 128, DK-2200, Copenhagen N, Denmark}

\author[0000-0002-1025-7569]{Axel Runnholm}
\affil{Stockholm University, Department of Astronomy and Oskar Klein Centre for Cosmoparticle Physics, AlbaNova University Centre, SE-10691,
Stockholm, Sweden}

\author[0000-0002-8505-4678]{E. Christian Herenz}
\affil{European Southern Observatory, Av. Alonso de Córdova 3107, 763 0355 Vitacura, Santiago, Chile}

\author[0000-0003-1427-2456]{Matteo Messa}
\affil{Stockholm University, Department of Astronomy and Oskar Klein Centre for Cosmoparticle Physics, AlbaNova University Centre, SE-10691,
Stockholm, Sweden}
\affil{Geneva Observatory, University of Geneva, 51 Chemin des Maillettes, CH-1290 Versoix, Switzerland}

\author[0000-0001-7144-7182]{Daniel Schaerer}
\affil{Geneva Observatory, University of Geneva, 51 Chemin des Maillettes, CH-1290 Versoix, Switzerland}

\author{Anne Verhamme}
\affil{Geneva Observatory, University of Geneva, 51 Chemin des Maillettes, CH-1290 Versoix, Switzerland}

\author[0000-0002-9204-3256]{T. Emil Rivera-Thorsen}
\affil{Stockholm University, Department of Astronomy and Oskar Klein Centre for Cosmoparticle Physics, AlbaNova University Centre, SE-10691,
Stockholm, Sweden}

\author[0000-0002-4902-0075]{Lucia Guaita}
\affil{Departamento de Ciencias Fisicas, Facultad de Ciencias Exactas, Universidad Andres Bello, Fernandez Concha 700, Las Condes, Santiago, Chile}
\affil{Núcleo de Astronomía, Facultad de Ingeniería y Ciencia, Universidad Diego Portales, Av. Ejército 441, Santiago, Chile}

\author[0000-0002-2244-9920]{Thomas Marquart}
\affil{Department of Astronomy and Space Physics, Box 515, 75120 Uppsala, Sweden}

\author[0000-0003-1111-3951]{Johannes Puschnig}
\affil{Universit{\"a}t Bonn, Argelander-Institut f{\"u}r Astronomie, Auf dem H{\"u}gel 71, D-53121 Bonn, Germany}

\author[0000-0003-1767-6421]{Alexandra Le Reste}
\affil{Stockholm University, Department of Astronomy and Oskar Klein Centre for Cosmoparticle Physics, AlbaNova University Centre, SE-10691,
Stockholm, Sweden}

\author{Andreas Sandberg}
\affil{Stockholm University, Department of Astronomy and Oskar Klein Centre for Cosmoparticle Physics, AlbaNova University Centre, SE-10691,
Stockholm, Sweden}

\author{Emily Freeland}
\affil{Stockholm University, Department of Astronomy and Oskar Klein Centre for Cosmoparticle Physics, AlbaNova University Centre, SE-10691,
Stockholm, Sweden}

\author{Joanna Bridge}



\begin{abstract}
We present \lya imaging of 45 low redshift star-forming galaxies observed with
	the Hubble Space Telescope.  The galaxies have been selected to have
	moderate to high star formation rates using far-ultraviolet (FUV)
	luminosity and \ha equivalent width criteria, but no constraints on
	\lya luminosity. We employ a pixel stellar continuum fitting
	code to obtain accurate continuum subtracted \lya, \ha and \hb maps. We
	find that \lya is less concentrated than FUV and optical line emission
	in almost all galaxies with significant \lya emission. We present
	global measurements of \lya and other quantities measured in apertures
	designed to capture all of the \lya emission.  We then show how the
	escape fraction of \lya relates to a number of other measured
	quantities (mass, metallicity, star formation, ionisation parameter,
	and extinction). We find that the escape fraction is strongly
	anti-correlated with nebular and stellar extinction, weakly
	anti-correlated with stellar mass, but no conclusive evidence for
	correlations with other quantities. We show that \lya escape fractions
	are inconsistent with common dust extinction laws, and discuss how a
	combination of radiative transfer effects and clumpy dust models can
	help resolve the discrepancies. We present a star formation rate
	calibration based on \lya luminosity, where the equivalent width of
	\lya is used to correct for non-unity escape fraction, and show that
	this relation provides a reasonably accurate SFR estimate.  
	We also show stacked growth curves of \lya for the galaxies
that can be used to find aperture loss fractions at a given physical radius.
\end{abstract}

\keywords{cosmology: observations --- galaxies: evolution --- galaxies: starburst ---
radiative transfer --- ultraviolet: galaxes --- techniques: image processing --- methods: data analysis}

\section{Introduction} 
\label{sec:intro}

Photons emitted in the \lya line at 1216 \AA\ (carrying an energy of 10.2 eV)
come from a radiative transition in the hydrogen (\hi) atom from the second
lowest energy state to the ground state ($2p$ to $1s$). The flux emitted in
this line comes from two processes, recombination and collisional excitation.
Recombination, when ionized atoms recombine with free electrons which go
through a cascade of radiative transitions to reach the ground state, is the
more important of these (responsible for $\sim$90\% of the line flux).  In this
way ionized gas  will emit photons in a number of emission lines at different
wavelengths, with energies corresponding to the energy levels involved. \lya is
a resonance line and has the highest transition probability of all radiative
transitions in the hydrogen atom --- roughly 68\% of all recombination events
result in the emission of \lya photon \citep[assuming case B recombination and
a gas temperature of $10^4$ K, ][]{2014PASA...31...40D}. Because of the
resonant nature of the line a large majority of the \lya photons emitted from
the nebular regions will typically be re-absorbed and re-emitted many times by
the neutral interstellar medium (ISM) on their way out from the galaxies. This
means that ISM properties affecting radiative transfer --- such as kinematics,
neutral gas geometry, and dust column densities --- will be very important, and
needs to be taken into account when estimating the output \lya luminosity
\citep[e.g.][]{2009ApJ...704.1640L,2012A&A...546A.111V,2016ApJ...826...14G}.

Given that hydrogen is the most common element in the Universe,
and is the fundamental ingredient in the star formation process, studying 
molecular, neutral, and ionized hydrogen in 
both absorption and emission is an important window to the astrophysics 
of galaxies. Young and massive stars dominate the radiation output of
star-forming galaxies, and are mostly distributed in stellar clusters. 
The regions close to these intense sites of star
formation are mostly ionized (\hii regions), and
thus emit strongly in recombination lines. This radiation is also known as
nebular line emission, and for non-resonant hydrogen transitions (such as the
optical Balmer recombination lines \ha and \hb) measurements of the nebular
line strength can be used to accurately determine the luminosity of ionizing
photons from massive stars in star-forming regions inside galaxies, 
and thereby infer the star formation rate (SFR) in them.  

Because of the strength and rest wavelength of
the \lya line it was suggested as a probe of high redshift galaxies already by
\citet{1967ApJ...147..868P}.  However, the first searches for \lya emission in
nearby star-forming galaxies were largely unsuccessful. Instead of a strong
emission line, \lya was found to be in absorption or much weaker than expected
\citep{1981ApJ...246L.109M}. Absorption of continuum and \lya photons by the 
surrounding neutral medium is of course expected, and even in strong emitters 
the line is often superposed on an
absorption trough. The measured \lya flux from these galaxies did thus not
match the expectations from the photon production as estimated from stellar
continuum or \ha observations. Furthermore, the unexpected weakness of the \lya
emission was also observed for active galactic nuclei \citep{1977Natur.269..203D}.

With more sensitive space-based instruments (i.e., International Ultraviolet 
Explorer --- IUE, Hopkins Ultraviolet Telescope --- HUT, Galaxy Evolution Explorer --- 
GALEX, Hubble Space Telescope --- HST) and larger
galaxy samples extragalactic spectroscopic \lya detections in star-forming
galaxies became common-place \citep[e.g.,][]{1996ApJ...466..831G}. These 
single aperture/slit observations did not have any information on the spatial 
distribution of \lya. For the HST Cosmic Origins Spectrograph (COS) observations 
the physical size of the aperture was also quite small and aperture losses of scattered
\lya emission were thus likely large and impossible
to constrain. The first \lya imaging at high resolution of low redshift galaxies 
showed the effects of scattering clearly \citep{2003ApJ...597..263K,
2005A&A...438...71H, 2009AJ....138..923O}, but the observations were hampered 
by the strong background of geocoronal \lya emission present in the line filter. 
The \emph{Lyman-$\alpha$ Reference Sample} (LARS) was designed
to simultaneously measure \lya emission on small scales (using HST/COS
spectroscopy) and large scales (using far-ultraviolet ---FUV--- imaging) 
for an unbiased sample of highly star-forming galaxies. The first part of the
sample, which contains 14 galaxies, was presented in 
\citet{2014ApJ...782....6H} and \citet{2014ApJ...797...11O}.
By combining data from many different
instruments and wavelength ranges this study made it possible to perform a
systematic investigation of global \lya emission and how it relates to other
galaxy properties. In addition, the relatively large field of view of the FUV
imaging data allowed measuring and characterizing the \lya halos (scattered
emission) in a sample of well-characterized galaxies for the first time
\citep{2015A&A...576A..51G, 2022A&A...662A..64R}. These studies, and others
\citep[e.g.,][]{2017ApJ...838....4Y}, have strongly confirmed the theoretical
predictions \citep[e.g.,][]{2009ApJ...704.1640L} that the radiative transfer
causes \lya radiation to be spatially scattered and thus significantly more
extended than both the stellar continuum and the non-resonant nebular line
emission.

Studies of star-forming galaxies in the nearby universe have shown that strong
net \lya emission is associated with: high specific SFRs
\citep[e.g.,][]{2011ApJ...738..136C,2014ApJ...782....6H}, low extinction
\citep[e.g.,][]{1996ApJ...466..831G,2014ApJ...782....6H}, low metallicity
\citep[e.g.,][]{1996ApJ...466..831G}, low neutral gas covering fraction
\citep[e.g.,][]{2015ApJ...805...14R}, high ionisation parameter
\citep[e.g.,][]{2017ApJ...838....4Y}, turbulent ionised gas kinematics
\citep[e.g.,][]{2016A&A...587A..78H}, and significant outflow velocities
\citep[e.g.,][]{1998A&A...334...11K,2011ApJ...730....5H}.  However, while these
conclusions hold when comparing strong emitters to non-emitters, they should
not be mistaken for direct correlations. Comparisons of \lya relative output
(equivalent widths, \wlya, or escape fractions, \fesclya) with these
observables for galaxies with non-zero escape show either no (sSFR,
metallicity), weak (\hi covering fraction, outflow velocity), or fairly strong
correlations (extinction) --- and with significant scatter that cannot be
explained by measurement uncertainties \citep{2015PASA...32...27H}.  Recently,
\citet{2020ApJ...892...48R} published a statistical study of the LARS galaxies.
By using multivariate regression techniques they show that a combination of
observables can be used to predict \lya output of star-forming galaxies to
within a factor of $\sim$50\% (0.19 dex). In this study the most important
observables for predicting \lya luminosity were found to be FUV luminosity and
extinction.  The distribution and surface brightness profile derived properties
of \lya for the sample used in this paper were studied in
\citet{2022A&A...662A..64R}. The main results from this paper were that
brighter \lya emitters have smaller halo fractions (a smaller fraction of their
\lya emission is detected far from star-forming regions) and that the spatial
distribution of \lya from them is more circularly symmetric. However, the
correlations found in this work are quite weak and show significant scatter.
Investigations of the neutral gas (\hi) in the LARS galaxies are currently
on-going and results from part of the sample have been published in
\citet{2014ApJ...794..101P}  and \citet{2022ApJ...934...69L}. A comparison of
the neutral gas and \lya properties of the full sample will be presented in Le
Reste et al. (in prep.).  

The dependence on extinction has been further explored by, e.g., 
\citet{2009ApJ...704L..98S}, \citet{2014A&A...561A..89A}, and
\citet{2018ApJ...852....9B}. These studies show that even with radiative
transfer effects a simple dust screen model is insufficient to explain the
observations. Most galaxies have escape fractions that are lower than expected,
which is a natural consequence of resonant radiative scattering
increasing the path lengths, and thus causing \lya radiation to suffer from
more extinction than nebular extinction measurements using other (non-resonant)
recombination lines would indicate.  However, many galaxies with quite substantial
nebular extinction have \lya escape fractions that far exceed the fractions
inferred from observed \hahb ratios and using standard extinction curves. 

At high redshift ($z=2-7$) \lya can be used to detect galaxies, and to
spectroscopically confirm their redshifts \citep[e.g.,][]{2004ApJ...617L...5M}.
The first samples of \lya emitters (LAEs) were found using narrow-band
techniques \citep[e.g., ][]{1998AJ....115.1319C}, which enables surveys
covering large areas at specific high redshift slices, but provides no
information on the \lya line profile. Narrow band \lya observations ---
combined with broad band imaging to reject low redshift interlopers --- thus
provides a powerful and secure method to find high redshift galaxies
\citep[e.g.,][]{2018PASJ...70S..13O}, but it is very challenging to
characterize them further using only low spatial resolution imaging. Follow-up
\lya spectroscopy of high-$z$ galaxies
\citep[e.g.,][]{2014ApJ...795...33E,2018MNRAS.477.2817S} and direct \lya
spectroscopy using Integral Field Unit (IFU) spectroscopy
\citep[e.g.,][]{2011ApJS..192....5A, 2017A&A...606A..12H} have provided a
wealth of data on high redshift LAEs. \citet{2019ApJ...887...85T} present
optical spectroscopy for a large sample ($\sim 700$) of redshift 2--3 LAEs,
finding that the equivalent width of metal line absorption (related to neutral
gas density or geometry) and the \oiiihb ratio (indicating overall ionisation
state) can be used to predict \wlya, albeit with large scatter. With simple
assumptions on the stellar population age and star formation history, and
without accounting for radiative scattering in a dusty medium, the escape
fraction of \lya should be directly proportional to the equivalent width
\citep{2019A&A...623A.157S}.  \citet{2019A&A...623A.157S} use this to construct
an empirical estimator of escape which relies only on \wlya. However, many of
the high redshift \lya surveys are limited to either massive galaxies (when
pre-selected from broad-band surveys) or to strong \lya emitters (narrow-band
or direct spectral detections), and it is uncertain if the same correlations
and conditions also apply to low-mass galaxies or galaxies with low \fesclya
and high extinction.

The first observations of large scale \lya halos at high redshift were done by
\citet{2011ApJ...736..160S} in a stack of Lyman break galaxies at $z\sim 3$.
The stacked narrow-band images showed extended \lya halos surrounding the
galaxies. The spatial scales are very large compared to observations of
extended \lya in nearby galaxies \citep{2015A&A...576A..51G,2022A&A...662A..64R}, but the 
sample properties and resolution/field-of-view are very different. Other narrow-band and IFU spectroscopy
surveys also show the same, \lya is invariably more extended than the FUV
\citep[e.g.,][]{2014MNRAS.442..110M, 2017A&A...608A...8L} or \ha emission \citep[e.g.,][]{2016MNRAS.458..449M}. 

With observations of low redshifts LAEs giving a more comprehensive
understanding of the effects that radiative transfer has on the \lya output, current
and future \lya surveys will give important constraints on re-ionisation and
the very first galaxies \citep[e.g.,][]{2013MNRAS.428.1366J, 2015MNRAS.449.4336S, 
2020ApJ...892...48R}.  In this work, we present the extension of LARS (eLARS), 
increasing the number of nearby
star-forming galaxies imaged in \lya from 14 to 45 and widening the sample
window to include also galaxies with lower star formation rates.  This
paper is organized as follows. In Section~\ref{sec:data} we describe the data
set used in this work along with details on the data reduction. 
In Section~\ref{sec:methods} we outline
the \lya continuum subtraction method used and also specify the assumptions and
limitations of the code. In addition, this section contains a description on how 
we determine sizes and perform photometry on the various line, continuum, and 
property maps. In Section~\ref{sec:spatial} we present maps of \lya,
FUV continuum and \ha for the 45 galaxies, along with surface brightness
profiles, and demonstrate that the \lya is more extended than FUV/\ha emission
in all galaxies with significant \lya escape.  Section~\ref{sec:globphot}
presents global measurements of \lya and other quantities from our data and
also investigates possible correlations between them. In Section~\ref{sec:disc}
we discuss some implications of our results and how they compare to previous work.
Finally, Section~\ref{sec:summary} summarizes our conclusions and results.
Together with this paper we also release \lya images and accompanying reduced
HST data for the 45 galaxies to the community.

Throughout this paper we use a $\Lambda$CDM cosmology with 
$\{H_0,\Omega_m,\Omega_{\Lambda}\} = 
\{70\mbox{km s}^{-1}\mbox{Mpc}^{-1},0.3,0.7\}$. 

\section{The Data} 
\label{sec:data}

\subsection{Overall survey strategy} 
\label{subsec:strat} 
The observational strategy for LARS is described in detail by
\citet{2014ApJ...797...11O}, and the same strategy for imaging \lya emission
from a combination of broad and narrow-band HST filters is employed for the
extended sample. For clarity we give a brief description of the strategy here.
The basic idea is to use a combination of FUV filters of the
Solar Blind Channel of the Advanced Camera for Surveys (ACS/SBC) on the Hubble
Space Telescope (HST) to measure the flux emitted in nebular \lya emission and
the FUV stellar continuum, and then isolate the emission from the \lya emission
line using continuum subtraction methods. This strategy was employed by
\citet{2003ApJ...597..263K} to obtain the first HST \lya image of a low
redshift galaxy, and later for a sample of six blue compact galaxies
\citep{2005A&A...438...71H, 2007MNRAS.382.1465H, 2009AJ....138..923O}. These
earlier surveys used the F122M filter in the SBC as the \lya on-line filter.
For the LARS galaxies, which are at slightly higher redshift, it was possible
to use a different filter setup, utilizing multiple long-pass filters of the
SBC. In addition to the much lower geocoronal background the higher throughput 
of the long-pass filters combined with having sharp
cut-on profiles on the blue side, means that continuum subtraction using these
filters are more contrast sensitive and can reach lower \wlya levels.

The slope of the stellar continuum spectrum at the wavelength of \lya is highly
sensitive to the age of the stellar population and effective attenuation from
dust (to a lesser extent, also to stellar metallicity). The slope 
for a stellar population producing \lya is not very sensitive to age, but the 
continuum subtraction needs to be perfomed also in regions that do not produce 
\lya photons (due to scattering). 
Hence, without access to multiple filters spanning both blue- and redward of the \lya line, spectral
energy distribution (SED) fitting is needed to accurately estimate the FUV
continuum flux. \citet{2009AJ....138..911H} showed that to securely constrain
the effect on the FUV continuum of stellar ages, mass, and stellar attenuation
at least four continuum data points are needed, spanning the FUV to optical
wavelength range. A further refinement to the method is possible by adding
narrow-band observations of \ha which makes it possible to quantify the
contribution of nebular continuum to the SED (along with \hb to extinction
correct the \ha map). The details of SED fitting and continuum subtraction is
further described in Sec.~\ref{sec:contsub}

To summarize, the imaging observations obtained for the original 14
galaxies and the extended sample are three FUV filters (one containing
\lya emission), three ultraviolet/optical broad band filters, and two optical
narrow-band filters (for \ha and \hb). For three of the galaxies (LARS~13, 
LARS~14, and Tololo~1247-232) only one FUV continuum filter was used.

\subsection{Sample selection} 
\label{subsec:sample}
\begin{figure*}[t!]
\centering
\includegraphics*[width=0.99\textwidth]{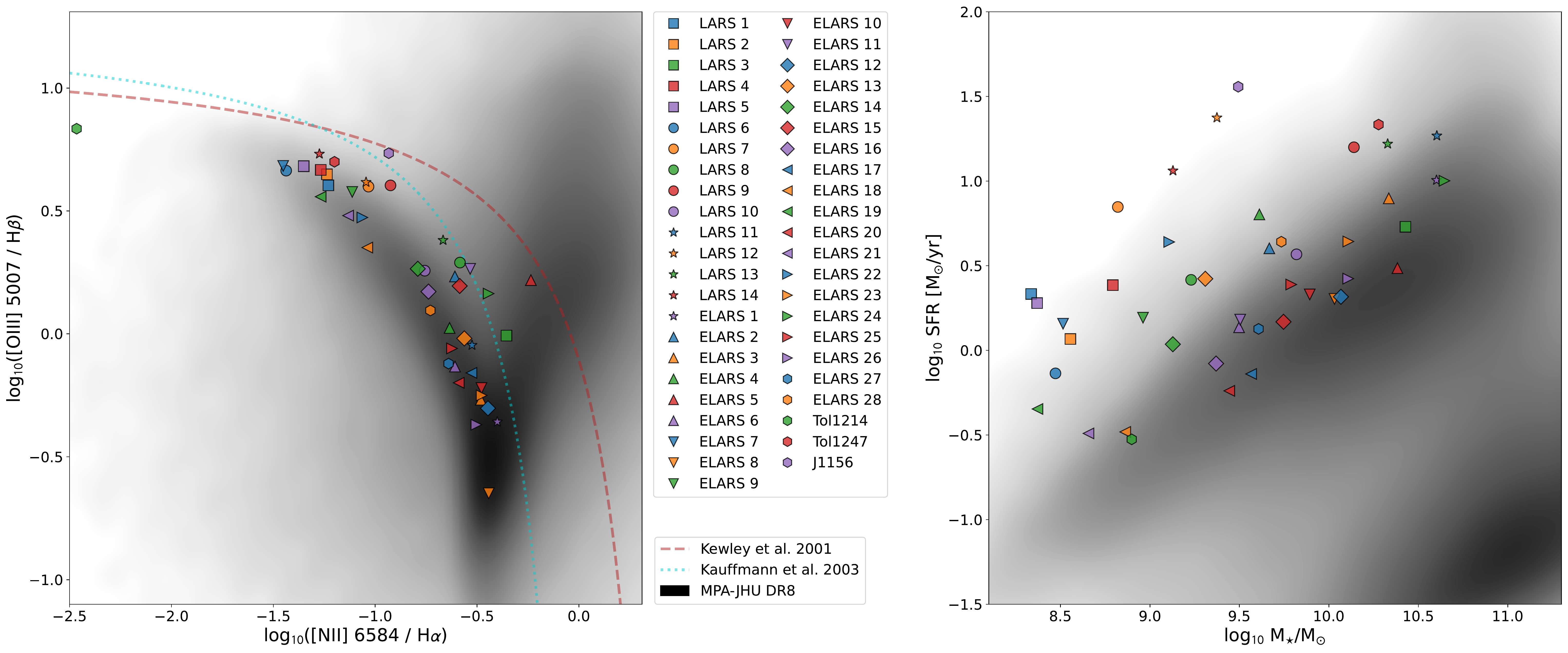}
\caption{Left panel: Emission line diagnostics for the sample. The grey-shaded
region shows the density of galaxies from SDSS DR8. All but one of the galaxies
fall in the region dominated by emission from \hii regions. Right panel: Galaxy
main sequence diagram using data from the JHU-DR8 derivation of stellar masses
and SFRs. Note that Tol~1214 and Tol~1247 do not have SDSS data and emission line
data from other sources (see Table~\ref{tab:sample}) are used. 
LARS~9 is not included
in the MPA-JHU DR8 catalogues, for this galaxy we use SFR and stellar mass from the
Granada Flexible Stellar Population Synthesis analysis of DR14 data
\citep{2016MNRAS.461.1131M}. Note the majority of galaxies 
in the sample are not included in the DR9+ stellar mass and emission 
line property catalogues, which is why we chose to use DR8 for this comparison.\label{fig:sample}}
\end{figure*}

The availability of FUV long-pass filters in SBC sets the upper limit on
redshift of the galaxies that can be targeted with our technique. The longest
wavelength SBC long-pass filter is F165LP, and at least one continuum FUV
filter is needed for accurate continuum subtraction.  This means that the
reddest \lya on-line filter that can be used is F150LP, giving an upper
redshift limit of 0.3 (but note that the longer wavelength SBC filters are less
efficient, which makes it hard to target faint galaxies at high redshift).
Geocoronal \lya emission and Milky Way absorption of \lya puts a lower limit to
the redshift range at around 0.01. The selection of galaxies included in the
original LARS sample is described in detail by \citet{2014ApJ...797...11O}. The
main astrophysical selection criteria are star formation rate (based
FUV luminosity) and stellar age (using \ha equivalent width, \wha).  Given that
the \ha line fluxes are measured in SDSS apertures they will underestimate the
total SFR of larger galaxies. The \wha cut of 100 \AA\xspace in particular
means that the sample only includes galaxies with young stellar
populations that produce \lya photons.  This selection also leads to the sample
being biased towards high specific SFR (irregular and merging systems), and
against the inclusion of high mass galaxies.

For the extended sample we decided to broaden the criteria and include galaxies
with lower specific star formation rate (\wha $>40$ \AA\xspace). Most of these
galaxies are also thus naturally fainter in the FUV and to reach acceptable
signal to noise while optimizing the use of HST exposure time we elected to
observe targets in the continuous viewing zone (CVZ) (at
$\delta \sim 60 \deg$). The longer HST orbit time makes the
observations more efficient, but has the drawback that contamination from the
\oi geocoronal emission is harder to avoid (see Section~\ref{subsec:obs}).
Furthermore, given the overall lower surface brightness of the galaxies we also
chose to only include lower redshift galaxies ($z \lesssim 0.05$) in the
extended sample. The lower redshift cut-off was 0.028 which was selected to get
a throughput for the \lya line of $>85$\% in the F125LP filter.  The redshift
band $0.036<z<0.044$ was excluded to avoid contamination of the
\oiii~4959+5007~\AA\xspace lines in the \hb narrow-band filter. 

With these considerations taken into account we started off by selecting
galaxies from the Sloan Digital Sky Survey (SDSS, DR8) with \wha$>40$\AA\xspace
and matched the coordinates to the GALEX (DR3) catalogues to find FUV fluxes
for the sources. Similar to the original LARS selection we excluded active
galactic nucleus (AGN) galaxies by rejecting galaxies with \ha line widths
(FWHM) larger than 300 \kms (or line of sight velocity dispersion larger than
130 \kms), and with emission line ratio diagnostics from SDSS spectra
(\oiii/\hb and \nii/\ha). In addition we required the galaxies to have Milky
Way extinction of less than 0.03. We then selected galaxies from the remaining
list trying to cover a range of FUV luminosities --- going down to
$\log(L_\mathrm{FUV}/L_\odot)=9.0$ (with $L_\mathrm{FUV}$ from the GALEX FUV
channel, defined as $\lambda\times L_{\lambda}$ at a rest wavelength of
1524~\AA\xspace $/(1+z)$), and morphologies --- compactness, clumpiness, disk
inclination, and spiral arm structure. The final number of galaxies selected
using this method is 26 and are all part of the extended LARS observational
programme.  Compared to the original 14 galaxies this includes more
continuously star-forming spiral galaxies. 

In addition to these 40 galaxies we include five more galaxies in the sample,
which all have the HST imaging required to perform the continuum subtraction.
ELARS~1 (NGC~6090) is a luminous infrared galaxy which was also part of the
\citet{2009AJ....138..923O} sample, but was missing \hb narrow-band imaging.
This galaxy and ELARS~2 (Markarian 65), a compact face-on spiral starburst
galaxy, both fulfil the selection criteria and were already observed in the SBC
long-pass filters as part of HST program 11110 (PI: S. McCandliss). Adding
these galaxies to the sample was thus a cheap addition in terms of
observational time, and they were both part of the extended LARS observational
program. Tololo~1247-232 is a Lyman continuum (LyC) leaking galaxy
\citep{2017MNRAS.469.3252P} and Tololo~1214-277 is a LyC leaker candidate
(\"{O}stlin et al., in prep.). They were both observed with the LARS filter
setup to study the connection between the escape of ionizing photons and \lya.
SDSS~J115630.63+500822.1 is an extreme star burst galaxy at higher redshift
than the rest of the sample and was observed in multiple FUV and optical
filters by \citet{2016ApJ...828...49H} in order to study extended O\,{\sc vi}
emission.

In Table~\ref{tab:sample} we show the basic properties that were used to select
the sample. This table also includes the Galactic extinction
\citep{2011ApJ...737..103S} from the Infrared Science Archive,
NASA/IPAC (DOI:10.26131/IRSA537) along the sightline to the galaxies,
these are very low for all of the galaxies by design.  The physical properties
of the galaxies, derived from the SDSS optical spectra, are given in
Table~\ref{tab:samplespec}.  The SDSS star formation rates (SFR) and masses are
taken from the MPA--JHU catalogue
\citep{2004MNRAS.351.1151B,2007ApJS..173..267S}. These are derived from \ha and
continuum spectral fitting, respectively, and we show the total estimates for
SFR and stellar mass.  For the emission line derived quantities (oxygen
abundance and \oiii~$\lambda5007/$\oii~$\lambda 3727$ ratio), we use the SDSS
based estimates from \citet{2020ApJ...892...48R}. For the three galaxies not
included there we use the references given in the table. Note that, apart from
O32 and metallicity, we do not use the SDSS measurements for the analysis and
comparisons done in this paper. They are presented here for completeness and to
help characterize the sample properties.

The left panel of Figure~\ref{fig:sample} shows the location of the complete
sample of galaxies in a standard BPT diagram \citep{1981PASP...93....5B}. With
the exception of one galaxy, the chosen sources all lie along the star-forming
locus. The outlier (ELARS05) lies in a region of the diagram where galaxies with
a low-ionisation nuclear emission-line region (LINER) normally fall. The
inclusion of this source in the sample was not intended, and was likely caused
by uncertain line fluxes in the selection process. The right panel of
Figure~\ref{fig:sample} shows where the sample falls in a galaxy main sequence
diagram \citep[e.g.,][]{2004MNRAS.351.1151B,2011ApJ...739L..40R}.  The
grey-shaded region is a density map of SFR and stellar mass for nearby galaxies
($z\lesssim 0.2$) from the MPA-JHU analysis of emission line data in SDSS
DR8. As noted above, the values for the LARS galaxies were pulled from the same
catalogue (with a few exceptions given in the figure caption).
While the final sample is not randomly selected it should be noted that no
information on actual \lya emission was used in the process. The complete
sample thus provides an unbiased data set of extended \lya emission from star
forming galaxies. 

\begin{deluxetable*}{lccccccc}[b!]
\tabletypesize{\scriptsize}
\tablecaption{Sample characteristics\label{tab:sample}}
\tablehead{
    \colhead{ID} &
        \colhead{SDSS ID} & 
        \colhead{Redshift} & 
        \colhead{\wha} & 
        \colhead{log(L$_{\mathrm{H}\alpha}$)} & 
        \colhead{log(\lfuv)} & 
        \colhead{$E(B-V)$(MW)} &
        \colhead{Common name} \\ 
    \colhead{} &
        \colhead{} &
        \colhead{} &
        \colhead{(\AA\xspace)} &
	\colhead{(erg/s)} &
        \colhead{(\lsun)} &
        \colhead{} &
        \colhead{} \\
    \colhead{(1)} & 
	    \colhead{(2)} & 
	    \colhead{(3)} & 
	    \colhead{(4)} & 
	    \colhead{(5)} &
	    \colhead{(6)} &  
	    \colhead{(7)} &  
	    \colhead{(8)} 
}
\startdata
LARS01   & SDSS-J132844.05+435550.5 & 0.028 & 560 & 41.43&  9.92 & 0.0144& 	Mrk 259 \\   
LARS02   & SDSS-J090704.88+532656.6 & 0.030 & 312 & 40.77&  9.48 & 0.0152& 	Shoc 240 \\	    
LARS03   & SDSS-J131535.06+620728.6 & 0.031 & 238 & 41.53&  9.52 & 0.0193& 	Arp 238 	\\    
LARS04   & SDSS-J130728.45+542652.3 & 0.033 & 234 & 40.93&  9.93 & 0.0168& 		          \\  
LARS05   & SDSS J135950.92+572622.9 & 0.034 & 333 & 41.36& 10.01 & 0.0099& 	Mrk 1486 	    \\
LARS06   & SDSS-J154544.52+441551.8 & 0.034 & 455 & 40.56&  9.20 & 0.0164& 	KISSR 2019 	\\    
LARS07   & SDSS-J131603.91+292254.0 & 0.038 & 423 & 41.50&  9.75 & 0.0090& 	IRAS 13136+2938 \\
LARS08   & SDSS-J125013.50+073441.5 & 0.038 & 167 & 40.78& 10.15 & 0.0288& 		        \\    
LARS09   & SDSS-J082354.96+280621.6 & 0.047 & 505 & 41.81& 10.46 & 0.0275&  IRAS 08208+2816\tablenotemark{a}\\
LARS10  & SDSS-J130141.52+292252.8 & 0.057 & 99  & 41.15&  9.74 & 0.0101& 	Mrk 0061 	  \\  
LARS11  & SDSS-J140347.22+062812.1 & 0.084 & 105 & 41.54& 10.70 & 0.0216& 		            \\
LARS12  & SDSS J093813.44+542824.9 & 0.102 & 408 & 42.11& 10.53 & 0.0154& 	LEDA 27453 	  \\  
LARS13  & SDSS-J015028.39+130858.4 & 0.147 & 201 & 42.07& 10.60 & 0.0619&  IRAS 01477+1254 \\
LARS14  & SDSS-J092600.41+442736.1 & 0.181 & 578 & 42.21& 10.69 & 0.0161& 		            \\
ELARS01  & SDSS-J161140.81+522727.0 & 0.029 & 215 & 41.68& 10.08 & 0.0172& 	NGC 6090 	   \\ 
ELARS02  & SDSS-J132109.07+590605.3 & 0.043 & 92  & 41.05& 10.03 & 0.0091& 	Mrk 65 	        \\
ELARS03  & SDSS-J114511.42+614230.6 & 0.035 & 65  & 40.99& 10.17 & 0.0186&  UGC 06727 	   \\ 
ELARS04  & SDSS-J172823.84+573243.4 & 0.029 & 103 & 41.06& 10.08 & 0.0333&  NGC 6387 	   \\ 
ELARS05  & SDSS-J110501.98+594103.5 & 0.034 & 41  & 40.95& 9.986 & 0.0062&  SBS 1102+599A \\	
ELARS06  & SDSS-J115840.58+645753.0 & 0.034 & 47  & 40.34& 9.680 & 0.0132&  	\\	            
ELARS07  & SDSS-J101242.97+613302.8 & 0.035 & 268 & 40.92& 9.595 & 0.0105&  		\\            
ELARS08  & SDSS-J103020.91+611549.3 & 0.031 & 48  & 40.49& 9.568 & 0.0088& 		     \\       
ELARS09  & SDSS-J132734.70+664516.6 & 0.030 & 94  & 40.68& 9.563 & 0.0126&  		  \\          
ELARS10 & SDSS-J110504.41+593957.3 & 0.033 & 75  & 40.66& 9.556 & 0.0062&  SBS 1102+599B \\	
ELARS11 & SDSS-J092715.48+583654.4 & 0.030 & 58  & 40.55& 9.507 & 0.0300&  MRK 1417\\ 	    
ELARS12 & SDSS-J130607.16+591302.4 & 0.032 & 46  & 40.57& 9.494 & 0.0136&  SBS 1304+594\\ 	
ELARS13 & SDSS-J105100.65+655940.5 & 0.032 & 125 & 40.85& 9.468 & 0.0103&  VII Zw 348\\ 	    
ELARS14 & SDSS-J112722.07+604454.1 & 0.033 & 82  & 40.58& 9.367 & 0.0103&  SBS 1124+610\\ 	
ELARS15 & SDSS-J134437.82+611424.3 & 0.035 & 54  & 40.44& 9.391 & 0.0148&  MCG +10-20-017\\ 	
ELARS16 & SDSS-J135209.13+560631.4 & 0.035 & 64  & 40.24& 9.333 & 0.0091&  \\		            
ELARS17 & SDSS-J102656.29+584941.6 & 0.031 & 44  & 40.19& 9.243 & 0.0061&  \\	        	    
ELARS18 & SDSS-J152053.59+571122.1 & 0.029 & 59  & 39.93& 9.029 & 0.0105&  \\	        	    
ELARS19 & SDSS-J120506.35+563330.9 & 0.031 & 133 & 40.30& 9.028 & 0.0131&  \\	        	    
ELARS20 & SDSS-J133858.25+614957.6 & 0.031 & 73  & 40.34& 9.021 & 0.0156&  \\	        	
ELARS21 & SDSS-J141911.37+654946.3 & 0.033 & 61  & 39.82  & 9.005 & 0.0110&  \\	        	    
ELARS22 & SDSS-J170912.73+604950.0 & 0.047 & 153 & 41.10& 10.09 & 0.0213& 	SBS 1708+608 \\	
ELARS23 & SDSS-J144827.20+630210.4 & 0.051 & 49  & 40.79& 10.08 & 0.0179& 	\\	            
ELARS24 & SDSS-J144305.31+611838.6 & 0.048 & 83  & 41.51& 9.996 & 0.0193&  \\	             	
ELARS25 & SDSS-J101201.93+603720.3 & 0.045 & 56  & 40.48& 9.833 & 0.0100&  KUG 1008+608 \\	
ELARS26 & SDSS-J115241.69+661827.2 & 0.046 & 43  & 40.50& 9.693 & 0.0100&  \\		            
ELARS27 & SDSS-J150247.62+622018.9 & 0.045 & 47  & 40.32& 9.734 & 0.0106&  \\		            
ELARS28 & SDSS-J115218.74+585657.0 & 0.046 & 131 & 40.86& 9.710 & 0.0255&  \\		
Tol1214 & ---  & 0.026\tablenotemark{b} &1644\tablenotemark{c} & 40.74& 9.008 & 0.0555 & Tololo 1214-277\\
Tol1247 & ---  & 0.049\tablenotemark{d} & 530\tablenotemark{e} & 42.44 & 10.36 & 0.0750 & Tololo 1247-232\\
J1156    & SDSS-J115630.63+500822.1 & 0.236 & 323 & 42.36 & 11.05 & 0.0187&  \\
\enddata
\tablecomments{(3) SDSS spectroscopic redshift (DR8). (4--5) SDSS (DR8). (6) GALEX FUV. 
\lfuv$=\lambda \times L_{\lambda}$/\lsun at $\lambda = 1524/(1+z)$ \AA\xspace. 
(6) Galactic extinction from the Infrared Science Archive, NASA/IPAC (DOI:10.26131/IRSA537) and
\citep{2011ApJ...737..103S}
}
\tablenotetext{a}{Data from SDSS (DR12)}
\tablenotetext{b}{\citet{2001AJ....121..169F}}
\tablenotetext{c}{{\citet{2011AA...529A.149G}}}
\tablenotetext{d}{\citet{2017MNRAS.469.3252P}}
\tablenotetext{e}{{\citet{1991A\string&AS...91..285T}}}
\end{deluxetable*}

\begin{deluxetable*}{lccccc}[b!]
\tabletypesize{\scriptsize}
\tablecaption{SDSS derived properties\label{tab:samplespec}}
\tablehead{
    \colhead{ID} &
        \colhead{log(M$_{\star}$)} &
        \colhead{SFR(\ha)} &
        \colhead{\ebvn} & 
        \colhead{O32} &
        \colhead{$12+\mathrm{log}(O/H)$} \\
    \colhead{} &
        \colhead{(\msun)} &
        \colhead{(\msunyr)} &
        \colhead{} &
        \colhead{} &
        \colhead{} \\
    \colhead{(1)} & 
	    \colhead{(2)} & 
	    \colhead{(3)} & 
	    \colhead{(4)} & 
	    \colhead{(5)} &  
	    \colhead{(6)}   
}
\startdata
LARS01   &  8.34 &   2.15  & 0.149 & $4.1215\pm0.0006$ & $8.2182\pm0.0001$   \\
LARS02   &  8.56 &   1.17  & 0.058 & $4.5481\pm0.0004$ & $8.2186\pm0.0040$   \\
LARS03   &  10.4 &   5.37  & 0.656 & $1.3374\pm0.0001$ & $8.4163\pm0.00001$  \\
LARS04   &  8.79 &   2.43  & 0.116 & $4.3706\pm0.0002$ & $8.1787\pm0.0003$   \\
LARS05   &  8.37 &   1.90  & 0.085 & $6.5808\pm0.0008$ & $8.0747\pm0.0004$   \\
LARS06   &  8.47 &   0.731 & 0.010 & $4.9972\pm0.0009$ & $8.0638\pm0.0351$   \\
LARS07   &  8.82 &   7.04  & 0.235 & $4.3694\pm0.0003$ & $8.3396\pm0.0001$   \\
LARS08   &  9.23 &   2.61  & 0.310 & $1.8424\pm0.0009$ & $8.5055\pm0.00003$  \\
LARS09   &  10.1 &   15.8  & 0.231 & $5.9102\pm0.0003$ & $8.3679\pm0.00003$  \\
LARS10  &  9.82 &   3.69  & 0.237 & $1.4772\pm0.0006$ & $8.5049\pm0.0001$   \\
LARS11  &  10.6 &   18.5  & 0.330 & $0.8637\pm0.0017$ & $8.4277\pm0.0004$   \\
LARS12  &  9.37 &   23.7  & 0.152 & $5.6296\pm0.0004$ & $8.3121\pm0.0006$   \\
LARS13  &  10.3 &   16.6  & 0.298 & $2.6281\pm0.0032$ & $8.5019\pm0.0003$   \\
LARS14  &  9.13 &   11.5  & 0.084 & $8.0942\pm0.0039$ & $7.9919\pm0.0416$   \\
ELARS01  &  10.6 &   10.1  & 0.446 & $0.6193\pm0.0114$ & $8.2278\pm0.0054$   \\
ELARS02  &  9.67 &   4.01  & 0.342 & $1.3851\pm0.0371$ & $8.5022\pm0.0009$   \\
ELARS03  &  10.3 &   7.91  & 0.381 & $0.6950\pm0.0254$ & $8.3670\pm0.0056$   \\
ELARS04  &  9.61 &   6.36  & 0.277 & $0.8170\pm0.0144$ & $8.4838\pm0.0015$   \\
ELARS05  &  10.4 &   3.06  & 0.300 & $1.3531\pm0.0520$ & $8.4560\pm0.0031$   \\
ELARS06  &  9.50 &   1.37  & 0.249 & $0.6497\pm0.0255$ & $8.4442\pm0.0047$   \\
ELARS07  &  8.51 &   1.44  & 0.131 & $5.7574\pm0.1313$ & $8.0152\pm0.0103$   \\
ELARS08  &  10.0 &   2.01  & 0.365 & $0.3456\pm0.0266$ & $8.1079\pm0.0195$   \\
ELARS09  &  8.96 &   1.56  & 0.109 & $2.3410\pm0.0458$ & $8.3176\pm0.0069$   \\
ELARS10 &  9.89 &   2.14  & 0.536 & $0.4988\pm0.0201$ & $8.3688\pm0.0071$   \\
ELARS11 &  9.50 &   1.52  & 0.129 & $1.1299\pm0.0262$ & $8.5014\pm0.0010$   \\
ELARS12 &  10.1 &   2.07  & 0.436 & $0.3831\pm0.0171$ & $8.2901\pm0.0091$   \\
ELARS13 &  9.31 &   2.65  & 0.154 & $1.4057\pm0.0265$ & $8.4589\pm0.0016$   \\
ELARS14 &  9.13 &   1.09  & 0.201 & $1.4566\pm0.0340$ & $8.4981\pm0.0012$   \\
ELARS15 &  9.75 &   1.48  & 0.235 & $1.0751\pm0.0416$ & $8.4984\pm0.0017$   \\
ELARS16 &  9.37 &   0.836 & 0.205 & $1.0130\pm0.0398$ & $8.5064\pm0.0002$   \\
ELARS17 &  9.57 &   0.726 & 0.257 & $0.7062\pm0.0369$ & $8.4000\pm0.0075$   \\
ELARS18 &  8.86 &   0.330 & 0.123 & $1.6726\pm0.0727$ & $8.4300\pm0.0076$   \\
ELARS19 &  8.37 &   0.451 & 0.164 & $3.7944\pm0.0864$ & $8.2358\pm0.0077$   \\
ELARS20 &  9.44 &   0.577 & 0.217 & $0.5578\pm0.0186$ & $8.4096\pm0.0052$   \\
ELARS21 &  8.66 &   0.323 & 0.120 & $2.2992\pm0.1015$ & $8.3374\pm0.0159$   \\
ELARS22 &  9.11 &   4.37  & 0.108 & $2.0325\pm0.0376$ & $8.3930\pm0.0042$   \\
ELARS23 &  10.1 &   4.40  & 0.310 & $0.5118\pm0.0210$ & $8.3488\pm0.0086$   \\
ELARS24 &  10.6 &   10.0  & 0.296 & $0.9139\pm0.0146$ & $8.4588\pm0.0018$   \\
ELARS25 &  9.78 &   2.45  & 0.150 & $0.6726\pm0.0230$ & $8.4685\pm0.0038$   \\
ELARS26 &  10.1 &   2.65  & 0.437 & $0.5232\pm0.0358$ & $8.2764\pm0.0132$   \\
ELARS27 &  9.61 &   1.34  & 0.072 & $0.6335\pm0.0266$ & $8.4553\pm0.0051$   \\
ELARS28 &  9.73 &   4.38  & 0.211 & $1.0799\pm0.0310$ & $8.5035\pm0.0008$   \\ 
Tol1214 &   8.90\tablenotemark{a} & 0.298  &0.000\tablenotemark{b} &  16.607$\pm0.0421$ & 7.2102$\pm0.0200$\tablenotemark{c} \\
Tol1247 &   10.3\tablenotemark{a} & 21.6   &0.158\tablenotemark{d} &  3.7224$\pm0.0209$ & 8.3983$\pm0.0239$\tablenotemark{e} \\
J1156    &  9.49 &  36.1  & 0.130 & 2.1096$\pm0.0191$& 8.4451$\pm0.0076$ \\
\enddata
\tablecomments{(2--3) Total estimates from the SDSS MPA/JHU catalogues \citep{2004MNRAS.351.1151B,2007ApJS..173..267S}. (4) Derived using \hahb from SDSS (DR8). 
(5) O32 = \oiii$\lambda$5007\AA/(\oii$\lambda$3727\AA + \oii$\lambda$3729\AA) from \citet{2020ApJ...892...48R}. (6) Derived using O3N2 strong line metallicity calibration of \citet{2007AA...462..535Y} \citep{2020ApJ...892...48R}.}
\tablenotetext{a}{From HST SED fitting (LaXs)}
\tablenotetext{b}{\citet{2001AJ....121..169F}}
\tablenotetext{c}{based on line fluxes from \citet{2011AA...529A.149G}}
\tablenotetext{d}{\citet{2017MNRAS.469.3252P}}
\tablenotetext{e}{based on line fluxes from \citet{1993MNRAS.260....3T}}
\end{deluxetable*}

\subsection{Observations} 
\label{subsec:obs}
All observations were carried out using HST with three different cameras:
ACS/SBC, ACS/Wide Field Camera (WFC), and Wide Field Camera 3 (WFC3)/UVIS. 
A list of which filters were used for the
individual targets can be found in Table~\ref{tab:obs}.  While we used 
archival data when available
the bulk of the data set consists of observations from HST program 12310
\citep[][the original LARS, observed from 2010--2012]{2014ApJ...797...11O},
and HST program 13483 (ELARS, observed from 2013--2014).  The observations for
Tol 1214-277 come from HST program 14923 ({\"O}stlin et al., in
prep.), for Tol 1247-232 from HST program 13027
\citep{2017MNRAS.469.3252P}, and for J1156 from HST program 13656 \citep[][
Rutkowski et al., in prep]{2016ApJ...828...49H}. All of the data presented 
in this paper were obtained from the Mikulski Archive for Space Telescopes 
(MAST) at the Space Telescope Science Institute. The specific observations 
analyzed can be accessed here:
\dataset[10.17909/4s6m-7255]{http://dx.doi.org/10.17909/4s6m-7255}.

The FUV continuum and \lya images were obtained with SBC. The FUV filter set
was chosen to avoid contamination from geocoronal \lya, but the O\,{\sc
i}~$\lambda\,1302$\,\AA\xspace line is strong enough to severely compromise the
F125LP observations. The exposures in this filter were thus done with the
telescope in Earth shadow. For the extended sample, which used a CVZ
observational setup, this is harder to manage but the timing of the F125LP
exposures was planned to mimimize the time spent observing outside of shadow.
We do not detect any significant change in the background between the original
and the extended sample.

The typical exposure times in the \lya filter were $3 \times
600$s, and $3 \times 400$ in the continuum filters.  
Using three sub-exposures with small offsets for the SBC observations was done to
cover the gap in the middle of the detector, and, to a smaller extent, limiting the effect of
bad pixels in the detector.

The optical continuum of the galaxies was imaged in three filters with
WFC3/UVIS (for some of the archival observations the ACS/WFC was used instead),
in general using two blue (F336W and F438W) and one red (F775W) filter. For the
ELARS observations with UVIS post-flash was used to limit the effect of charge
transfer efficiency (CTE) losses. The typical exposure time for the optical
continuum filters was $2 \times 400$s.  The choice of using only two dithers
was motivated by data transfer limitations when packing observations in
multiple filters into a single HST orbit. It should be noted that this limits
what pixel scale can be achieved when drizzling the data, and how well cosmic
rays are rejected (see Section~\ref{subsec:red}). Given that the individual
galaxies are at different redshifts the narrow-band filters used to capture \ha
and \hb photons vary. When possible we used UVIS narrow-band filters, but for
some of the targets (the higher redshift galaxies in particular) ACS/WCS ramp
filters were used instead. The typical exposure time in \ha is $2\times
400$s., and $2\times 600$s. in \hb. 

Most of the galaxies are at similar redshift, which means that the cosmological
surface brightness dimming is a small effect. The necessary
exposure times are thus very similar despite a quite large range in distances.
However, three of the targets (LARS~13, 14, and J1156) are at significantly higher
redshift. For these galaxies the exposure times were scaled up to account for
part of the surface brightness dimming, but the rest-frame depth reached is somewhat reduced.

\subsection{Data reduction}
\label{subsec:red}
\begin{figure*}[t!]
\centering
\includegraphics*[width=0.95\textwidth]{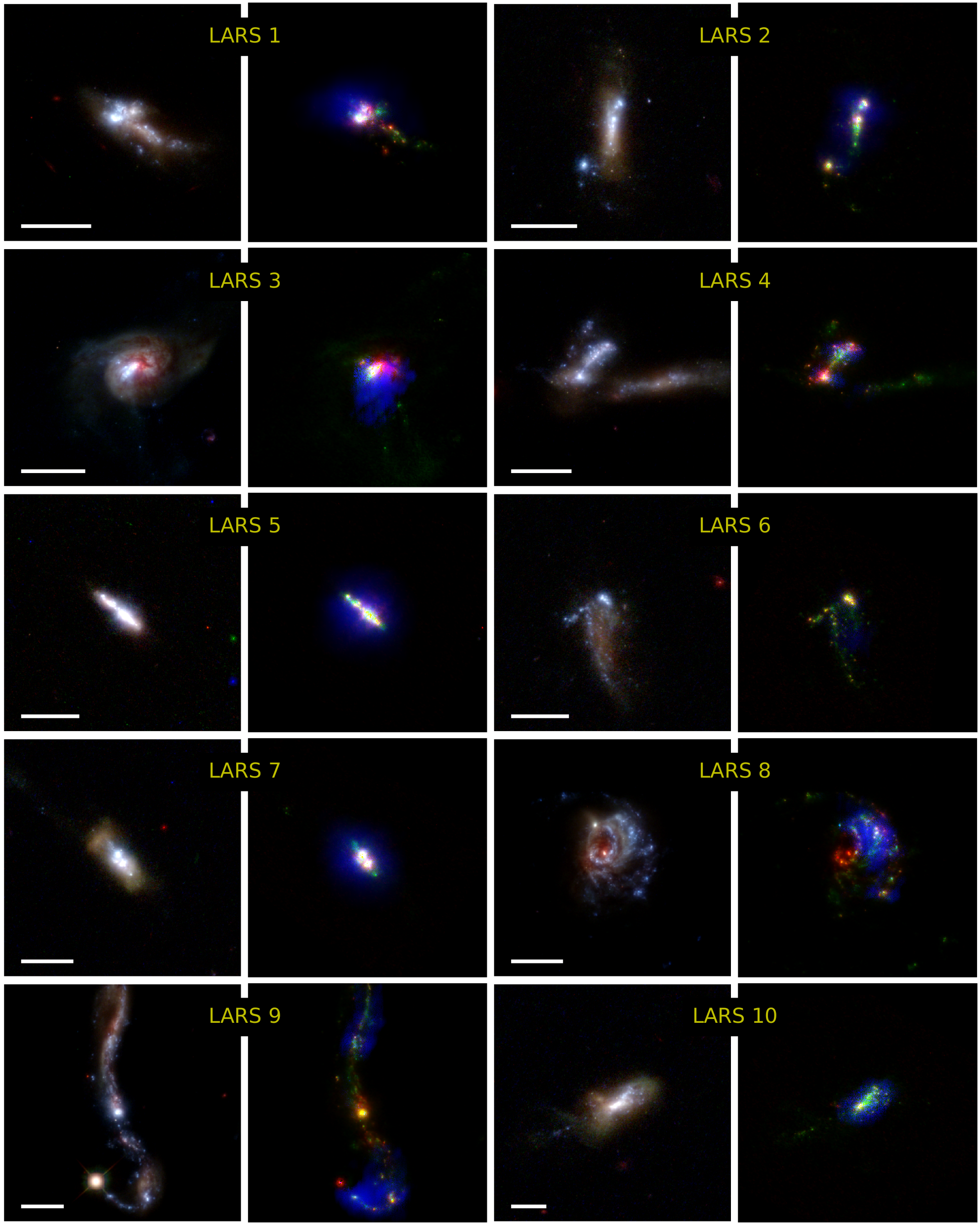}
\caption{Color composites for the LARS~1--10 galaxies. Two panels are shown for each 
galaxy with the ID number connecting the corresponding panels. The left sub-panel 
for each galaxy shows a stellar continuum composite using the HST I, B, and U band 
images for the RGB channels. The scale bar in these panels has a length of 5 kpc.
The right panel for each galaxy shows a composite of red (H$\alpha$), blue (Ly$\alpha$), 
and green (stellar continuum in the FUV at the wavelength of Ly$\alpha$).  
The Ly$\alpha$ maps have been adaptively smoothed to enhance the
contrast and be able to show the low surface brightness parts of the galaxy.
\label{fig:rgbs1}}
\end{figure*}

\begin{figure*}[pt]
\centering
\includegraphics*[width=0.95\textwidth]{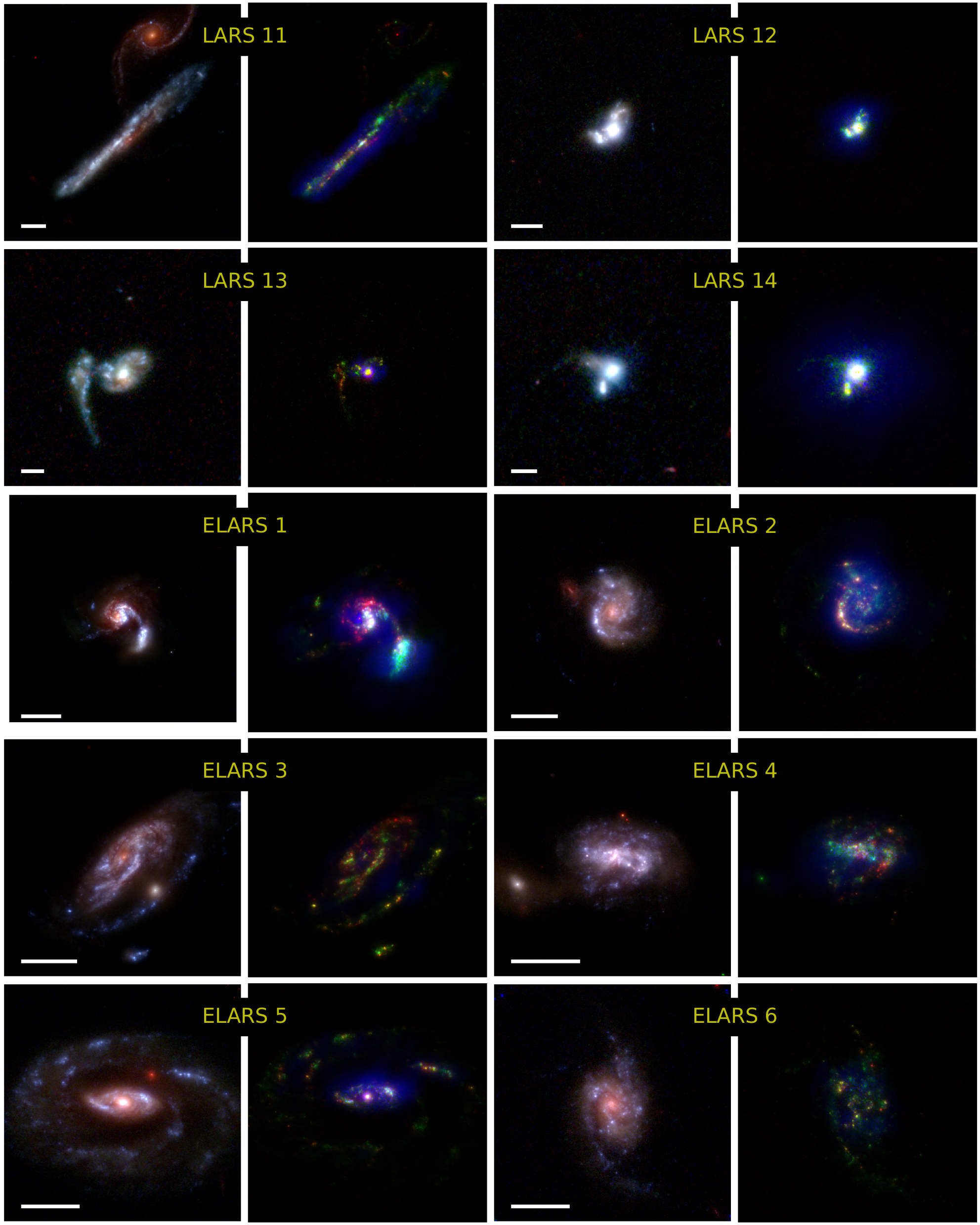}
\caption{Color composites for the LARS~11--14 and ELARS~1--6 galaxies. See the caption for 
Figure~\ref{fig:rgbs1} for details. 
\label{fig:rgbs2}}
\end{figure*}

\begin{figure*}[pt]
\centering
\includegraphics*[width=0.95\textwidth]{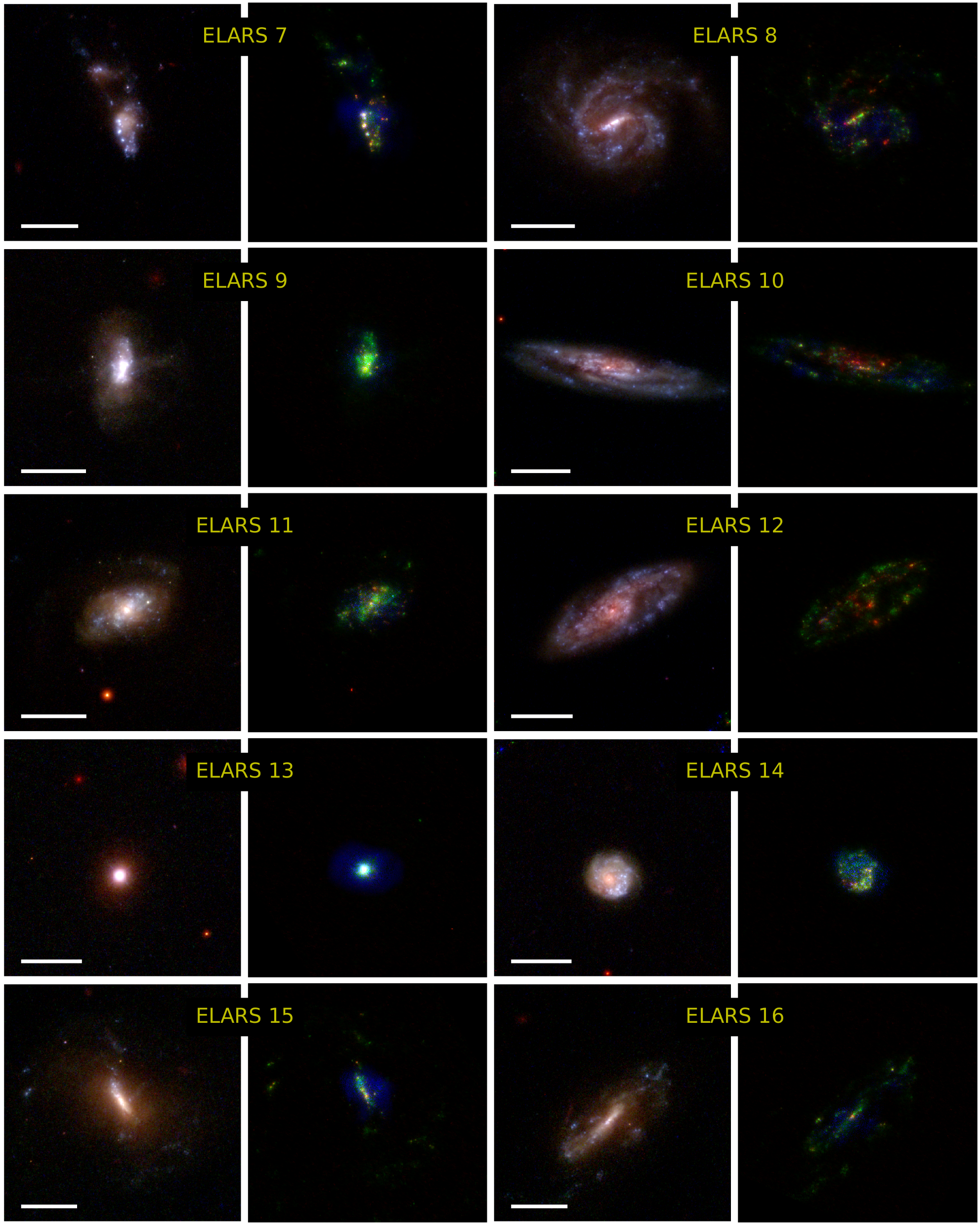}
\caption{Color composites for the ELARS~7-16 galaxies. See the caption for 
Figure~\ref{fig:rgbs1} for details. 
\label{fig:rgbs3}}
\end{figure*}

\begin{figure*}[pt]
\centering
\includegraphics*[width=0.95\textwidth]{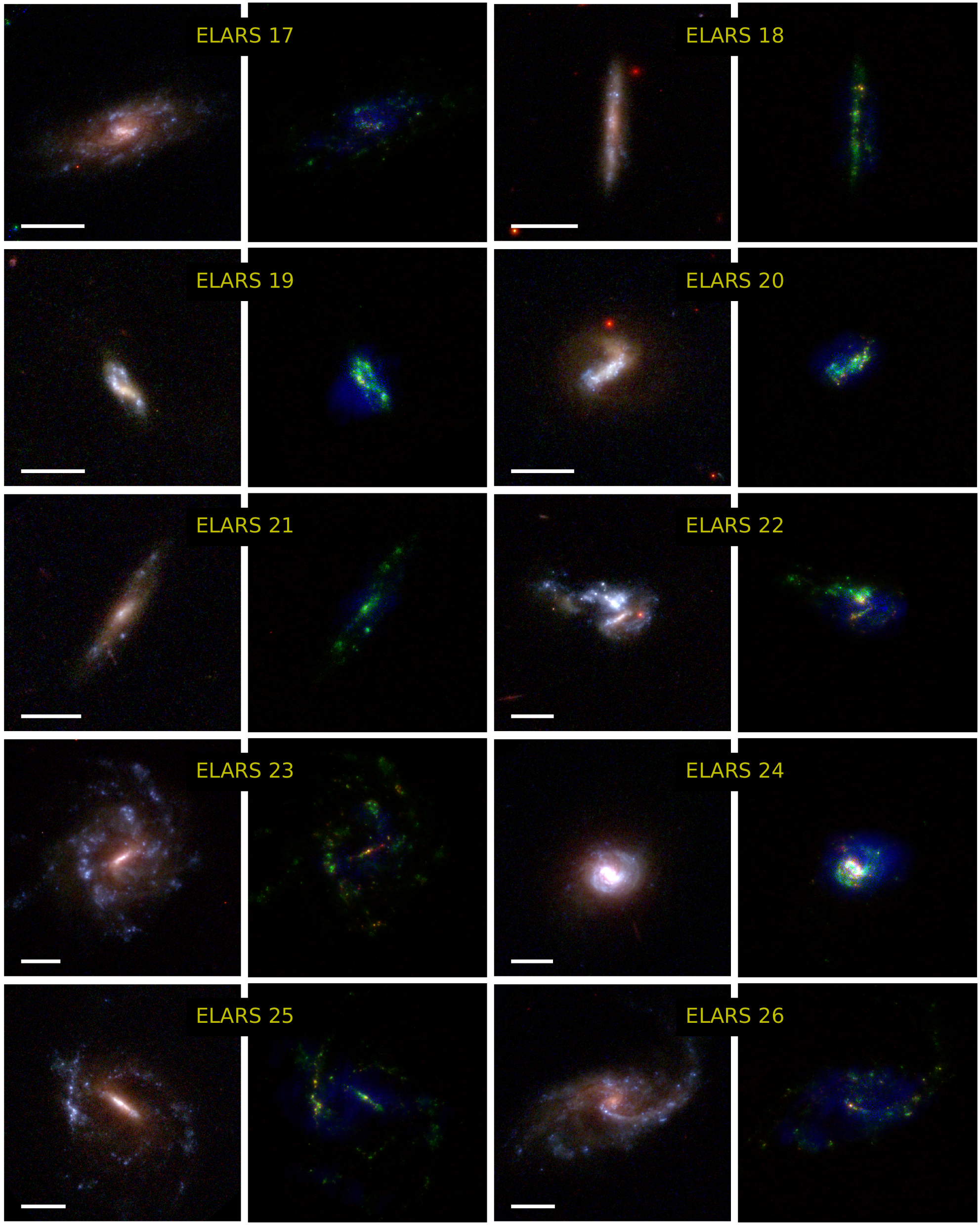}
\caption{Color composites for the ELARS~17-26 galaxies. See the caption for 
Figure~\ref{fig:rgbs1} for details. 
\label{fig:rgbs4}}
\end{figure*}

\begin{figure*}[t]
\centering
\includegraphics*[width=0.95\textwidth]{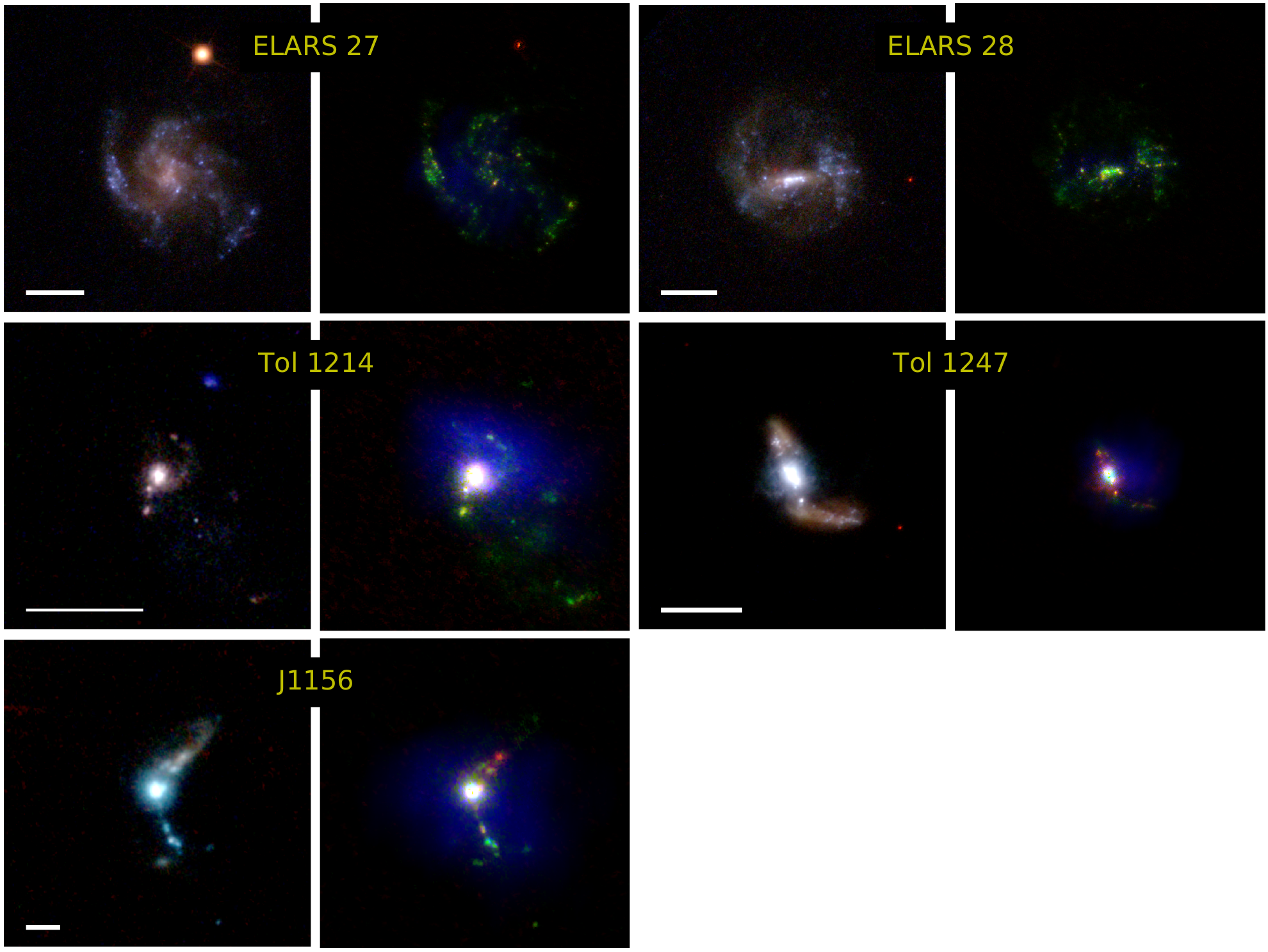}
\caption{Color composites for the ELARS~27--28, Tol~1214, Tol~1247, and J1156 galaxies. See the caption for 
Figure~\ref{fig:rgbs1} for details. 
\label{fig:rgbs5}}
\end{figure*}

Basic image reduction (bias and dark current subtraction, flat-fielding, and
post-flash subtraction where applicable) was done with the standard HST
reduction pipelines, and individual reduced image frames were downloaded from
the Mikulski Archive for Space Telescopes (MAST).  In \citet{2014ApJ...797...11O} full-frame CTE
correction was done with the stand-alone tool available at the time, and only
for the ACS/WFC images. From 2016 full-frame CTE corrected data has been available
directly from MAST and we used this corrected data for both WFC3/UVIS and
ACS/WFC data (ISR WFC3 2016-002).

The optical observations use only two dithers per filter which has one major
consequence for data reduction. The cosmic ray rejection performed during
drizzling of the data fails to mask out many of the affected pixels. Before
drizzling we thus performed an additional rejection using the LACosmic software
\citep{2001PASP..113.1420V}. We implemented the LACosmic routines in a python
wrapping script that flags the found pixels in the data quality layer of each
individual frame.  The flags are then recognized during drizzling and data
affected by cosmic rays are excluded.  The individual filters were registered
and stacked using the standard HST drizzling software
\citep[Drizzlepac;][]{2012drzp.book.....G} to a common pixel
scale of 0.04~\arcsec/pxl. We also produced drizzled weight maps for each 
filter that correctly reflect the total noise per pixel (inverse variance maps).

Sky subtraction is normally handled by the drizzle routine, but for the ACS/SBC
data this subtraction is not able to fully remove large scale backgrounds. 
For the FUV filters we thus remove sky
from the drizzled frames by estimating an average sky level well outside the
galaxy (usually in the corners of the images). In some cases the sky background 
in the F125LP filter (because of geocoronal \oi emission, see Section~\ref{subsec:obs})
varies substantially (by factors up to 10) between exposures, in this 
case we subtract the sky from each
individual frame before drizzling. It should be noted that we perform this
background subtraction also for the galaxies that may have \lya emission to the
edge of the chip (discussed further in Section~\ref{subsec:syserr}).  For the
optical images, the sky subtraction done during drizzling is normally
sufficient. However, for some of the galaxies the ACS/WFC ramp filters --- and
the WFC3/UVIS quad filters --- show a residual background gradient which may be
caused by problems due to reflected background light in the detector. 
For the frames affected by this we fitted and removed a spatially varying 
background from each drizzled frame.

The observations in the different filters for each galaxy are done in multiple
HST orbits at different times (however, all observations for a given galaxy and
filter are observed in the same orbit).  Because of this the guide star used is
not necessarily the same for all filters which leads to world coordinate system (WCS)
offsets of up to 0.5\arcsec\ between the drizzled images. Registering images to
a common WCS before drizzling using standard techniques (e.g., Drizzlepac
TweakReg) does not work for the SBC images. Instead we chose to register the
images by using cross-correlation techniques ({\sc pyraf/xregister}). The
reference image for the registration was the F438W filter (or equivalent). The
average registration uncertainty is around 0.2 pixels ($\sim 0.008$\arcsec),
but never exceeds 0.5 pixels. After registration we also inspect all of the
images to make sure that the reduction steps have been successful, and to
construct a mask for each galaxy which marks areas not covered by all filters
(particularly important for SBC which has a smaller field of view than the
optical filters). These masks are then used for the continuum subtraction and
photometry. Color composites of the galaxies using the U,
B, and I (HST equivalent bands), showing the stellar continuum are shown in
Figures~\ref{fig:rgbs1}, \ref{fig:rgbs2}, \ref{fig:rgbs3}, \ref{fig:rgbs4}, and
\ref{fig:rgbs5}.
 
In order to perform accurate continuum subtraction from pixel-based
SED fitting) the effect of the points spread
function (PSF) varying between filters needs to be handled. While the PSF
differences in the optical filters are quite small, the PSFs of the SBC filters
(among these the \lya filter) have very extended wings \citep[see, e.g.][ISR
ACS 2016-05]{2016ApJ...828...49H} which makes them differ substantially from
the optical filters. If PSF matching is performed without including the
extended wings, the stellar continuum from the optical filters is
under-estimated relative to the FUV filters. When subtracting the continuum from the 
\lya filter too little flux will be subtracted in the wings, causing the \lya surface 
brightness to be over-estimated in regions of the galaxy not dominated by a
strong point-like source. To study low surface brightness \lya halos it is thus
critical to correctly constrain not only the core of the PSF but also the
wings, and match the PSFs of all filters to a common reference. 

We start by constructing 2D PSF models for all filters and galaxies. For the
optical filters we use {\sc TinyTim} \citep{2011SPIE.8127E..0JK} to make
models, but for the SBC images these models are not accurate enough. The SBC
PSFs were instead constructed by stacking stars observed in the stellar cluster
NGC~6681 during calibrations. For each galaxy we then made a maximum width 2D
PSF model by normalizing the models for each filters to the same peak flux and
then stacking them by maximum pixel value. Because of the non-standard shape of
the SBC PSF matching cannot be done with a gaussian-based convolution kernel,
which makes this a non-trivial task. Instead we match the images in individual
filters to the maximum width model by convolving with a delta-function-based
kernel \citep[similar to the technique described in][]{2012MNRAS.425.1341B}.  A
detailed description of how we modeled the PSFs for the different cameras and
filters, and how we performed PSF matching for the data is given in
Appendix~\ref{app:psfmatch}.

In addition to the drizzled science images in each filter, the data reduction
pipeline also produces weight frames. We use these weight frames to estimate
the noise per pixel in each filter: $\sigma = 1 / \sqrt(W)$, with $\sigma$
denoting the standard deviation map (i.e., noise) and $W$ the weight map.
Because of the drizzling the pixel-to-pixel noise in the images is correlated
\citep{2002PASP..114..144F}, but the noise map still represents the best
possible estimate of the uncertainty for each individual pixel we can find.
Nevertheless, it is important to note that errors on spatially integrated
quantities estimated directly from the noise maps will underestimate the true
noise. We perform a full Monte Carlo (MC) simulation of the continuum
subtraction and photometry, using the noise maps as input (see
Section~\ref{sec:staterr}). However, it should be noted that Voronoi binning
will create binned noise maps that will underestimate the noise per bin and
that the resulting errors used in the MC are thus still underestimated. Because
of the complicated nature of the how the Voronoi variable binning and noise
correlations interact we have not tried to rescale the noise, thus the
resulting uncertainties are underestimated by factors ranging between $\sim 0$
(for the maximum $25^2$ Voronoi binsize) to $\sim 40$\% (for the minimum
binsize of 1 pixel). The $\chi^2$ fitting performed by LaXs (see
Section~\ref{sec:contsub}) will be affected by this, with the effect that the
fits appear to be slightly worse than they are. The global apertures (and
annuli used for surface brightness profiles) are much larger than the maximum
bin size and the error estimates for them will thus not be directly affected by
the correlated noise.

\section{Analysis methods}
\label{sec:methods}
\subsection{Ly$\alpha$ continuum subtraction} 
\label{sec:contsub}
With the drizzled and PSF-matched images in hand we can perform the pixel SED
fitting necessary to obtain emission line maps for \lya, \ha, and \hb. In
addition to modeling the continuum underneath these lines the fitting
software also produces maps of the best-fit parameters (stellar mass, age, and
stellar extinction) and derived properties (e.g., nebular extinction,
mechanical energy yields, ionizing flux) for the stellar populations and gas in
each pixel. The continuum subtraction is done with the Lyman $\alpha$
eXtraction Software, written in \texttt{python}. This code was used for the
results presented in \citet{2014ApJ...782....6H} and \citet{2014ApJ...797...11O},
but development of the code has continued since then and we therefore describe
the code in detail below. LaXs is run on both un-binned and Voronoi tessellated 
data but below we will use the term pixel also to describe the individual bins.

\subsubsection{The Lyman $\alpha$ eXtraction Software, LaXS}
\label{subsec:laxs}
We fit models to the measured continuum using $\chi^2$ minimisation (i.e.,
weighted least-squares minimisation) on a semi-discrete grid of four
parameters. 
\begin{equation}
    \label{eq:xi2}
    \chi^2_j= \sum_{i}^{N} \frac{\left(F_{i,j} - 
        M_{i}(m_y,m_o,t,E(B-V))\right)^2}
        {\sigma_{i,j}^2}
\end{equation}
For each pixel, denoted by $j$
above, we have measured continuum fluxes ($F_{i,j}$) and associated errors 
($\sigma_{i,j}$) from four or five filters, denoted by $i$ with $N=4$ or 5.  
The model fluxes for each of
these filters ($M_{i}$) are calculated  from stellar population
models that depend on a number of parameters, some that are varied in the
fitting, and some that are not. 

The standard models used here are Starburst99 \citep{1999ApJS..123....3L} single stellar
population (SSP) templates which use a Kroupa IMF \citep{1993MNRAS.262..545K}. 
For the model SEDs we use the low resolution 
($\Delta \lambda \sim 10$ \AA)
templates at an assumed metallicity and without nebular continuum. The metallicity 
used is limited by the available Starburst99 templates and we choose the one closest to the 
measured SDSS gas metallicities (given in Table~\ref{tab:samplespec}). 
We then calculate a grid of template SEDs for a discrete set of stellar ages
($t$) and extinction ($E(B-V)_s$). We perform testing to find the most appropriate
extinction curve for each galaxy (further detailed in Section~\ref{subsec:syserr}).
The parameter ranges and step sizes can be
found in  Table~\ref{tab:laxsparams}. The spectral templates for each age step
is first reddened using the $E(B-V)_s$ steps and the flux in each observed
filter calculated by using \texttt{pysynphot} and filter transmission curves
for each continuum filter. The resulting template library contains
$\mathrm{N}_{\mathrm{age}}\times \mathrm{N}_{\mathrm{ext}}$ SEDs that are
normalized to a stellar mass of $10^6 M_{\odot}$. The observed fluxes in the 
Balmer line filters are also corrected for stellar absorption depending on the 
age of the models. This absorption is calculated using the Starburst99 
high resolution stellar templates.

The model SED that is compared to the observations, $M_i$, is built from
two stellar populations of this type, one old with fixed age of 1 Gyr and no
extinction, and one young with a varying age and extinction. The final two free
parameters in the fitting are the masses for the two components which are
continuous parameters. The full model SED is thus:
\begin{equation}
    \label{eq:sedmodel}
    M_i = m_o \times S(t_o,0) + m_y \times S(t,E(B-V)_s), 
\end{equation}
where $m_o$ and $m_y$ are the old and young population masses, respectively, in
units of $10^6 M_{\odot}$, $S(t,E(B-V))$ is the individual SED template for a
given age and extinction, and $t_o$ is the age of the old stellar population.

The minimisation is then performed by calculating the $\chi^2$ for each SED
model (each $\{t, E(B-V)_s \}$ pair) where we also solve a linear least squares
problem to get the optimum normalisation (i.e., stellar masses). The best-fit
stellar masses are thus not restricted to a discrete grid. The code is
parallellized using the python package
SCOOP~\footnote{\url{https://github.com/soravux/scoop}}
\citep{10.1145/2616498.2616565}, which --- due to each pixel being independent
from another --- offers a speedup in execution time close to the number of
threads available. The initial best-fitting SED model is then selected as the
one which minimizes the $\chi^2$.

From this initial run we use the \ha and \hb line maps to produce a dust
corrected \ha map.  The nebular continuum and line contribution to all observed
continuum filters and pixels is then estimated by scaling the nebular continuum
spectrum from \citet{1984ASSL..112.....A} combined with nebular lines from 
\citet{2003A&A...401.1063A} with this \ha map, again using
\texttt{pysynphot} to calculate the flux in each filter. The resulting nebular
continuum fluxes are subtracted from the observed filters and the entire
fitting redone. The fraction of nebular continuum to observed flux in \ha-detected 
regions range from $\sim 1$\% in the FUV to $\sim 10$\% in the red optical filter, 
but for the brightest line emitting regions (and particularly in the galaxies 
with highest specific SFR) the nebular contribution can be even higher.
By forcing the nebular continuum to scale with the dust
corrected \ha this method provides the most secure determination of nebular
continuum for the limited wavelength range.  Furthermore, by virtue of directly
observing both \ha and \hb we can treat the nebular continuum as a semi-independent
quantity that does not have to be directly determined from the stellar continuum
fitting (but note that the \ha+\hb continuum does come from the fitting).

\begin{deluxetable*}{lcccc}[t!]
\tabletypesize{\scriptsize}
\tablecaption{LaXs parameters\label{tab:laxsparams}}
\tablehead{
    \colhead{Type} &
        \colhead{Parameter} & 
        \colhead{Range} & 
        \colhead{Steps/\# tests} & 
        \colhead{Uncertainty, \lya/FUV} \\
    \colhead{(1)} & 
	    \colhead{(2)} & 
	    \colhead{(3)} & 
	    \colhead{(4)} & 
	    \colhead{(5)}  
}
\startdata
Free  & $m_o$   & $>0$   & Continuous   & N/A   \\
{   } & $m_y$   & $>0$   & Continuous   & N/A   \\
{   } & $t$     & $10^6$--$10^{10}$ yrs & 200 & N/A \\ 
{   } & $E(B-V)$& 0--1.0 & 200  & N/A \\
\hline
	Fixed & Attenuation & CCM, SMC, Calz.\tablenotemark{a} & 3 & 10/1 \% \\
{    }& Z$_\star$      & 0.001, 0.004, 0.008, 0.020\tablenotemark{b} & 2/3 & 5/0.5 \% \\
{    }& $t_o$          & $10^9$--$10^{10}$ yrs & 10 & $\ll$1/0.1 \% \\
{    }& SFH            & SSP+SSP, CSF+SSP, SSP+CSF, CSF+CSF& 4 & $\ll$1/0.1 \% \\
{    }& SP library     & Starburst99, BPASS & 2 & $\ll$1/0.1 \% \\
\enddata
\tablecomments{(1) The type is either Free, for parameters that are fitted, or
Fixed, for the parameters which are kept constant. (3) For the Free parameters
this is the number of steps in the fitting grid, for the Fixed parameters it is the
number of tests performed on the systematic uncertainties. (5) This is the standard
deviation of the total flux in the high-SNR region for \lya and the \lya
continuum. We have not explored the systematic uncertainties on using a particular range for 
the free parameters and they are thus listed as not available. Statistical uncertainties of 
the free parameters are available from the Monte Carlo runs.}
\tablenotetext{a}{CCM --- {\citet{1989ApJ...345..245C}}, SMC --- {\citet{1984AA...132..389P}}, 
	Calz. --- {\citet{2000ApJ...533..682C}}.}
\tablenotetext{b}{The stellar metallicities tested are given by the available
	stellar population spectra in Starburst 99. For most galaxies we test
	three different metallicities, the nominal one (the closest match
	between Starburst 99 models and SDSS gas metallicities), and the ones
	just below and above.}

\end{deluxetable*}

\subsubsection{Statistical error estimates from Monte Carlo simulations}
\label{sec:staterr} 
To investigate the statistical errors of the continuum subtracted emission line
maps (and the SED fit parameter maps) we perform Monte Carlo simulations using
100 realisations. The input data for each of these (fluxes in all filters and
pixels) are generated from a Gaussian distribution with a standard deviation
from the noise maps. LaXs is then run for each of these realisations and produces cubes
of maps (containing the full output of all the 100 Monte Carlo realisations), 
along with maps of standard deviation, skewness and other
statistical diagnostics. As previously mentioned, the continuum subtraction is
computationally expensive, which means that we can only estimate statistical
uncertainties for the binned maps, and that the number of Monte Carlo
realisations that can be used is limited.  However, tests on individual
galaxies show that increasing the number of realisations to 1000 does not
change the estimated uncertainties significantly.
The full Monte Carlo cubes are used to estimate uncertainties in all integrated 
quantities presented in this paper (see Sections~\ref{subsec:profiles} and \ref{sec:globphot}). 

\subsubsection{Systematic error estimates}
\label{subsec:syserr}
In addition to statistical errors the continuum subtraction will also suffer
from systematic uncertainties which originate in the assumptions made when
building the SED templates. In addition to the four free parameters, there are
five major fixed parameters (or assumptions) that go into the modeling. These
are star formation histories (SFHs) for the two populations, stellar attenuation law,
age of the old stellar population, metallicity, and stellar population template library.

We estimate the systematic uncertainties from these five fixed parameters by
running LaXs in a high signal-to-noise region in each galaxy (SNR$>10$ in all
filters). We vary each of these parameters on a grid while keeping all the
other parameters fixed. In Table~\ref{tab:laxsparams} we show the test ranges.
Old stellar population age ($t_o$) can be chosen freely but only ages that are
older than the young population component are tested.  It should be noted that 
because the UV brightest parts of the galaxies are used as test regions the old population
age assumption will have very little effect on the output, but given that we are interested in the
FUV and \lya systematic uncertainties this is the chosen method.
The nominal SFHs used
are two single stellar populations (burst models, SSPs) for both the young
(with varying age) and old populations. We also test combinations of a burst
model and a constant star formation (CSF) model. 
The extinction curves tested are 
CCM \citep{1989ApJ...345..245C}, SMC \citep{1984AA...132..389P}, and Calzetti \citep{2000ApJ...533..682C}.

The resulting uncertainties
are reported in Table~\ref{tab:laxsparams}, where the numbers given are the
relative standard deviations (i.e., the standard deviation divided by the
nominal case fluxes), averaged over the full sample, of the total FUV and \lya flux 
in the high SNR region for the different
assumed cases. We find that the assumption giving rise to the highest
systematic uncertainty is the choice of extinction curve, which has an
associated uncertainty of $\sim 10$\% and $\sim 1$\% for \lya and the modeled
FUV continuum, respectively. The assumed metallicity is the second most
important assumption with associated \lya/FUV uncertainties of $\sim 5$\% and
$\sim 0.5$\%. It should be noted that this error is a conservative upper limit,
given that the metallicity grid is very sparse and that we have a robust
metallicity determination from the SDSS spectroscopy. Of course, this case
could also be interpreted as testing the effects of metallicity gradients
within the galaxies on the continuum subtraction. The remaining fixed parameters
show much smaller \lya\ systematic uncertainties ($\ll 1$\%).

The systematic uncertainty tests are also used to choose the extinction curve to
use for each galaxy. For each galaxy, the results for the three curves 
are compared by co-adding the best-fit $\chi^2$ inside the high SNR region
in each case. We then do a simple model selection, choosing the attenuation law
that has the lowest co-added $\chi^2$.

For the faintest levels of \lya emission there is an additional systematic
error resulting from the subtraction of geocoronal emission in the F125LP
filter (see Section~\ref{subsec:red}). For galaxies with wide-spread \lya
emission that fills the SBC chip the estimated background will thus include
\lya, which causes our maps to systematically underestimate the flux to an
uncertain degree. However, comparing the removed background to average levels
of geocoronal emission (using the ACS/SBC exposure time calculation tool and
assuming a low level of geocoronal contamination) shows that geocoronal \oi
emission is dominating the background for most of the galaxies (31 out of 45).
The remaining 14 galaxies have larger (and unknown) systematic uncertainties on
\lya luminosities, \wlya, and \fesc; and have been marked in
Table~\ref{tab:globprop}.  However, it should be noted that inspecting the
individual raw SBC data frames shows that most of these images have a flat
background that varies significantly from frame to frame, which would not be
the case for true \lya emission from the galaxy. Three of the galaxies (LARS04,
ELARS04, and ELARS08) are large, have a subtracted background significantly
higher than the ETC estimate, and do not show any variation in the background.
These three galaxies could thus suffer from substantial oversubtraction of
\lya. Assuming that 100\%/10\% of the subtracted background is \lya emission
for these three galaxies results in the \lya fluxes (as well as \wlya and
\fesc) being underestimated by factors 2.43/1.14, 1.86/1.09, and 3.94/1.29 for
LARS04, ELARS04, and ELARS08, respectively.  More details on how the unknown
background affects size and \lya distribution measurements in galaxies is given
in \citet[][]{2022A&A...662A..64R}. Furthermore, the uncertainty of the removed
background is also a source of systematic error but is negligible compared to
the statistical errors in the binned maps.

\begin{figure*}[t!]
\centering
\includegraphics*[width=0.95\textwidth]{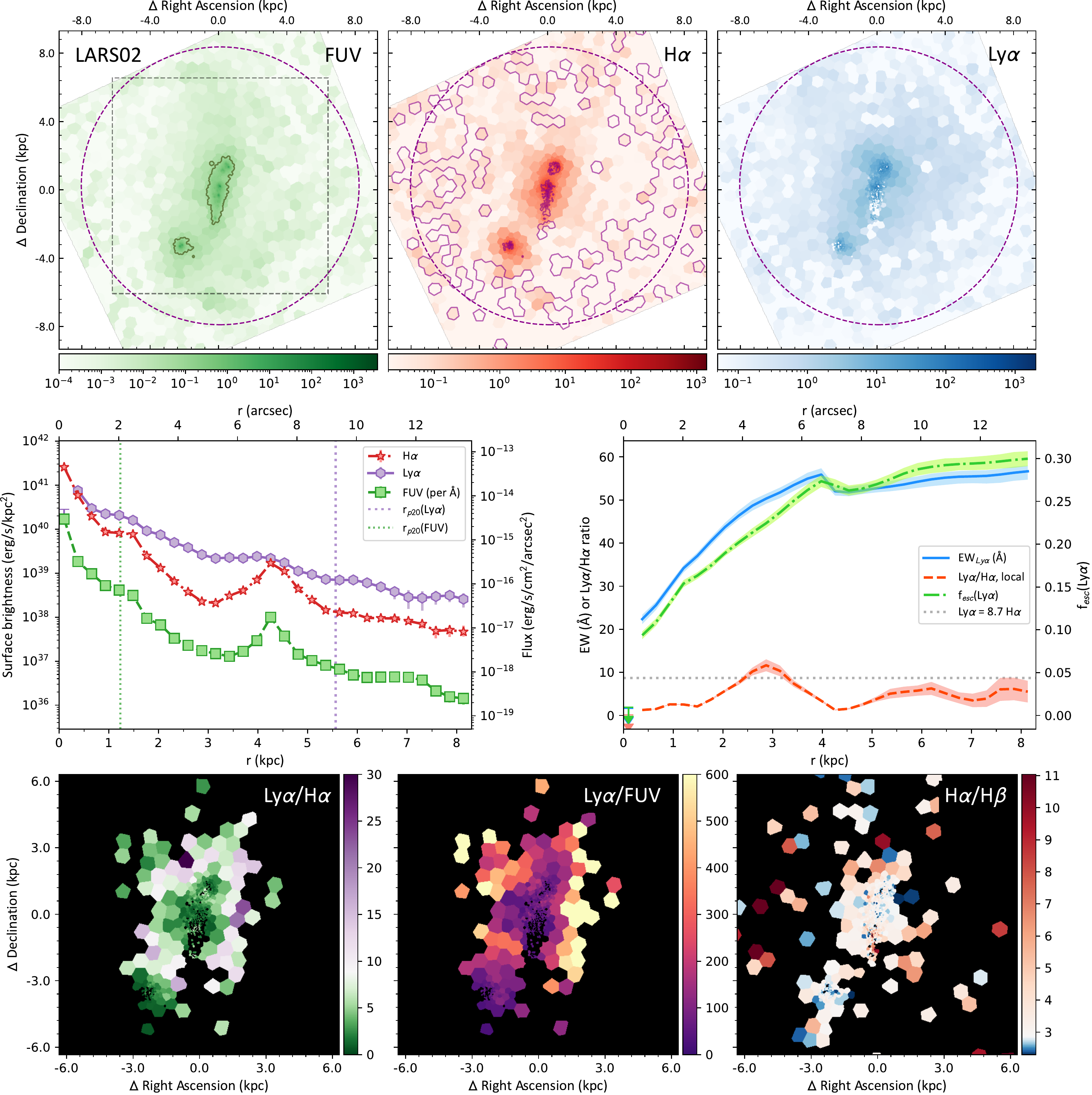}
\caption{Compilation of measurements on LARS02 (example of a strong \lya emitter). 
The maps and profiles are all
constructed from Voronoi tesselated data. The top three panels show maps
of (from left to right): stellar FUV continuum, \ha emission, and \lya
emission.  In the FUV map the dashed grey box shows the cut-out box used
for the line ratio maps and the dark-green contour shows the Petrosian
isophotal limit (corresponding to $r_{\mathrm{p, FUV}}$).  The purple contours in
the \ha map shows three logarithmically spaced \lya levels. The purple dashed
circle in the maps indicate the aperture (with radius
$r_{\mathrm{tot}}$ used to obtain the total estimates of fluxes and other
properties. The center of this circle is the central point for the surface
brightness profile. The color bar for these panels is in units of $10^{-17}$ \ergc 
(for FUV) and \ergl (for line maps).
The middle left panel shows surface brightness
profiles in FUV, \ha, and \lya. The upper limit symbols are placed at a
surface brightness of two times the positive error (estimated from the 
Monte Carlo generated maps) for each data point.
The middle right panel shows line ratio profiles. Note that these are
cumulative profiles in all cases but for \lya / \ha.  The lower panels
contain line ratio maps for a cut-out region, from left to right: \lya / \ha,
\lya / FUV (\wlya), and \ha / \hb.  Bins with SNR$<2$ in the input line maps 
have been masked in the ratio maps and are shown in black. The complete figure 
set (45 images) is available in the online journal.}
\label{fig:mpanel_l02}
\end{figure*}

\begin{figure*}[pt!]
\centering
\includegraphics*[width=0.95\textwidth]{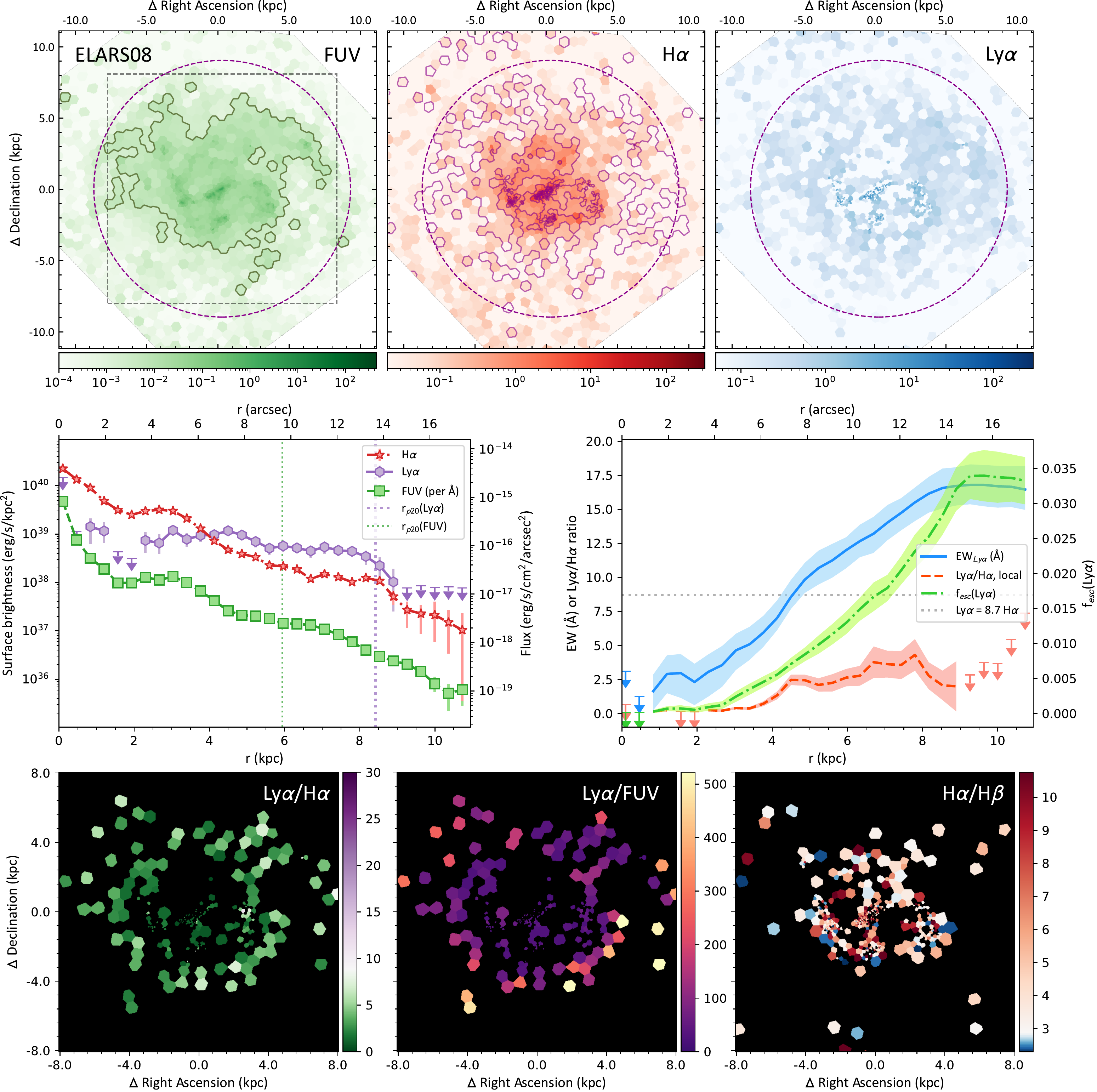}
\caption{Compilation of measurements on ELARS08 (example of an intermediate 
	\lya emitter). A description of the elements in this figure is given in 
	Figure~\ref{fig:mpanel_l02}. The complete figure 
	set (45 images) is available in the online journal.}
\label{fig:mpanel_el08}
\end{figure*}

\begin{figure*}[pt!]
\centering
\includegraphics*[width=0.95\textwidth]{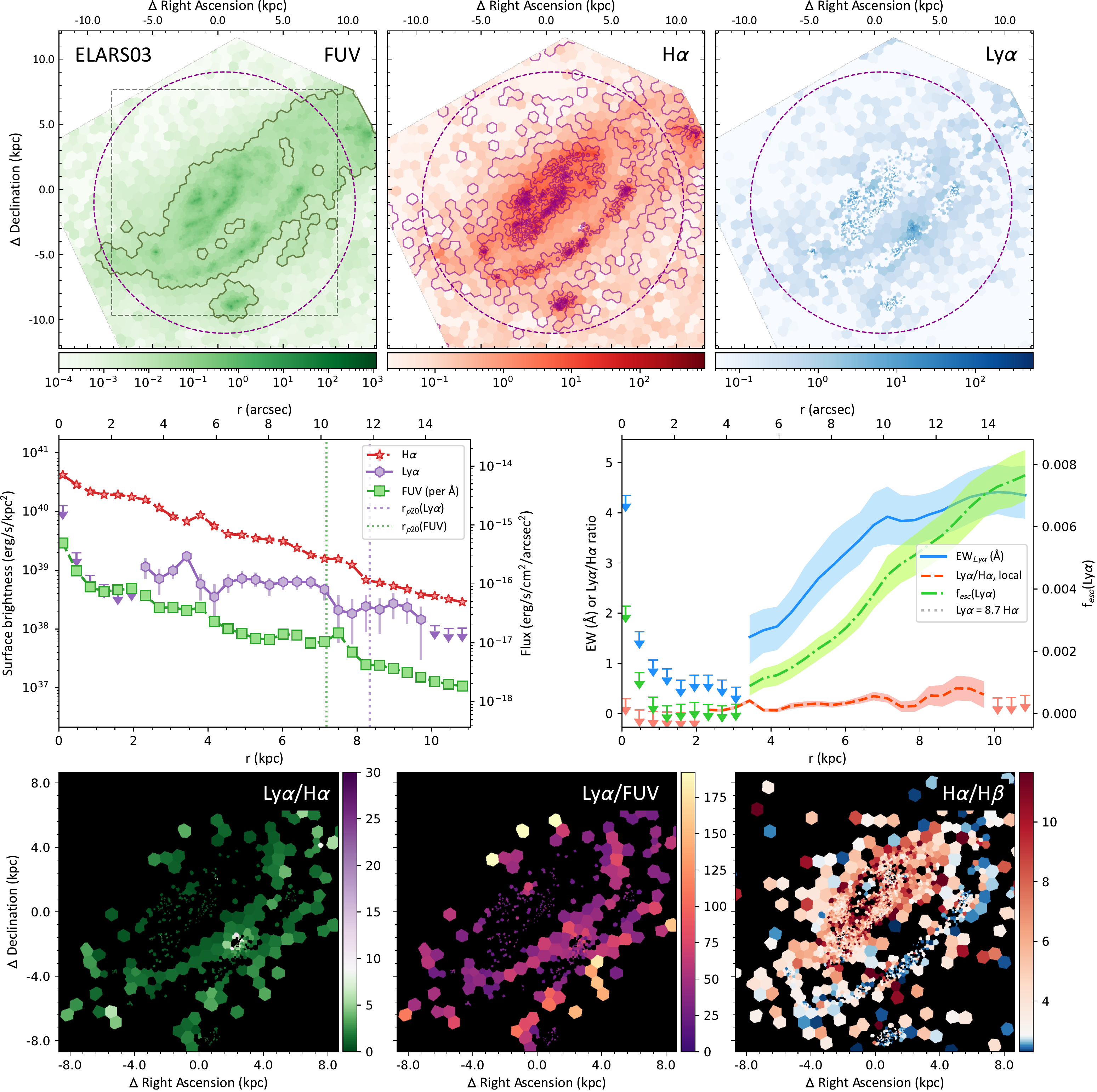}
\caption{Compilation of measurements on ELARS03 (example of a weak
	\lya emitter). A description of the elements in this figure is given in 
	Figure~\ref{fig:mpanel_l02}. The complete figure 
	set (45 images) is available in the online journal.} 
\label{fig:mpanel_el03}
\end{figure*}

\subsection{Photometry and size measurements}
\label{sec:phot}
We measure the fluxes in the \lya, \ha, and FUV maps, as well as in some of the
parameter maps output from LaXs (young population stellar mass, young
population stellar age, $E(B-V)_s$). The parameter maps do strictly speaking
not contain fluxes, but in the following we will refer to them as such to
facilitate the description. All of the measurements are done on the Voronoi binned maps.
To define the regions used for the surface
brightness profile (or local estimates of the parameters above) we use two
different methods. 

\subsubsection{Circular photometry}
\label{sec:circ}
The first method is to use circular annuli centered on the brightest pixel in the FUV
map. The largest radii for this estimate is given by the largest possible
circle that can be fit within the field-of-view (also using the mask described
in \ref{subsec:red}), and the smallest radius is 0.5 kpc.  The number of bins
is set to 30 which is a good compromise between SNR per bin and resolution. For
the most distant galaxies (LARS13, LARS14, and J1156) we use logarithmic
spacing of the bins, while for the rest of the galaxies a linear spacing is
used. We have varied the number of bins (between 25 and 35 bins) to check the
systematic errors on sizes and other spatial measurements, but the changes are
small. Using a significantly smaller number of bins will of course increase the
SNR per bin greatly, but this comes at the cost of not being able to
characterize the spatial variation.

With the bins defined we use the python package \texttt{photutils} to measure
fluxes in the annuli of all maps. The measured fluxes are all converted into
luminosities using the redshifts from SDSS (see Table~\ref{tab:sample}).  We
compute rest frame surface brightness for \lya, \ha, and FUV in each bin by
dividing the measured luminosity by the annulus area.  For the surface
brightness profiles the radius associated to each bin is the radius that divides
the annulus into two equal areas ($r = \sqrt{r_{\mathrm{inner}}^2 +
r_{\mathrm{outer}}^2}/2$). For the measurements in the stellar age and
$E(B-V)_s$ maps we use luminosity weighted (from the FUV images) averages to
make sure that low-SNR Voronoi bins are down-weighted. Note that this weighting
scheme means that there is a bias towards lower average ages and extinction. 

We also compute curves (annulus and aperture) for derived quantities \wlya,
\lyaha, $E(B-V)_n$, and \fesc. Nebular extinction is calculated assuming
an intrinsic \hahb ratio of 2.86 (case B recombination, $T=10^4 K$, $n_e =
100$cm$^{-3}$), and using the \citet[][hereafter CCM]{1989ApJ...345..245C}
extinction law.  
The escape fraction is calculated by assuming an intrinsic \lyaha ratio of 8.7
\citep[][and references therein]{2015ApJ...809...19H} and using the measured
$E(B-V)_n$ for the same region: 
\begin{equation} 
f_{\mathrm{esc}} = \frac{F_{\mathrm{Ly\alpha}}}{8.7 F_{\mathrm{H\alpha}} 
	\cdot 10^{-0.4 k(\mathrm{H\alpha}) E(B-V)_n}  },  
\label{eq:fesc}
\end{equation} 
where $k(\mathrm{H\alpha})$ is the extinction coefficient at the \ha wavelength
The CCM extinction curve assumes a dust screen model for the nebular emission.
Opting to use the attenuation curve of \citet{2000ApJ...533..682C} instead for
both the calculation and application of $E(B-V)_n$ in the expression above
leads to somewhat lower escape fractions. For example, a \hahb ratio of $\sim$4
(corresponding to a Calzetti $E(B-V)_n$ of 0.29) leads to a factor 1.06 lower
escape fraction, but this rises with \hahb ratio.  The difference is quite
small for our galaxy sample and we adopt CCM extinction for all nebular line
measurements in this paper. Note that the estimated \fesc (or $E(B-V)_n$) does not 
change if the SMC extinction curve is used because the CCM and SMC curves are very similar at the 
\ha and \hb wavelengths.

For determining the global nebular extinction for the galaxies the \hb image 
limits how large an aperture can be used. Because of the dependence on the
\hahb ratio of $E(B-V)_n$, including non-detected data points for \hb leads to
huge errors on the extinction. To obtain a safer estimate of $E(B-V)_n$ we find
the radius where the \hb SNR drops below one. For
calculating cumulative \fesc profiles we set the $E(B-V)_n$ outside of
this radius to the value at that radius. For the edge-on galaxy
LARS11, \hb is too faint all over the galaxy, and we cannot obtain a meaningful
$E(B-V)_n$ from HST imaging. For this galaxy we instead use a constant
nebular extinction from SDSS (see Table~\ref{tab:sample}) for all radii.

The local escape fraction (i.e. escape fraction within an annulus) of \lya
photons is very hard to interpret because of the resonant scattering of the
photons.  The same applies to \wlya, although this is a direct ratio of
observed quantities and is thus less model-dependent. 
Hence we do not measure
\fesc and \wlya for each annulus but only for the apertures. 
However, we do
measure \lyaha ratios for each annulus which can be used to show where
scattered \lya dominates.

\subsubsection{Isophotal photometry}
\label{sec:iso}
We also perform isophotal photometry on the binned \lya, \ha, and FUV maps. The
number of isophotes used is 30 (variations between 25 and 35 still show little
effect on the found size estimates), and we use logarithmic spacing from the
maximum flux in the map to the 1$\sigma$ surface brightness limit. The 1$\sigma$
limit is computed from the Monte Carlo error simulations for each map using a
sigma-rejection averaging routine (to find the limiting surface brightness well outside
the source). 

Each isophotal bin is then treated as a radial bin for surface brightness
profiles with the effective radius for the region calculated as the radius of a
circle with the same area as the isophotal region (i.e. the area for all pixels
with flux higher than the lower limit of the current bin).  The isophotal
measurements are used to measure the Petrosian radii in the maps, but are
otherwise not used in this paper. 

\subsubsection{Size measurements}
\label{sec:sizes}

\begin{deluxetable}{lcccc}
\tabletypesize{\scriptsize}
\renewcommand{\arraystretch}{1.3}
\tablecaption{Size measurements\label{tab:sizes}}
\tablehead{
    	\colhead{ID} &
	\colhead{$r_{tot}$} & 
	\colhead{$r_{p, FUV}$} & 
	\colhead{$r_{p,\mathrm{H}\alpha}$} & 
	\colhead{$r_{p,\mathrm{Ly}\alpha}$} \\    
	\colhead{} &
	\colhead{(kpc)} & 
	\colhead{(kpc)} & 
	\colhead{(kpc)} & 
	\colhead{(kpc)} \\
	\colhead{(1)} & 
	\colhead{(2)} & 
	\colhead{(3)} & 
	\colhead{(4)} & 
	\colhead{(5)} 
}
\startdata
LARS01  & $\ge$6.70 & 1.490$^{0.006}_{0.005}$ & 1.570$^{0.008}_{0.006}$ & 3.520$^{0.038}_{0.072}$ \\   
LARS02  & $\ge$8.13 & 1.230$^{0.019}_{0.021}$ & 1.620$^{0.406}_{0.022}$ & 5.560$^{0.171}_{0.388}$ \\   
LARS03  & $\ge$7.63 & 1.180$^{0.097}_{0.054}$ & 1.100$^{0.004}_{0.004}$ & 8.610$^{0.197}_{0.346}$ \\   
LARS04	& 3.28 & 4.200$^{0.056}_{0.063}$ & 2.000$^{0.021}_{0.017}$ & 1.680$^{2.502}_{0.174}$ \\        
LARS05  & $\ge$8.37 & 1.090$^{0.002}_{0.003}$ & 1.490$^{0.009}_{0.006}$ & 3.140$^{0.092}_{0.138}$ \\   
LARS06  & $5.45$\tablenotemark{a} & 3.660$^{0.293}_{0.144}$ &2.630$^{0.737}_{0.074}$ & 0.540$^{0.488}_{0.854}$ \\    
LARS07  & $\ge$7.75 & 1.080$^{0.008}_{0.011}$ & 1.060$^{0.005}_{0.004}$ & 4.250$^{1.186}_{0.130}$ \\   
LARS08  & $\ge$7.45 & 5.230$^{0.088}_{0.052}$ & 4.460$^{0.069}_{0.069}$ & 10.100$^{0.563}_{4.616}$ \\  
LARS09  & $\ge$12.86 & 6.840$^{0.055}_{0.042}$ & 5.590$^{0.041}_{0.038}$ & 10.740$^{0.202}_{0.825}$ \\ 
LARS10  & 5.72 & 2.770$^{0.234}_{0.192}$ & 2.680$^{0.037}_{0.049}$ & $\ge$10.960 \\                    
LARS11  & 20.22 & 9.010$^{0.160}_{0.069}$ & 8.770$^{0.050}_{0.044}$ & 19.170$^{6.353}_{6.154}$ \\      
LARS12  & 19.39 & 2.000$^{0.009}_{0.006}$ & 2.520$^{0.035}_{0.027}$ & 5.720$^{0.434}_{0.333}$ \\       
LARS13  & $\ge$26.59 & 0.800$^{0.172}_{0.048}$ & 7.190$^{0.016}_{0.017}$ & 28.050$^{4.534}_{7.637}$ \\         
LARS14  & $\ge$37.62 & 0.930$^{0.010}_{0.018}$ & 2.370$^{0.012}_{0.010}$ & 4.340$^{0.270}_{0.607}$ \\  
ELARS01 & $\ge$6.85 & 1.530$^{0.004}_{0.004}$ & 2.510$^{0.017}_{0.023}$ & 2.070$^{0.061}_{0.054}$ \\   
ELARS02 & 7.40 & 4.190$^{0.018}_{0.022}$ & 3.310$^{0.051}_{0.065}$ & 4.060$^{0.313}_{0.337}$ \\        
ELARS03 & 10.03 & 7.180$^{0.035}_{0.046}$ & 6.660$^{0.020}_{0.023}$ & 8.340$^{0.276}_{0.334}$ \\       
ELARS04 & $\ge$9.65 & 3.170$^{0.017}_{0.012}$ & 3.000$^{0.007}_{0.008}$ & 7.430$^{0.082}_{0.123}$ \\   
ELARS05 & $\ge$10.74 & 8.280$^{0.149}_{0.134}$ & 8.600$^{0.000}_{0.001}$ & 8.850$^{0.129}_{0.079}$ \\  
ELARS06 & 9.08 & 5.360$^{0.179}_{0.167}$ & 6.960$^{0.120}_{0.198}$ & 9.120$^{0.105}_{0.303}$ \\        
ELARS07 & 10.23 & 3.700$^{0.104}_{0.121}$ & 0.430$^{0.021}_{0.019}$ & 8.050$^{0.913}_{3.982}$ \\       
ELARS08 & 9.00 & 5.940$^{0.095}_{0.161}$ & 7.660$^{0.033}_{0.031}$ & 8.420$^{0.506}_{0.386}$ \\        
ELARS09 & 5.63 & 1.390$^{0.014}_{0.013}$ & 1.360$^{0.020}_{0.015}$ & 7.270$^{0.440}_{0.135}$ \\        
ELARS10 & $\ge$8.59 & $\ge$9.340 & 3.720$^{0.020}_{0.020}$ & 8.860$^{3.757}_{0.270}$ \\                
ELARS11 & 5.52 & 2.770$^{0.053}_{0.017}$ & 2.650$^{0.011}_{0.013}$ & 6.530$^{0.546}_{0.285}$ \\        
ELARS12 & 6.28 & 4.100$^{0.170}_{0.069}$ & 4.140$^{0.005}_{0.003}$ & 6.870$^{0.638}_{0.133}$ \\        
ELARS13 & 3.40 & 0.110$^{0.000}_{0.000}$ & 0.530$^{0.014}_{0.010}$ & 0.210$^{0.018}_{0.011}$ \\        
ELARS14 & 4.14 & 2.080$^{0.054}_{0.079}$ & 1.750$^{0.002}_{0.003}$ & $\ge$7.280 \\                     
ELARS15 & 8.07 & $\ge$9.460 & 8.790$^{0.000}_{0.000}$ & 8.610$^{4.129}_{1.398}$ \\                     
ELARS16 & 4.13 & 5.180$^{1.371}_{0.541}$ & 7.300$^{0.000}_{0.000}$ & 0.280$^{1.944}_{0.780}$ \\        
ELARS17 & 7.49 & 6.090$^{0.000}_{0.000}$ & 6.850$^{0.000}_{0.000}$ & 8.480$^{0.121}_{0.267}$ \\        
ELARS18 & 5.77 & 2.560$^{0.143}_{0.058}$ & 6.240$^{0.000}_{0.000}$ & $\ge$6.750 \\                     
ELARS19 & 4.33 & 1.800$^{0.082}_{0.077}$ & 6.700$^{0.038}_{0.163}$ & 7.030$^{0.417}_{0.099}$ \\        
ELARS20 & 4.04 & 1.890$^{0.025}_{0.023}$ & 7.510$^{0.000}_{0.000}$ & $\ge$6.400 \\                     
ELARS21 & 3.23 & 3.360$^{0.000}_{0.000}$ & 7.220$^{0.000}_{0.000}$ & $\ge$6.660 \\                     
ELARS22 & 9.74 & 4.090$^{0.138}_{0.134}$ & 10.100$^{0.000}_{0.000}$ & 10.230$^{0.232}_{1.546}$ \\      
ELARS23 & 12.87 & 10.640$^{0.298}_{0.207}$ & 10.680$^{0.071}_{0.090}$ & 11.780$^{0.406}_{0.219}$ \\    
ELARS24 & 11.05 & 2.720$^{0.077}_{0.057}$ & 2.070$^{0.258}_{0.028}$ & 5.390$^{0.262}_{0.567}$ \\       
ELARS25 & $\ge$12.31 & 8.830$^{0.415}_{0.270}$ & 9.110$^{0.000}_{0.000}$ & 11.980$^{0.356}_{2.611}$ \\ 
ELARS26 & 11.87 & 8.190$^{0.460}_{0.198}$ & 14.320$^{0.000}_{0.000}$ & 11.040$^{0.389}_{0.709}$ \\     
ELARS27 & $\ge$8.18 & 6.620$^{0.150}_{0.214}$ & 9.110$^{0.000}_{0.000}$ & 9.200$^{0.336}_{0.230}$ \\   
ELARS28 & 7.38 & 6.170$^{0.084}_{0.087}$ & 9.040$^{0.000}_{0.000}$ & 9.730$^{1.437}_{0.154}$ \\        
Tol1214 & $\ge$5.92 & 0.470$^{0.007}_{0.008}$ & 0.520$^{0.013}_{0.010}$ & 1.960$^{0.805}_{0.070}$ \\   
Tol1247 & $\ge$10.50 & 0.690$^{0.003}_{0.004}$ & 2.460$^{0.008}_{0.007}$ & 4.560$^{0.117}_{0.099}$ \\  
J1156   & 26.64 & 1.470$^{0.020}_{0.248}$ & 5.930$^{0.078}_{0.022}$ & 26.060$^{6.400}_{0.540}$ \\      
\enddata
\tablenotetext{a}{\lya is undetected in all circular annuli in this galaxy, the radius used for photometry and shown here is the maximum radius to the edge of the chip.}
\renewcommand{\arraystretch}{1}
\end{deluxetable}
We measure the sizes of the galaxies in two different ways. Apart from giving
information on the spatial extent of the emission, these sizes can also be used
to define global apertures. The size measurements for all galaxies are shown in 
Table~\ref{tab:sizes}.

\textbf{Petrosian radii, $r_{\mathrm{p}}$:} We calculate 20\% Petrosian radii
\citep{1976ApJ...209L...1P} in \lya, \ha, and FUV using the SDSS definition
from \citet{2001AJ....121.2358B}.  To get a more precise measurement of the
radius we perform spline interpolation on the surface brightness profiles. As
mentioned above these radii are calculated using isophotal photometry and are
used to compare the size of \lya emission and FUV
stellar continuum of the galaxies, and to 
estimate specific star formation rates and star formation rate surface densities.

\textbf{Total \lya aperture, $r_{\mathrm{tot}}$:} Given that \lya is often more extended
than \ha and FUV, global apertures based on petrosian radii in those maps fail
to capture all of the \lya emission from the galaxies. Apertures based on the
isophotal photometry are problematic to use when based on \lya, because of \lya
absorbed regions in the galaxies (see Section~\ref{subsec:images}). Another option 
would be to use elliptical apertures which would possibly capture more of the \lya 
emission from galaxies with asymmetric \lya halos. We have performed some tests, 
but, because of the irregular nature of the \lya spatial distribution in the galaxies, 
getting acceptable elliptical annuli requires extensive finetuning of the fitted 
elliptical isophotes. Also, given that we want to compare \lya to other flux maps 
which often show quite different morphology, measurements from elliptical apertures 
will have have similar systematic uncertainties as circular apertures.  
We thus conclude that the safest estimate of global \lya flux is found by using a large circular
aperture centered on the FUV brightest pixel as described above. 

The radius of this global aperture is defined as the radius where the annulus
signal-to-noise ratio (SNR) of \lya surface brightness drops below 1. If this
happens at more than one radii we chose the largest radii where this condition
is fulfilled. This condition is necessary because of the patchiness of \lya
emission and the problems with continuum absorption described above.
Furthermore, if the \lya SNR stays above one all the way out to the maximum
radius, the total aperture is defined as the circle with this radius. In these
cases the total radius is a lower limit to how far out the detectable \lya
extends. Given that the Monte Carlo derived uncertainties are used to derive 
$r_{\mathrm{tot}}$ we cannot estimate errors for this radius.

\subsection{Line ratio maps}
\label{subsec:linerats}
For the discussion on spatial \lya properties of the sample we also make
\lyaha, \wlya, and \hahb maps for the galaxies. These are created by using the
2D binned line maps from LaXs. None of the used line maps are extinction
corrected.  By using the Monte Carlo derived uncertainties for each line map we
also create masks marking bins where SNR$<2$ in any of the involved line maps.
These masks are used to indicate non-detected regions in the line ratio plots
below. Note that, because of non-zero extinction, the masked region sometimes
extend into the center of the galaxies when \hb becomes undetected. In this
situation we display an upper limit to the ratio using the $1\sigma$ error on
\hb in the ratio plots. Further out in the galaxies, both \ha and \hb becomes
too faint and the ratio is unconstrained.

\section{The spatial distribution of Ly$\alpha$ emission, dust and stars}
\label{sec:spatial}
The continuum subtraction code produces line maps for \lya, \ha, and \hb, along
with best fit stellar continuum maps at the respective wavelengths. In addition
to the flux maps the code also outputs maps of the best fit parameters (stellar
masses, $E(B-V)_s$, and stellar population age). In the following sections we
use the map of stellar continuum at \lya as the nominal FUV image. While the
Voronoi tesselated maps yield the most secure representation of line fluxes and
stellar population properties (and also include the full error estimate), we
also run LaXs on the un-binned data.

For the descriptions of the binned maps we also divide up the galaxies into
three sub-samples and define: i) \emph{weak \lya emitters}, with \wlya$<5$ \AA;
\emph{intermediate \lya emitters}, with $5\leq$\wlya$\leq 20$ \AA; \emph{strong
\lya emitters}, with \wlya$>20$ \AA. The \wlya values used here are the global
measurements (see Section~\ref{sec:globphot}), and the weak \lya emitters also
include net absorbers.  With this definition there are 9 weak \lya emitters
among the galaxies (LARS04, LARS06, LARS10, LARS13, ELARS03, ELARS12, ELARS14,
ELARS16, ELARS28), 21 intermediate \lya emitters (LARS08-09, LARS12, ELARS02, ELARS04,
ELARS06-11, ELARS15, ELARS17-23, ELARS25, ELARS27), and 15 strong \lya
emitters (LARS01-03, LARS05, LARS07, LARS11, LARS14, ELARS01, ELARS05,
ELARS13, ELARS24, ELARS26, Tol~1214, Tol~1247, J1156).

\subsection{Ly$\alpha$, H$\alpha$, and FUV color composites}
\label{subsec:images}
In Figures~\ref{fig:rgbs1}, \ref{fig:rgbs2}, \ref{fig:rgbs3}, \ref{fig:rgbs4},
and \ref{fig:rgbs5} we show color composites of \ha (red), FUV (green), and
\lya (blue) for all of the galaxies. These composites are constructed from
un-binned data, but applying mild smoothing with a bilateral denoising filter
\citep[\texttt{python/scikit-image}][]{scikit-image}. The smoothing makes it
possible to show the extended, but somewhat patchy, \lya emission together with
the more concentrated emission from stars and \ha.  The scaling (arcsinh) and
cut-levels for the respective color channels and different galaxies have been
chosen to increase contrast. From inspecting these images it is clear that \lya
is almost always more extended than the \ha emission and stellar continuum. In
many of the galaxies the central parts are dominated by red and green,
indicating that \lya is weak or even in absorption there. The main limitation
of the pseudo-narrowband technique used to isolate the \lya emission is that it
does not resolve emission and absorption of the line. In regions where the FUV
continuum is strong and the column density of \hi high, absorption may very
well dominate the measurement which results in negative \lya fluxes in the
maps. As seen in the color composites (as well as in the binned images and
surface brightness profiles) this turns out to happen in many of
the galaxies in our sample. 

The \lya-FUV-\ha color composites for LARS01-LARS14 differs from 
the previously published images in \citet{2013ApJ...765L..27H}. Most of this comes from the
change in calibration of the ACS/SBC instrument, but also from updates to 
the reduction process and continuum subtraction, and the choice of channel scaling 
in putting the new images together. We note that the earlier images and measurements were 
obtained with an incorrect calibration for the SBC and blue UVIS filters
\footnote{The changes in calibration are documented in ISR ACS 2019-05, and ISR 2021-04}. 
The updated results in this paper thus replaces the earlier measurements (see also Section~\ref{sec:globphot}).
Because of the arbitrary nature of how the smoothing and scaling was
done we do not perform any detailed measurements on these images, but
instead work with the binned maps for the following analysis.

\subsection{Spatial \lya properties of the sample.}
\label{subsec:profiles}
In this paper we present detailed measurements of surface brightness profiles
of the galaxies in the sample, and give an overall description of the profile
properties for the different \wlya groups. An extensive statistical study of
the \lya profiles and how they relate to other properties of the galaxies is
given in \citet{2022A&A...662A..64R}. In the following section we will describe
the spatial \lya properties of the three sub-samples. How the global properties
relate to one another is described in Section~\ref{sec:globphot}.
Figures~\ref{fig:mpanel_l02}, \ref{fig:mpanel_el08}, and \ref{fig:mpanel_el03}
show maps of the fluxes and derived quantities, as well as surface
brightness profiles of these. These three galaxies
are shown as examples of the three sub-sample groups, 
figures for the full sample are available in the online journal. Surface 
brightness profiles measured in \lya and FUV for the full sample are given 
in Table~\ref{tab:sbprofs} in Appendix~\ref{app:sbtables}.

\subsubsection{Weak \lya emitters}
\label{subsec:weaklaes}
This subgroup is characterized by weak and very patchy \lya emission. The small
amounts of emission that manages to escape comes out in select regions. The
effect of continuum absorption is also strong in these galaxies, the \lya flux
measured in larger apertures tend to be close to zero or even negative.  This
has the effect that the radial profiles are quite messy, they jump up and down
as single spots of emission enters the annuli.

Most of these galaxies are quite dusty (with the exception of LARS04), and from
inspecting the \hahb maps, the regions where \lya escapes seems to have lower
nebular extinctions. The \wlya maps are noisy and shows low values  all over
the galaxies, with the possible exception of ELARS03, which does show somewhat
higher \wlya on the side inclined towards the observer, and particularly close
to a star-forming knot in one of spiral arms. The \lyaha maps show a similar
picture with values predominantly $\ll$ 8. It should be noted that the 
intrinsic ratios may be larger than this, 
but obtaining an accurate local extinction corrected \lya flux is not possible, 
because of the longer path lengths for the resonantly scattered \lya photons. We 
use the non-extinction corrected \lyaha maps to signify where in the galaxies 
\lya has been scattered away from the production sites. The specific distribution of
dust may also cause variations of the \lyaha ratios, but unless there are significant 
dust column densities outside the \ha or UV bright region this is unlikely to cause 
the high \lyaha ratios we observe in the outskirts of some of the galaxies from the stronger LAEs.
In addition, obtaining dust corrected \lyaha maps that cover the full galaxies is not possible, 
because of low S/N in the \hahb map.

In general these galaxies also lack \lya halo emission (meaning emission
visible outside the FUV/\ha emitting areas), the emission is quite concentrated
to small patches close to the UV center of the galaxies. Two of the galaxies
(ELARS03 and ELARS28) do have measurable halo emission, and for ELARS03 this
might extend outside the field-of-view.

\subsubsection{Intermediate \lya emitters}
\label{subsec:intermlaes}
The intermediate emitters also have patchy \lya emission, but the emission is
stronger (or continuum absorption weaker). Some of these galaxies show quite
asymmetric emission when inspecting the \lya map (ELARS19--21, ELARS23).  This
is particularly true for the galaxies with a visibly inclined disk, where the
emission comes out from the side closest to us (determined from the \hahb ratio
map). Also galaxies without a well-defined disk show very little \lya emission in 
the regions with high \hahb ratio. 

Inspection of the surface brightness profiles show that \lya is more extended
than FUV/\ha in all but a few cases (exceptions are LARS08, ELARS02, ELARS06,
and ELARS18). The other galaxies have extended \lya halos, and six of them have
emission extending outside the field-of-view.  In the inclined disk cases, the
emission is more extended in the off-axis direction (which is not visible in
the profile plots).

The \lyaha ratio is below 8.7 in the galaxies, but higher than for the weak
emitter sub-group. The strongest spots of \lya in some of the galaxies do reach
this limit, and are situated outside the FUV emitting part of the galaxies.
ELARS04 and 11 shows some annulus regions with \lyaha$>8.7$, but with large
errors.

\subsubsection{Strong \lya emitters}
\label{subsec:stronglaes}
The strong emitters show \lya in large portions of the image, although some
absorption is visible, in particular in regions with high FUV surface
brightness and/or high \hahb ratio. The spatial distribution of \lya is mostly
symmetric, but in some galaxies there appears to be directions that are
unfavored for escape (LARS01, LARS03, LARS11, ELARS01, ELARS24, Tol~1214).

Most of the galaxies in this group have low global $E(B-V)_n$, but, somewhat
surprisingly, a few dusty galaxies are also included.  Even though \wlya is
higher than 20 \AA\ globally the escape fractions are very low, indicating that
we see \lya emission from low-dust regions where the relative escape of the
photons can be quite high.  In these galaxies (LARS03, LARS11, ELARS01, and
ELARS24) the \lya emission shows up in distinct regions that likely indicates
channels where the dust content is lower.

The profile plots show the that the fraction of \lya emission coming out in the
halo varies among the galaxies. A majority of them (11 out of 15) show
\lyaha$>8.7$ in parts of the image, preferably in the halo, indicating that
resonant scattering of the \lya photons is in full effect.  Most (13) also have
\lya emission extending all the way to the edge of the field-of-view.

\subsubsection{Escape fraction and equivalent width variations with aperture size}
\label{subsec:fescvar}
The cumulative escape fraction profiles generally rise as you go further out,
but many galaxies in all sub-groups show a dip in the profile. One explanation
to this behaviour is strong gradients in the extinction profile (which directly
affects the \fesc profile), but could also be due to \lya continuum absorption
in off-center regions with strong FUV emission and high \hi column densities.
The \fesc profile shape of these galaxies thus indicates that that both direct
(with weak resonant scattering) and indirect (via strong scattering) escape
takes place in the galaxies.

In twelve of the galaxies we find that the shapes of the \fesc and \wlya
cumulative profile differ from one another. This is somewhat unexpected given
that they both provide a normalized estimate of escaping \lya. In some of these
cases (e.g., LARS01, LARS03) the escape fraction is highest in the center of
the galaxies,  decreases as the aperture grows bigger, and then starts to
increase again further out. This is very different from the \wlya cumulative
profile in these galaxies that just increases with radius. The cause of this
difference lies in that \wlya provides an estimate of \lya escape relative to
the non-extinction corrected FUV luminosity. The measured FUV luminosity will
thus not give the total budget of ionizing radiation, and will in particular
differ from an absolute measure (\fesc) in dusty regions of the galaxies. We
see this effect in dusty galaxies and in galaxies with strong extinction
gradients in the sample.  The specific case of high central escape with a sharp
drop is likely the effect of choosing the brightest FUV coordinates to be the
centrum for the apertures. This point is likely to have low extinction which
leads to a high escape fraction. If the extinction then goes up with radius
\fesc will drop. This effect is strongest in the dusty galaxies that often only
have small relatively dust-free regions,

\subsection{Average \lya growth curves}
\label{sec:aplosses}
\begin{figure}[pt!]
\centering
\epsscale{1.2}
\plotone{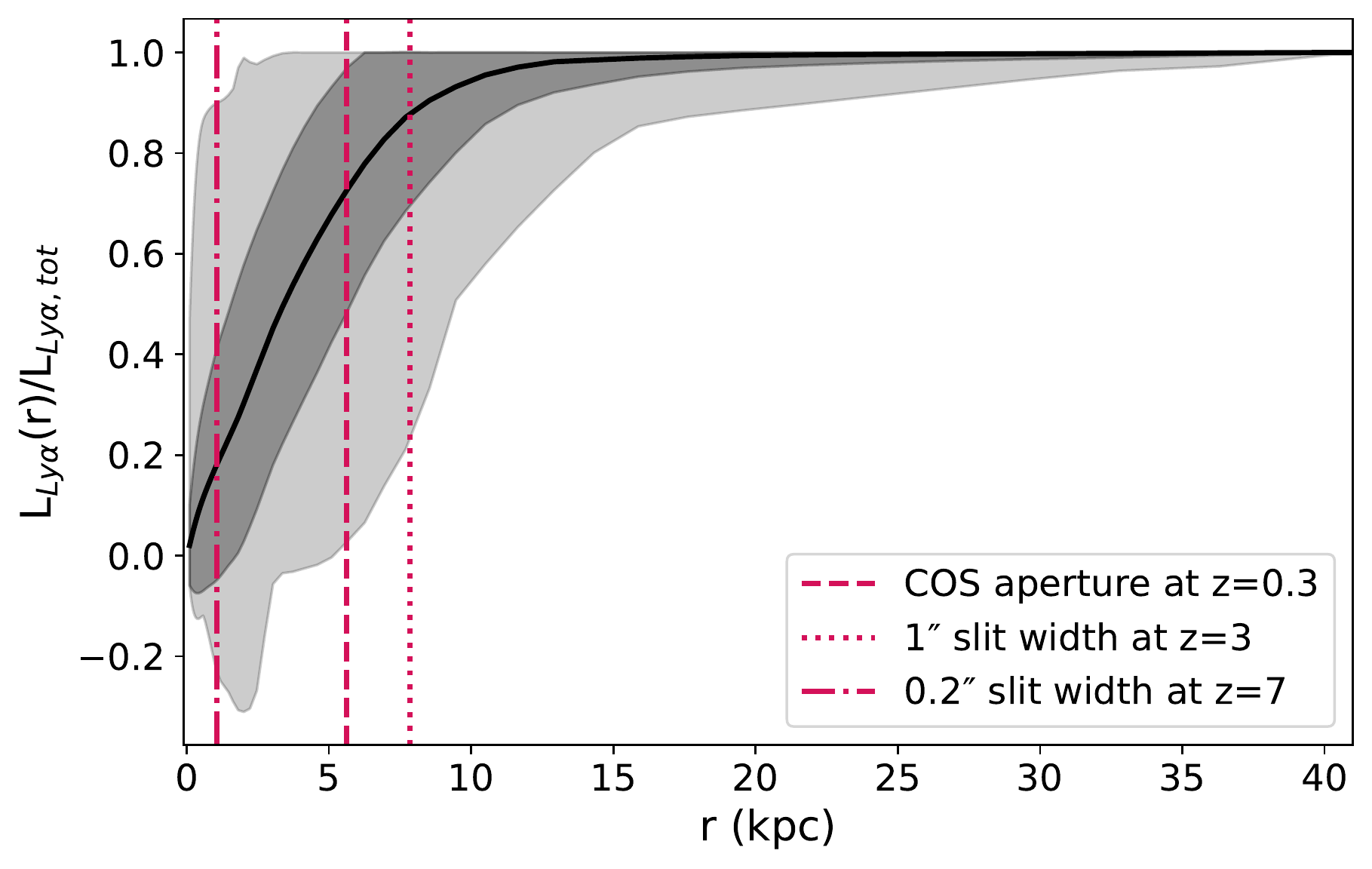}
\caption{Average growth curve of \lya luminosity for the galaxies in the 
	sample with \wlya$>0$ \AA. The dark grey shaded area shows the 
	standard deviation (capped at a ratio of 1) over the sample at each radius and the light 
	grey shade shows the full envelope of the sample.}
\label{fig:lyagrowth}
\end{figure}
Spectroscopic measurements of the \lya line using a limited slit or aperture
size can potentially miss part of the extended emission. Given the effects of
resonant scattering this impacts \lya much more than the other nebular
lines or stellar continuum. With our \lya imaging data we can follow the
spatial emission profile further out and estimate slit losses for an aperture
of a given physical size. 
It should be noted that
observations of high redshift LAEs with VLT/MUSE (see Section~\ref{sec:lumlya})
do not suffer from any slit losses given that this instrument is an integral
field unit spectrograph. The same is true for slitless spectroscopy (e.g., 
HST/WFC3 or JWST NIRCAM/NIRISS grism spectroscopy).

In Figure~\ref{fig:lyagrowth} we show the average growth curve of \lya
luminosity for the galaxies with confirmed global \lya emission (\wlya$>0$ \AA)
together with physical sizes for three different spectroscopic slit (or
aperture) sizes. Depending on the redshift and angular slit width these
physical sizes will vary. The total luminosity used for this plot is measured
within the total \lya aperture described above. It should be noted that many of
the galaxies (20/45) have detectable \lya all the way out to the edge of the
detector, which means that the total luminosity is a lower limit. The average
L$_{\mathrm{Ly}\alpha}(r) /$L$_{\mathrm{Ly}\alpha, tot}$ curve thus shows an
upper limit to the flux recovered at a given physical aperture size.

We have estimated slit losses (defined as $1 -
\mathrm{L}_{\mathrm{Ly}\alpha}(r) /\mathrm{L}_{\mathrm{Ly}\alpha, tot}$) for
three typical cases of \lya spectroscopy. For these estimates we use the
average curve shown in Figure~\ref{fig:lyagrowth} (with the exception of our
own sample where we use the individual growth curves).  We also provide
estimates using infinite aperture \lya luminosities (L$_{\mathrm{Ly}\alpha,
exp}$), obtained by fitting an exponential function for the surface brightness
profile in the halo and extrapolating, from \citet{2022A&A...662A..64R}. These slit loss
estimates avoid the problem of \lya extending outside the detector but suffer
from unknown systematic uncertainties due to the assumption of purely
exponential halos.
\\
\textbf{HST/COS aperture:} For the data presented in this paper (with an
average redshift of 0.05) the COS aperture suffers from quite severe losses.
The average aperture loss for COS for our galaxies compared to
L$_{\mathrm{Ly}\alpha, tot}$ (L$_{\mathrm{Ly}\alpha, exp}$) is $0.5\pm 0.2$
($0.6\pm 0.3$), where the uncertainty is the standard deviation for the sample.
Inspecting the growth curves of our sample and looking at more distant galaxies
at $z\sim 0.3$ (by calculating the physical aperture size at that redshift) we
find that COS is able to recover more of the \lya flux. The average aperture
loss is then $0.27 \pm 0.23$ and $0.40\pm 0.26$ for L$_{\mathrm{Ly}\alpha, tot}$
and L$_{\mathrm{Ly}\alpha, exp}$ total flux estimates, respectively.
\\
\textbf{High redshift LAE observed with a 1\arcsec slit}: For galaxies at
$z=3.0$ and assuming a typical slit width of 1\arcsec for ground-based multi
object spectroscopy (MOS) studies of LAEs we find that the slit loss is $0.12
\pm 0.18$ and $0.25\pm 0.27$ for L$_{\mathrm{Ly}\alpha, tot}$ and
L$_{\mathrm{Ly}\alpha, exp}$ total flux estimates, respectively. The slit
losses from pure \lya scattering are not very significant in this case, but the
issue could be further compounded by PSF losses which need to be computed
assuming an extended source rather than a point source.
\\
\textbf{Extreme high redshift LAEs observed with JWST/NIRSPEC}:
With JWST/NIRSPEC it is now possible to observe extreme redshift LAEs
($z\gtrsim 7$) using the NIRSPEC MOS mode. However, the slit width of a single
micro-shutter array (MSA) shutter is very narrow (0.2\arcsec), which means that
slit losses may be very high. Indeed, calculating the average slit loss at
$z=7.0$ within a single MSA shutter we find losses of $0.80 \pm 0.21$ and $0.82\pm
0.21$ for L$_{\mathrm{Ly}\alpha, tot}$ and L$_{\mathrm{Ly}\alpha, exp}$ total
flux estimates, respectively.

\section{Global measurements of Ly$\alpha$ and other properties}
\label{sec:globphot}
\begin{deluxetable*}{lccccc}
\tabletypesize{\scriptsize}
\renewcommand{\arraystretch}{1.3}
\tablecaption{Global \lya, \ha, and FUV measurements\label{tab:globprop}}
\tablehead{
    	\colhead{ID} &
	\colhead{$L_{\mathrm{FUV}}$} &
	\colhead{$L_{\mathrm{H}\alpha}$} &
	\colhead{$L_{\mathrm{Ly}\alpha}$} &
	\colhead{\wlya} &
	\colhead{\fesc} \\
    	\colhead{} &
	\colhead{(10$^{40}$ \ergc)} &
	\colhead{(10$^{41}$ \ergl)} &
	\colhead{(10$^{41}$ \ergl)} &
	\colhead{(\AA)} &
	\colhead{} \\
	\colhead{(1)} & 
	\colhead{(2)} & 
	\colhead{(3)} & 
	\colhead{(4)} & 
	\colhead{(5)} &  
	\colhead{(6)} \\   
}
\startdata
LARS01  		 & 1.978$^{0.004}_{0.004}$ & 5.126$^{0.010}_{0.014}$ & 8.345$^{0.089}_{0.089}$ & 42.195$^{0.397}_{0.458}$ & 0.134$^{0.002}_{0.002}$ \\ 
LARS02   		 & 0.733$^{0.004}_{0.003}$ & 1.590$^{0.011}_{0.012}$ & 4.154$^{0.093}_{0.116}$ & 56.661$^{1.304}_{1.946}$ & 0.299$^{0.010}_{0.010}$ \\ 
LARS03   		 & 0.561$^{0.003}_{0.003}$ & 6.091$^{0.025}_{0.021}$ & 1.917$^{0.076}_{0.095}$ & 34.164$^{1.348}_{1.848}$ & 0.005$^{0.000}_{0.000}$ \\ 
LARS04\tablenotemark{a}  & 1.338$^{0.002}_{0.003}$ & 3.831$^{0.009}_{0.011}$ & 0.276$^{0.062}_{0.058}$ & 2.065$^{0.467}_{0.435}$ & 0.006$^{0.001}_{0.001}$ \\ 
LARS05  		 & 2.613$^{0.006}_{0.005}$ & 4.254$^{0.018}_{0.013}$ & 6.516$^{0.107}_{0.107}$ & 24.935$^{0.471}_{0.467}$ & 0.126$^{0.002}_{0.002}$ \\ 
LARS06   		 & 0.375$^{0.004}_{0.003}$ & 0.648$^{0.007}_{0.006}$ & $\le$0.089 & $\le$2.374 & $\le$0.012 \\ 
LARS07   		 & 1.717$^{0.006}_{0.008}$ & 3.789$^{0.016}_{0.015}$ & 6.661$^{0.132}_{0.144}$ & 38.801$^{0.935}_{1.000}$ & 0.111$^{0.002}_{0.002}$ \\ 
LARS08   		 & 2.316$^{0.009}_{0.016}$ & 11.072$^{0.345}_{0.124}$ & 4.017$^{0.267}_{0.193}$ & 17.345$^{1.302}_{0.915}$ & 0.006$^{0.000}_{0.000}$ \\ 
LARS09   		 & 5.913$^{0.015}_{0.011}$ & 18.962$^{0.035}_{0.027}$ & 5.810$^{0.284}_{0.272}$ & 9.827$^{0.477}_{0.474}$ & 0.016$^{0.001}_{0.001}$ \\ 
LARS10  		 & 1.120$^{0.005}_{0.007}$ & 2.005$^{0.012}_{0.022}$ & $\le$0.235 & $\le$2.100 & $\le$0.003 \\ 
LARS11  		 & 8.725$^{0.051}_{0.055}$ & 9.929$^{0.113}_{0.079}$ & 17.899$^{1.304}_{0.973}$ & 20.515$^{0.976}_{0.710}$ & 0.065$^{0.009}_{0.009}$ \\ 
LARS12  		 & 8.502$^{0.043}_{0.037}$ & 13.904$^{0.144}_{0.092}$ & 15.396$^{0.896}_{0.758}$ & 18.108$^{1.045}_{1.041}$ & 0.027$^{0.001}_{0.001}$ \\ 
LARS13  		 & 11.956$^{0.229}_{0.273}$ & 20.978$^{0.084}_{0.074}$ & $\le$2.989 & $\le$1.774 & $\le$0.004 \\ 
LARS14  		 & 11.358$^{0.194}_{0.129}$ & 18.908$^{0.239}_{0.238}$ & 55.602$^{1.914}_{1.982}$ & 48.955$^{2.281}_{2.102}$ & 0.263$^{0.010}_{0.010}$ \\ 
ELARS01  		 & 2.443$^{0.003}_{0.003}$ & 13.134$^{0.016}_{0.017}$ & 5.169$^{0.063}_{0.064}$ & 21.156$^{0.288}_{0.297}$ & 0.012$^{0.000}_{0.000}$ \\ 
ELARS02  		 & 2.660$^{0.005}_{0.004}$ & 6.188$^{0.036}_{0.035}$ & 3.311$^{0.116}_{0.082}$ & 12.444$^{0.450}_{0.327}$ & 0.061$^{0.002}_{0.002}$ \\ 
ELARS03\tablenotemark{a} & 2.779$^{0.008}_{0.007}$ & 10.127$^{0.032}_{0.025}$ & 1.227$^{0.145}_{0.119}$ & 4.414$^{0.535}_{0.428}$ & 0.005$^{0.001}_{0.001}$ \\ 
ELARS04  		 & 2.587$^{0.005}_{0.005}$ & 4.223$^{0.015}_{0.017}$ & 4.419$^{0.121}_{0.107}$ & 17.078$^{0.511}_{0.449}$ & 0.070$^{0.002}_{0.002}$ \\ 
ELARS05   		 & 2.217$^{0.009}_{0.009}$ & 3.679$^{0.023}_{0.018}$ & 5.881$^{0.217}_{0.180}$ & 26.523$^{1.006}_{0.890}$ & 0.148$^{0.006}_{0.006}$ \\ 
ELARS06  		 & 1.029$^{0.006}_{0.007}$ & 1.312$^{0.018}_{0.025}$ & 1.297$^{0.142}_{0.096}$ & 12.610$^{1.402}_{0.956}$ & 0.060$^{0.005}_{0.005}$ \\ 
ELARS07\tablenotemark{a} & 1.148$^{0.009}_{0.006}$ & 2.537$^{0.028}_{0.032}$ & 0.811$^{0.129}_{0.159}$ & 7.062$^{1.137}_{1.383}$ & 0.032$^{0.006}_{0.006}$ \\ 
ELARS08\tablenotemark{a} & 0.821$^{0.005}_{0.004}$ & 1.740$^{0.019}_{0.016}$ & 1.375$^{0.116}_{0.113}$ & 16.738$^{1.413}_{1.446}$ & 0.017$^{0.002}_{0.002}$ \\ 
ELARS09\tablenotemark{a} & 0.817$^{0.004}_{0.002}$ & 0.756$^{0.015}_{0.012}$ & 0.461$^{0.079}_{0.083}$ & 5.648$^{0.978}_{1.029}$ & 0.054$^{0.010}_{0.010}$ \\ 
ELARS10		 	 & 0.552$^{0.005}_{0.007}$ & 1.379$^{0.016}_{0.017}$ & 0.929$^{0.121}_{0.104}$ & 16.825$^{2.425}_{1.890}$ & 0.016$^{0.002}_{0.002}$ \\ 
ELARS11\tablenotemark{a} & 0.697$^{0.003}_{0.003}$ & 0.911$^{0.013}_{0.011}$ & 0.621$^{0.070}_{0.062}$ & 8.913$^{1.046}_{0.916}$ & 0.065$^{0.007}_{0.007}$ \\ 
ELARS12\tablenotemark{a} & 0.635$^{0.002}_{0.004}$ & 2.311$^{0.019}_{0.015}$ & $\le$0.082 & $\le$1.272 & $\le$0.002 \\ 
ELARS13 		 & 0.672$^{0.005}_{0.004}$ & 0.886$^{0.006}_{0.007}$ & 2.532$^{0.092}_{0.087}$ & 37.700$^{1.627}_{1.384}$ & 0.207$^{0.008}_{0.008}$ \\ 
ELARS14\tablenotemark{a} & 0.515$^{0.003}_{0.003}$ & 1.080$^{0.009}_{0.011}$ & $\le$0.066 & $\le$1.270 & $\le$0.005 \\ 
ELARS15			 & 0.404$^{0.004}_{0.005}$ & 0.646$^{0.020}_{0.018}$ & 0.807$^{0.116}_{0.104}$ & 19.969$^{3.002}_{2.643}$ & 0.124$^{0.018}_{0.018}$ \\ 
ELARS16			 & 0.267$^{0.002}_{0.003}$ & 0.345$^{0.009}_{0.012}$ & $\le$0.063 & $\le$2.377 & $\le$0.009 \\ 
ELARS17			 & 0.399$^{0.005}_{0.006}$ & 0.571$^{0.018}_{0.015}$ & 0.774$^{0.117}_{0.097}$ & 19.403$^{3.108}_{2.470}$ & 0.082$^{0.012}_{0.012}$ \\ 
ELARS18			 & 0.220$^{0.003}_{0.002}$ & 0.344$^{0.012}_{0.012}$ & 0.173$^{0.059}_{0.075}$ & 7.877$^{2.730}_{3.330}$ & 0.030$^{0.013}_{0.013}$ \\ 
ELARS19\tablenotemark{a} & 0.270$^{0.002}_{0.002}$ & 0.430$^{0.008}_{0.009}$ & 0.270$^{0.053}_{0.043}$ & 9.996$^{2.028}_{1.559}$ & 0.064$^{0.010}_{0.010}$ \\ 
ELARS20 		 & 0.273$^{0.002}_{0.002}$ & 0.522$^{0.009}_{0.006}$ & 0.215$^{0.049}_{0.052}$ & 7.867$^{1.844}_{1.964}$ & 0.031$^{0.008}_{0.008}$ \\ 
ELARS21			 & 0.124$^{0.002}_{0.002}$ & 0.118$^{0.010}_{0.008}$ & 0.111$^{0.042}_{0.039}$ & 8.961$^{3.473}_{3.117}$ & 0.047$^{0.033}_{0.033}$ \\ 
ELARS22\tablenotemark{a} & 2.723$^{0.011}_{0.006}$ & 3.644$^{0.054}_{0.055}$ & 1.854$^{0.139}_{0.152}$ & 6.808$^{0.530}_{0.562}$ & 0.056$^{0.005}_{0.005}$ \\ 
ELARS23			 & 2.295$^{0.014}_{0.011}$ & 4.222$^{0.081}_{0.073}$ & 1.493$^{0.214}_{0.241}$ & 6.507$^{0.958}_{1.073}$ & 0.040$^{0.007}_{0.007}$ \\ 
ELARS24			 & 2.007$^{0.010}_{0.010}$ & 6.929$^{0.064}_{0.071}$ & 4.472$^{0.180}_{0.184}$ & 22.283$^{0.982}_{0.927}$ & 0.004$^{0.000}_{0.000}$ \\ 
ELARS25\tablenotemark{a} & 1.718$^{0.011}_{0.011}$ & 1.954$^{0.040}_{0.060}$ & 1.643$^{0.183}_{0.203}$ & 9.564$^{1.098}_{1.173}$ & 0.094$^{0.012}_{0.012}$ \\          
ELARS26\tablenotemark{a} & 1.037$^{0.008}_{0.009}$ & 3.305$^{0.048}_{0.053}$ & 2.233$^{0.176}_{0.171}$ & 21.547$^{1.689}_{1.829}$ & 0.059$^{0.005}_{0.005}$ \\ 
ELARS27\tablenotemark{a} & 1.089$^{0.009}_{0.010}$ & 1.865$^{0.038}_{0.045}$ & 2.051$^{0.159}_{0.156}$ & 18.832$^{1.638}_{1.491}$ & 0.115$^{0.011}_{0.011}$ \\ 
ELARS28			 & 1.180$^{0.006}_{0.008}$ & 2.326$^{0.041}_{0.048}$ & 0.370$^{0.150}_{0.133}$ & 3.138$^{1.285}_{1.128}$ & 0.013$^{0.005}_{0.005}$ \\ 
Tol1214			 & 0.304$^{0.002}_{0.002}$ & 0.925$^{0.015}_{0.016}$ & 2.180$^{0.049}_{0.045}$ & 71.715$^{2.008}_{1.887}$ & 0.271$^{0.007}_{0.007}$ \\ 
Tol1247			 & 10.381$^{0.022}_{0.019}$ & 25.306$^{0.066}_{0.062}$ & 24.794$^{0.391}_{0.322}$ & 23.883$^{0.403}_{0.351}$ & 0.059$^{0.001}_{0.001}$ \\ 
J1156  			 & 24.012$^{0.423}_{0.429}$ & 31.913$^{0.763}_{0.706}$ & 63.696$^{4.799}_{4.555}$ & 26.527$^{2.124}_{2.230}$ & 0.115$^{0.009}_{0.009}$ \\ 
\enddata
\tablenotetext{a}{The \lya emission from these galaxies are potentially underestimated due to uncertain background subtraction, see Section~\ref{subsec:syserr}.}
\renewcommand{\arraystretch}{1}
\end{deluxetable*}

\begin{deluxetable*}{lcccccccc}
\tabletypesize{\scriptsize}
\renewcommand{\arraystretch}{1.3}
\tablecaption{Derived galaxy properties\label{tab:globderiv}}
\tablehead{
    	\colhead{ID} &
	\colhead{SFR$_{\mathrm{H}\alpha}$} &
	\colhead{SFR$_{FUV}$} &
	\colhead{sSFR$_{\mathrm{H}\alpha}$} &
	\colhead{$\Sigma_{\mathrm{SFR}}$} &
	\colhead{Mass} &
	\colhead{Age} &
	\colhead{$E(B-V)_s$} & 
	\colhead{$E(B-V)_n$} \\ 
    	\colhead{} &
	\colhead{(M$_{\odot}/\mathrm{yr}$)} &
	\colhead{(M$_{\odot}/\mathrm{yr}$)} &
	\colhead{($10^{-10}$yr$^{-1}$)} &
	\colhead{(M$_{\odot}/\mathrm{yr}/\mathrm{kpc}^2$)} &
	\colhead{($10^{10}\mathrm{M}_{\odot}$)} &
	\colhead{(Myr)} & 
	\colhead{} & 
	\colhead{} \\
	\colhead{(1)} & 
	\colhead{(2)} & 
	\colhead{(3)} & 
	\colhead{(4)} & 
	\colhead{(5)} &  
	\colhead{(6)} &   
	\colhead{(7)} &  
	\colhead{(8)} &  
	\colhead{(9)}   
}
\startdata
LARS 1   &3.86$^{0.01}_{0.01}$ & 3.04$^{0.01}_{0.01}$ & 2.83$^{0.03}_{0.03}$ & 0.472$^{0.003}_{0.003}$ & 1.65$^{0.04}_{0.04}$ & 7.95$^{0.71}_{0.50}$ & 0.09$^{0.00}_{0.00}$ & 0.140$^{0.002}_{0.002}$ \\  
LARS 2   &0.86$^{0.01}_{0.01}$ & 2.43$^{0.06}_{0.06}$ & 2.44$^{0.06}_{0.06}$ & 0.110$^{0.001}_{0.001}$ & 0.57$^{0.03}_{0.03}$ & 19.64$^{1.37}_{1.74}$ & 0.17$^{0.00}_{0.00}$ & 0.002$^{0.009}_{0.006}$ \\  
LARS 3   &25.52$^{0.09}_{0.09}$ & 16.68$^{0.65}_{0.44}$ & 26.70$^{2.19}_{2.19}$ & 8.566$^{0.297}_{0.284}$ & 2.15$^{0.10}_{0.10}$ & 77.46$^{1.37}_{1.14}$ & 0.39$^{0.00}_{0.00}$ & 0.850$^{0.011}_{0.015}$ \\    
LARS 4   &3.06$^{0.01}_{0.01}$ & 1.97$^{0.01}_{0.01}$ & 2.72$^{0.03}_{0.03}$ & 0.326$^{0.003}_{0.003}$ & 1.77$^{0.02}_{0.02}$ & 11.42$^{0.58}_{0.61}$ & 0.09$^{0.00}_{0.00}$ & 0.164$^{0.003}_{0.004}$ \\  
LARS 5   &3.19$^{0.01}_{0.01}$ & 3.35$^{0.01}_{0.01}$ & 4.05$^{0.04}_{0.04}$ & 0.455$^{0.003}_{0.002}$ & 0.84$^{0.01}_{0.01}$ & 4.38$^{0.61}_{0.49}$ & 0.07$^{0.00}_{0.00}$ & 0.138$^{0.002}_{0.002}$ \\  
LARS 6   &0.40$^{0.00}_{0.00}$ & 0.39$^{0.00}_{0.00}$ & 1.47$^{0.09}_{0.09}$ & 0.024$^{0.001}_{0.001}$ & 0.39$^{0.02}_{0.02}$ & 13.38$^{1.86}_{1.01}$ & 0.05$^{0.00}_{0.00}$ & 0.053$^{0.010}_{0.012}$ \\  
LARS 7   &3.70$^{0.01}_{0.01}$ & 2.06$^{0.01}_{0.01}$ & 6.97$^{0.07}_{0.07}$ & 1.086$^{0.004}_{0.008}$ & 0.86$^{0.02}_{0.02}$ & 24.68$^{0.74}_{0.90}$ & 0.06$^{0.00}_{0.00}$ & 0.248$^{0.003}_{0.002}$ \\  
LARS 8   &41.73$^{0.47}_{0.47}$ & 4.63$^{0.07}_{0.07}$ & 4.68$^{0.70}_{0.70}$ & 0.799$^{0.105}_{0.165}$ & 10.01$^{0.35}_{0.35}$ & 72.89$^{8.50}_{8.41}$ & 0.12$^{0.00}_{0.00}$ & 0.806$^{0.006}_{0.008}$ \\    
LARS 9   &23.08$^{0.03}_{0.03}$ & 9.71$^{0.13}_{0.09}$ & 4.59$^{0.04}_{0.04}$ & 0.299$^{0.002}_{0.002}$ & 5.85$^{0.07}_{0.07}$ & 20.40$^{6.72}_{4.24}$ & 0.10$^{0.00}_{0.00}$ & 0.338$^{0.002}_{0.002}$ \\     
LARS 10  &5.72$^{0.06}_{0.06}$ & 4.78$^{0.09}_{0.10}$ & 5.22$^{0.30}_{0.30}$ & 0.378$^{0.024}_{0.019}$ & 2.35$^{0.09}_{0.09}$ & 38.16$^{1.49}_{1.27}$ & 0.19$^{0.00}_{0.00}$ & 0.691$^{0.019}_{0.022}$ \\  
LARS 11  &16.90$^{0.13}_{0.13}$ & 36.45$^{1.79}_{1.82}$ & 1.36$^{0.05}_{0.05}$ & 0.090$^{0.001}_{0.001}$ & 15.24$^{0.37}_{0.37}$ & 32.26$^{0.52}_{0.50}$ & 0.19$^{0.00}_{0.00}$ & 0.477$^{0.050}_{0.050}$ \\    
LARS 12  &34.98$^{0.23}_{0.23}$ & 9.18$^{0.08}_{0.08}$ & 12.19$^{0.20}_{0.20}$ & 1.835$^{0.027}_{0.021}$ & 3.20$^{0.08}_{0.08}$ & 14.59$^{0.92}_{1.60}$ & 0.05$^{0.00}_{0.00}$ & 0.639$^{0.004}_{0.005}$ \\      
LARS 13  &40.47$^{0.14}_{0.14}$ & 65.27$^{1.30}_{1.99}$ & 28.27$^{4.72}_{4.72}$ & 0.528$^{0.009}_{0.008}$ & 3.83$^{0.45}_{0.45}$ & 35.42$^{1.92}_{2.72}$ & 0.22$^{0.00}_{0.01}$ & 0.529$^{0.008}_{0.007}$ \\    
LARS 14  &13.03$^{0.16}_{0.16}$ & 16.47$^{0.21}_{0.17}$ & 13.29$^{0.35}_{0.35}$ & 0.701$^{0.007}_{0.006}$ & 1.18$^{0.08}_{0.08}$ & 11.89$^{2.59}_{2.41}$ & 0.08$^{0.00}_{0.00}$ & 0.103$^{0.003}_{0.003}$ \\    
ELARS 1  &25.71$^{0.03}_{0.03}$ & 6.18$^{0.02}_{0.02}$ & 3.74$^{0.05}_{0.05}$ & 1.329$^{0.009}_{0.011}$ & 7.85$^{0.08}_{0.08}$ & 21.52$^{0.66}_{0.27}$ & 0.14$^{0.00}_{0.00}$ & 0.535$^{0.002}_{0.003}$ \\  
ELARS 2  &3.32$^{0.02}_{0.02}$ & 3.53$^{0.02}_{0.02}$ & 1.39$^{0.02}_{0.02}$ & 0.100$^{0.001}_{0.001}$ & 3.08$^{0.02}_{0.02}$ & 36.77$^{0.49}_{0.53}$ & 0.07$^{0.00}_{0.00}$ & 0.000$^{0.000}_{0.000}$ \\  
ELARS 3  &13.80$^{0.03}_{0.03}$ & 29.22$^{0.40}_{0.41}$ & 0.67$^{0.01}_{0.01}$ & 0.044$^{0.000}_{0.000}$ & 8.96$^{0.11}_{0.11}$ & 26.60$^{0.50}_{0.61}$ & 0.29$^{0.00}_{0.00}$ & 0.385$^{0.004}_{0.005}$ \\    
ELARS 4  &3.89$^{0.02}_{0.02}$ & 4.08$^{0.02}_{0.01}$ & 1.47$^{0.02}_{0.02}$ & 0.143$^{0.001}_{0.001}$ & 3.32$^{0.03}_{0.03}$ & 39.32$^{0.71}_{0.54}$ & 0.09$^{0.00}_{0.00}$ & 0.223$^{0.002}_{0.003}$ \\  
ELARS 5  &2.45$^{0.01}_{0.01}$ & 7.30$^{0.19}_{0.16}$ & 0.33$^{0.01}_{0.01}$ & 0.010$^{0.000}_{0.000}$ & 7.08$^{0.07}_{0.07}$ & 49.82$^{0.57}_{0.76}$ & 0.17$^{0.00}_{0.00}$ & 0.088$^{0.009}_{0.009}$ \\  
ELARS 6  &1.34$^{0.03}_{0.03}$ & 1.49$^{0.01}_{0.02}$ & 1.44$^{0.04}_{0.04}$ & 0.012$^{0.000}_{0.000}$ & 1.53$^{0.03}_{0.03}$ & 42.80$^{1.86}_{1.76}$ & 0.08$^{0.00}_{0.00}$ & 0.265$^{0.008}_{0.012}$ \\  
ELARS 7  &1.56$^{0.02}_{0.02}$ & 1.40$^{0.01}_{0.01}$ & 34.57$^{1.00}_{1.00}$ & 6.469$^{0.106}_{0.135}$ & 1.44$^{0.08}_{0.08}$ & 30.81$^{3.70}_{3.24}$ & 0.07$^{0.00}_{0.00}$ & 0.055$^{0.004}_{0.004}$ \\  
ELARS 8  &4.94$^{0.05}_{0.05}$ & 4.53$^{0.15}_{0.15}$ & 0.65$^{0.05}_{0.05}$ & 0.015$^{0.001}_{0.001}$ & 4.55$^{0.10}_{0.10}$ & 45.16$^{1.05}_{1.07}$ & 0.22$^{0.00}_{0.00}$ & 0.689$^{0.025}_{0.026}$ \\  
ELARS 9  &0.52$^{0.01}_{0.01}$ & 3.00$^{0.09}_{0.09}$ & 1.68$^{0.70}_{0.70}$ & 0.096$^{0.008}_{0.028}$ & 0.54$^{0.04}_{0.04}$ & 42.33$^{1.70}_{1.92}$ & 0.18$^{0.00}_{0.00}$ & 0.105$^{0.010}_{0.008}$ \\  
ELARS 10 &3.50$^{0.04}_{0.04}$ & 3.42$^{0.17}_{0.16}$ & 3.01$^{0.15}_{0.15}$ & 0.138$^{0.006}_{0.005}$ & 2.02$^{0.08}_{0.08}$ & 61.42$^{3.19}_{3.53}$ & 0.23$^{0.01}_{0.01}$ & 0.642$^{0.021}_{0.023}$ \\  
ELARS 11 &0.59$^{0.01}_{0.01}$ & 2.25$^{0.05}_{0.05}$ & 0.28$^{0.01}_{0.01}$ & 0.027$^{0.001}_{0.001}$ & 1.90$^{0.07}_{0.07}$ & 35.56$^{0.81}_{1.18}$ & 0.17$^{0.00}_{0.00}$ & 0.075$^{0.011}_{0.010}$ \\  
ELARS 12 &2.68$^{0.02}_{0.02}$ & 7.13$^{0.28}_{0.26}$ & 1.15$^{0.07}_{0.07}$ & 0.060$^{0.004}_{0.003}$ & 3.14$^{0.11}_{0.11}$ & 41.93$^{1.87}_{1.66}$ & 0.29$^{0.01}_{0.00}$ & 0.319$^{0.022}_{0.019}$ \\  
ELARS 13 &0.76$^{0.01}_{0.01}$ & 1.40$^{0.02}_{0.03}$ & 1.64$^{0.03}_{0.03}$ & 0.849$^{0.013}_{0.014}$ & 1.17$^{0.08}_{0.08}$ & 3.29$^{0.58}_{0.62}$ & 0.12$^{0.00}_{0.00}$ & 0.191$^{0.007}_{0.008}$ \\  
ELARS 14 &0.72$^{0.01}_{0.01}$ & 3.66$^{0.11}_{0.12}$ & 1.91$^{0.18}_{0.18}$ & 0.091$^{0.007}_{0.004}$ & 2.40$^{0.30}_{0.30}$ & 22.85$^{2.18}_{1.65}$ & 0.25$^{0.00}_{0.00}$ & 0.089$^{0.009}_{0.009}$ \\  
ELARS 15 &0.40$^{0.01}_{0.01}$ & 2.31$^{0.15}_{0.11}$ & 1.31$^{0.13}_{0.13}$ & 0.005$^{0.000}_{0.000}$ & 1.26$^{0.06}_{0.06}$ & 74.92$^{2.75}_{3.58}$ & 0.22$^{0.01}_{0.00}$ & 0.059$^{0.024}_{0.023}$ \\  
ELARS 16 &0.41$^{0.01}_{0.01}$ & 1.67$^{0.12}_{0.13}$ & 2.64$^{0.01}_{0.01}$ & 0.010$^{0.000}_{0.000}$ & 0.86$^{0.03}_{0.03}$ & 41.02$^{2.86}_{4.05}$ & 0.23$^{0.01}_{0.01}$ & 0.333$^{0.018}_{0.023}$ \\  
ELARS 17 &0.58$^{0.01}_{0.01}$ & 1.34$^{0.08}_{0.08}$ & 0.38$^{0.00}_{0.00}$ & 0.003$^{0.000}_{0.000}$ & 1.36$^{0.04}_{0.04}$ & 60.70$^{3.13}_{2.03}$ & 0.17$^{0.01}_{0.01}$ & 0.263$^{0.028}_{0.026}$ \\  
ELARS 18 &0.36$^{0.01}_{0.01}$ & 0.84$^{0.04}_{0.04}$ & 8.83$^{3.10}_{3.10}$ & 0.030$^{0.001}_{0.001}$ & 0.53$^{0.18}_{0.18}$ & 73.24$^{7.94}_{7.09}$ & 0.18$^{0.01}_{0.01}$ & 0.271$^{0.015}_{0.022}$ \\  
ELARS 19 &0.26$^{0.01}_{0.01}$ & 0.35$^{0.01}_{0.00}$ & 1.73$^{0.27}_{0.27}$ & 0.003$^{0.000}_{0.000}$ & 0.35$^{0.02}_{0.02}$ & 23.29$^{3.52}_{3.24}$ & 0.07$^{0.00}_{0.00}$ & 0.047$^{0.011}_{0.011}$ \\  
ELARS 20 &0.42$^{0.00}_{0.00}$ & 2.26$^{0.14}_{0.13}$ & 0.39$^{0.09}_{0.09}$ & 0.003$^{0.000}_{0.000}$ & 4.28$^{0.20}_{0.20}$ & 22.97$^{3.78}_{2.60}$ & 0.26$^{0.01}_{0.01}$ & 0.169$^{0.024}_{0.024}$ \\  
ELARS 21 &0.15$^{0.01}_{0.01}$ & 0.16$^{0.01}_{0.00}$ & 0.86$^{0.00}_{0.00}$ & 0.002$^{0.000}_{0.000}$ & 0.44$^{0.05}_{0.05}$ & 77.33$^{6.75}_{6.08}$ & 0.07$^{0.00}_{0.01}$ & 0.348$^{0.199}_{0.249}$ \\  
ELARS 22 &2.05$^{0.03}_{0.03}$ & 13.17$^{0.26}_{0.20}$ & 1.97$^{0.13}_{0.13}$ & 0.008$^{0.000}_{0.000}$ & 1.27$^{0.06}_{0.06}$ & 16.49$^{1.68}_{1.37}$ & 0.21$^{0.00}_{0.00}$ & 0.019$^{0.008}_{0.006}$ \\  
ELARS 23 &2.31$^{0.04}_{0.04}$ & 14.29$^{0.34}_{0.52}$ & 0.54$^{0.21}_{0.21}$ & 0.007$^{0.001}_{0.001}$ & 5.58$^{0.11}_{0.11}$ & 38.63$^{1.13}_{1.19}$ & 0.23$^{0.00}_{0.00}$ & 0.007$^{0.022}_{0.017}$ \\  
ELARS 24 &67.01$^{0.69}_{0.69}$ & 22.14$^{0.63}_{0.65}$ & 6.80$^{0.45}_{0.45}$ & 3.181$^{0.204}_{0.164}$ & 8.97$^{0.15}_{0.15}$ & 24.97$^{1.32}_{1.11}$ & 0.29$^{0.00}_{0.00}$ & 1.196$^{0.024}_{0.024}$ \\     
ELARS 25 &1.07$^{0.03}_{0.03}$ & 5.87$^{0.16}_{0.16}$ & 0.70$^{0.21}_{0.21}$ & 0.006$^{0.001}_{0.000}$ & 3.26$^{0.20}_{0.20}$ & 22.96$^{2.07}_{2.15}$ & 0.17$^{0.00}_{0.00}$ & 0.009$^{0.008}_{0.010}$ \\  
ELARS 26 &2.32$^{0.04}_{0.04}$ & 1.32$^{0.03}_{0.02}$ & 0.58$^{0.00}_{0.00}$ & 0.004$^{0.000}_{0.000}$ & 4.51$^{0.06}_{0.06}$ & 87.91$^{5.14}_{5.07}$ & 0.07$^{0.00}_{0.00}$ & 0.111$^{0.020}_{0.019}$ \\  
ELARS 27 &1.10$^{0.03}_{0.03}$ & 1.31$^{0.03}_{0.02}$ & 0.73$^{0.01}_{0.01}$ & 0.005$^{0.000}_{0.000}$ & 2.00$^{0.03}_{0.03}$ & 62.93$^{2.67}_{2.30}$ & 0.07$^{0.00}_{0.00}$ & 0.041$^{0.032}_{0.025}$ \\  
ELARS 28 &1.74$^{0.04}_{0.04}$ & 7.73$^{0.42}_{0.27}$ & 11.11$^{0.56}_{0.56}$ & 0.012$^{0.000}_{0.000}$ & 1.41$^{0.06}_{0.06}$ & 25.65$^{2.64}_{2.28}$ & 0.24$^{0.01}_{0.00}$ & 0.136$^{0.011}_{0.020}$ \\  
Tol 1214 &0.50$^{0.01}_{0.01}$ & 0.41$^{0.01}_{0.01}$ & 109.13$^{6.02}_{6.02}$ & 0.574$^{0.004}_{0.004}$ & 0.04$^{0.01}_{0.01}$ & 34.90$^{7.52}_{7.20}$ & 0.08$^{0.00}_{0.00}$ & 0.000$^{0.000}_{0.000}$ \\  
Tol 1247 &25.76$^{0.06}_{0.06}$ & 16.65$^{0.07}_{0.05}$ & 21.26$^{0.54}_{0.54}$ & 1.355$^{0.008}_{0.008}$ & 1.41$^{0.04}_{0.04}$ & 4.20$^{0.75}_{0.72}$ & 0.09$^{0.00}_{0.00}$ & 0.265$^{0.002}_{0.002}$ \\    
J1156    &34.22$^{0.76}_{0.76}$ & 30.03$^{0.47}_{0.45}$ & 1.62$^{0.01}_{0.01}$ & 0.316$^{0.003}_{0.002}$ & 3.45$^{0.30}_{0.30}$ & 13.58$^{3.32}_{5.34}$ & 0.06$^{0.00}_{0.00}$ & 0.286$^{0.008}_{0.011}$ \\  
\enddata
\renewcommand{\arraystretch}{1}
\tablecomments{(1--3) Extinction corrected SFR. (4--5) Based on extinction corrected \ha luminosity within a radius of $r_{\mathrm{p, H}\alpha}$.}
\end{deluxetable*}
Global luminosities and derived quantities are measured within
$r_{\mathrm{tot}}$, with three exceptions.  The first is $E(B-V)_n$, which uses
the smaller aperture size given by \hb (see Section~\ref{sec:circ}). The total
\lya aperture is sometimes quite small (in particular for weak or non-emitters)
and measured masses are therefore underestimated. To obtain total masses for
the galaxies we thus re-measure global masses within a circular aperture of
size determined from mass growth curves in the same way as for \hb (but with a
SNR threshold of 1). The HST imaging derived masses are in general much
higher than the masses from SDSS spectroscopy/imaging SED fits. This difference
come from the apertures used for the HST imaging being much larger than the
standard SDSS aperture.

The SFR density and specific SFR is determined within an isophotal region
determined from the \ha map with an effective radius of $r_{\mathrm{p,
H}\alpha}$ (given in Table~\ref{tab:sizes}).  This aperture is in general
smaller than the \lya total aperture, but provides a much safer estimate of
these observables due to large uncertainties in mass and SFR for large
apertures. It should be noted that the measurements from the HST \ha
observations sometimes show significantly higher SFR than the SDSS measurements
given in Table~\ref{tab:samplespec}. This difference comes from the
larger apertures used in our imaging measurements, when comparing
measurements for the same aperture size the HST SFR estimates are conistent 
(within the errors) with the SDSS measurements.
For consistency we use the imaging derived
masses, extinctions and SFRs throughout this paper.

In addition to the measurements obtained from HST maps we also use SDSS derived
values of $12 + \mathrm{log}(O/H)$ and \oiii~$\lambda5007/$\oii~$\lambda 3727$ ratio
(O32) from \citet{2020ApJ...892...48R}.  For the oxygen abundance the strong line N2P calibration
\citep{2007AA...462..535Y} is used, which can be estimated for all of the
galaxies in the sample.  The O32 ratio provides a diagnostic on the overall
ionisation state of the nebular gas \citep[e.g., ][]{2002ApJS..142...35K}.  The
luminosities and derived quantities for the full sample are given in
Tables~\ref{tab:globprop} and \ref{tab:globderiv}.

For the comparisons shown in this section we use the censored Kendall-$\tau$
test \citep[][as implemented in the R package \texttt{CENKEN}]{10.2307/2291140} to check for
correlations or anti-correlations. The censored test allows the inclusion of
upper limits in the calculations, so that also the \lya non-detected galaxies
can be included.  We use limits of $\left|\tau\right|>0.3$ and  $p_0<0.003$ to signify
a strong correlation (or anticorrelation), but also discuss weaker trends.  

\subsection{The luminosity, equivalent width, and escape fraction of \lya}
\label{sec:lumlya}
\begin{figure*}[pt!]
\centering
\includegraphics*[width=0.95\textwidth]{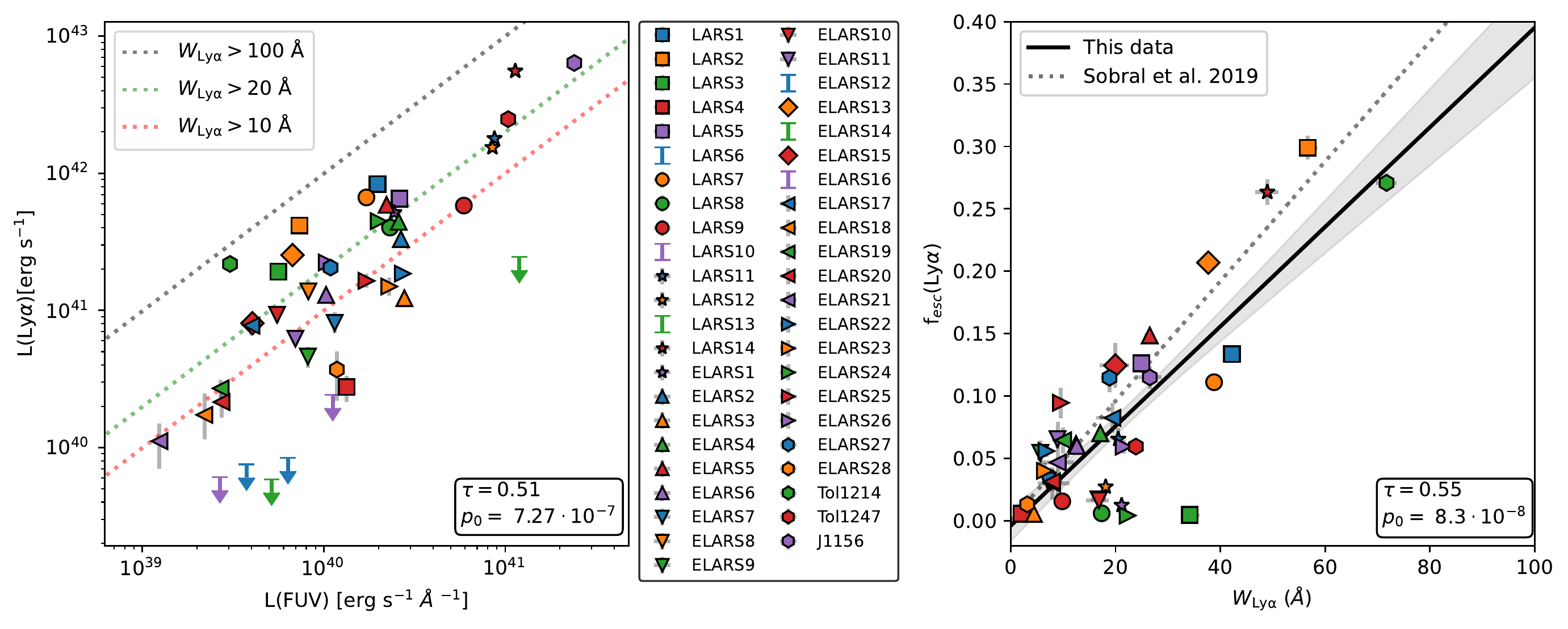}
\caption{Left panel: The \lya versus the FUV luminosity of the galaxies in this sample. 
	Diagonal dashed lines show different limits of \wlya. Right panel: 
	\fesc vs \wlya for the galaxies. In addition to a linear fit to the current data, the relation from
	\citet{2019A&A...623A.157S} is shown.}
\label{fig:llya_fesc_comb}
\end{figure*}

We detect global \lya emission from 39 out of 45 galaxies at a signal-to-noise ratio of 2 or more 
(the median ratio for the detected galaxies is $\sim$13). Three of the six 
non-detections are net absorbers and we measure a negative \lya flux for them.

The luminosity of \lya is expected to scale with FUV luminosity, primarily
through the connection to star formation with the production of \lya photons.
In Figure~\ref{fig:llya_fesc_comb} we show total luminosity of \lya versus the
FUV luminosity. There is indeed a strong correlation between the two ($\tau =
0.51$, $p_0 = 7.3\cdot 10^{-7}$), but also substantial scatter. The spread seen here
can be explained by the combination of dust extinction and resonant scattering of the 
\lya photons (see Section~\ref{sec:fescebv}). For an individual source this effect 
makes it hard to use \lya observations as a direct probe of star formation. 
The plot also shows the range of \wlya the sample covers, from a few to $\sim$100
\AA.

Figure~\ref{fig:llya_fesc_comb} also shows the relation between \fesc and
\wlya. Given that both of these quantities give \lya luminosity normalized with
measurements that are directly coupled to the star formation rate they are
expected to correlate, which is also what we find ($\tau = 0.55$, $p_0 =
8.3\cdot 10^{-8}$). However, also this relation shows significant scatter, in
particular at low \fesc.  The main difference between the two is that \fesc is
an absolute estimate (based on extinction corrected \ha) of the fraction of
\lya photons that escapes the galaxy, while \wlya provides a relative estimate
of the escape as compared to observed FUV continuum flux.  In galaxies where
much of the star formation is hidden by dust, the observed \wlya can still be
quite high because the global aperture measurements are dominated by the
non-dusty regions.  The extinction corrected \ha does not suffer from this
(modulo extinction uncertainties) and thus yields a more complete picture of
the ionizing photon budget. In these galaxies \fesc will sit below the locus in
the figure. In Section~\ref{sec:lyasfr} we fit \fesc as a linear function of \wlya and show how 
this allows \lya to be used as a rough SFR diagnostic.
Interestingly, all but one of the galaxies that fall well below the linear fit
of \fesc vs \wlya in the figure have high nebular and stellar extinction. 
In the single case which seems to have low nebular extinction but still
outlie the locus (ELARS22) the galaxy is inclined towards us and has a higher
stellar than nebular extinction (0.23 and 0.05, respectively), perhaps
indicating that the measured nebular extinction for this galaxy is under-estimated.

Another explanation that will cause a shift in this relation is the hardness of
the stellar continuum \citep{2019A&A...623A.157S}. A stellar population
generating more ionizing flux (due to, e.g., younger age, lower metallicity,
higher stellar rotation) will show a higher \lya equivalent width, at constant
\fesc. This effect can be measured by computing the ionizing production
efficiency, $\xi_{ion}$, which is the ratio of ionizing photons to the dust
corrected UV luminosity. For galaxies with multiple stellar populations a high
$\xi_{ion}$ value indicates that they are currently undergoing a starburst
phase, while lower values indicates a less active star formation phase. For our
sample we find an average $\log(\xi_{ion}/[\mathrm{Hz}\,\mathrm{erg}^{-1}]) =
24.9$ which is $\sim$0.5 dex lower than the typical value for high redshift
LAEs \citep[e.g.,][]{2018MNRAS.477.2098N}. This difference is not surprising
given that our sample includes many galaxies that have lower SFR than the high
redshift LAEs, and could explain the scattering of data points to lie to the
left of the fitted line (i.e. galaxies with lower than average $\xi_{ion}$). As
opposed to the effect of extinction, inspecting $\xi_{ion}$ for individual
galaxies to see where they lie in the plot does not show any clear effects. 

\begin{figure}[pt!]
\centering
\includegraphics[width=0.45\textwidth]{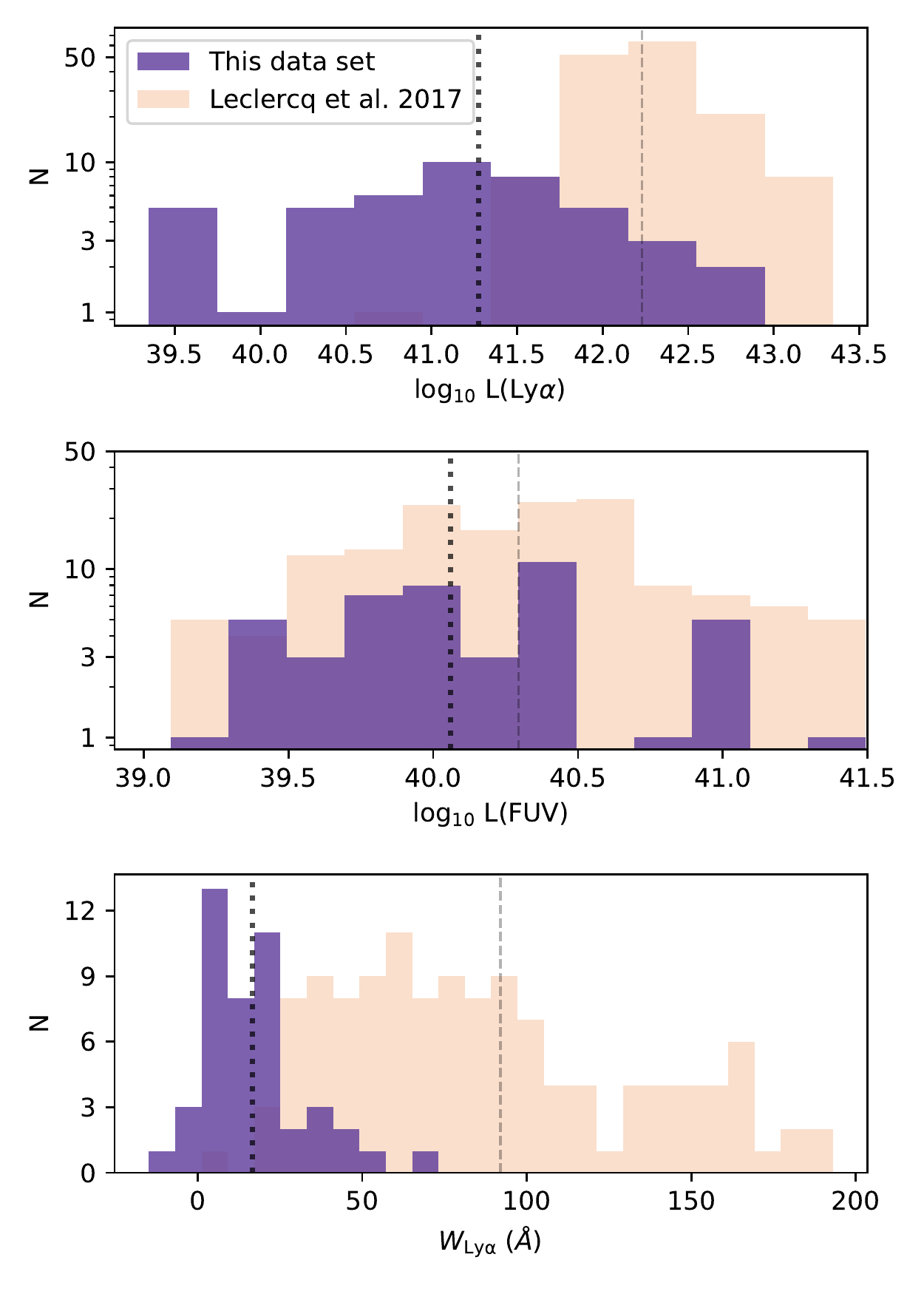}
\caption{Histograms of \llya (upper panel), \lfuv (middle panel), and 
\wlya (lower panel). In
addition we also show histograms for data on high redshift LAEs from
VLT/MUSE observations \citep{2017A&A...608A...8L}.  Thick dotted vertical lines
show the median \llya, \lfuv, and \wlya values for the net emitters 
in our data set and the thin dashed lines show the medians for the MUSE LAEs.}
\label{fig:llya_hist}
\end{figure}

In Figure~\ref{fig:llya_hist} we show the range of \lya luminosities, FUV
luminosities, and \wlya covered by our sample. Most of our galaxies have
$40<\log$(\llya)$<42$ with a tail towards higher luminosities. Note that the
non-emitters are included in the lowest bin. Comparing our data to the high
redshift LAEs from \citet{2017A&A...608A...8L} it is clear that the LARS galaxies are in
general fainter in \lya (median $\log$(\llya) for LARS and MUSE LAEs are
41.3, and 42.3, respectively), but with some overlap. This difference could be
due to luminosity bias, but is more likely to be caused by the quite different
sample selection in this case. The MUSE LAEs are selected based on their \lya
emission, while the LARS galaxies were selected based on SFR and stellar
population age. Indeed, when comparing the FUV luminosities of the two samples
the difference is markedly lower (median $\log$(\lfuv) for LARS and MUSE LAEs are
40.1, and 40.3, respectively).

The same difference is seen in the \wlya histogram where the two samples are
quite distinct (median \wlya for LARS and MUSE are 17 and 92 \AA,
respectively). At high redshift finding the most
extreme LAEs is more likely because of the larger volume covered, but also due to
observing galaxies at the peak of star formation. 
It should thus be noted that the results from this study have to be treated
with some care when comparing to high redshift studies, in particular if only
the highest \wlya emitters are observed.  However, the LARS+ELARS galaxies should be
more representative of the \lya properties of the general star forming galaxy population, with 
more dust and weaker \lya emission.

For the 14 galaxies previously presented in \citet{2014ApJ...782....6H} we note
that the \lya luminosities, equivalent widths, and escape fractions have
changed compared to the previous values. This is mainly due to the updated
photometric calibration of ACS/SBC (10-30\% change depending on filter) and WFC3/UVIS (10\% for the F336W filter), 
but is also affected by the difference in global aperture between the earlier paper and this work.

\subsection{The relation between \fesc and other properties}
\label{sec:fesc_scatter}
\begin{figure*}[pt!]
\centering
\includegraphics*[width=0.95\textwidth]{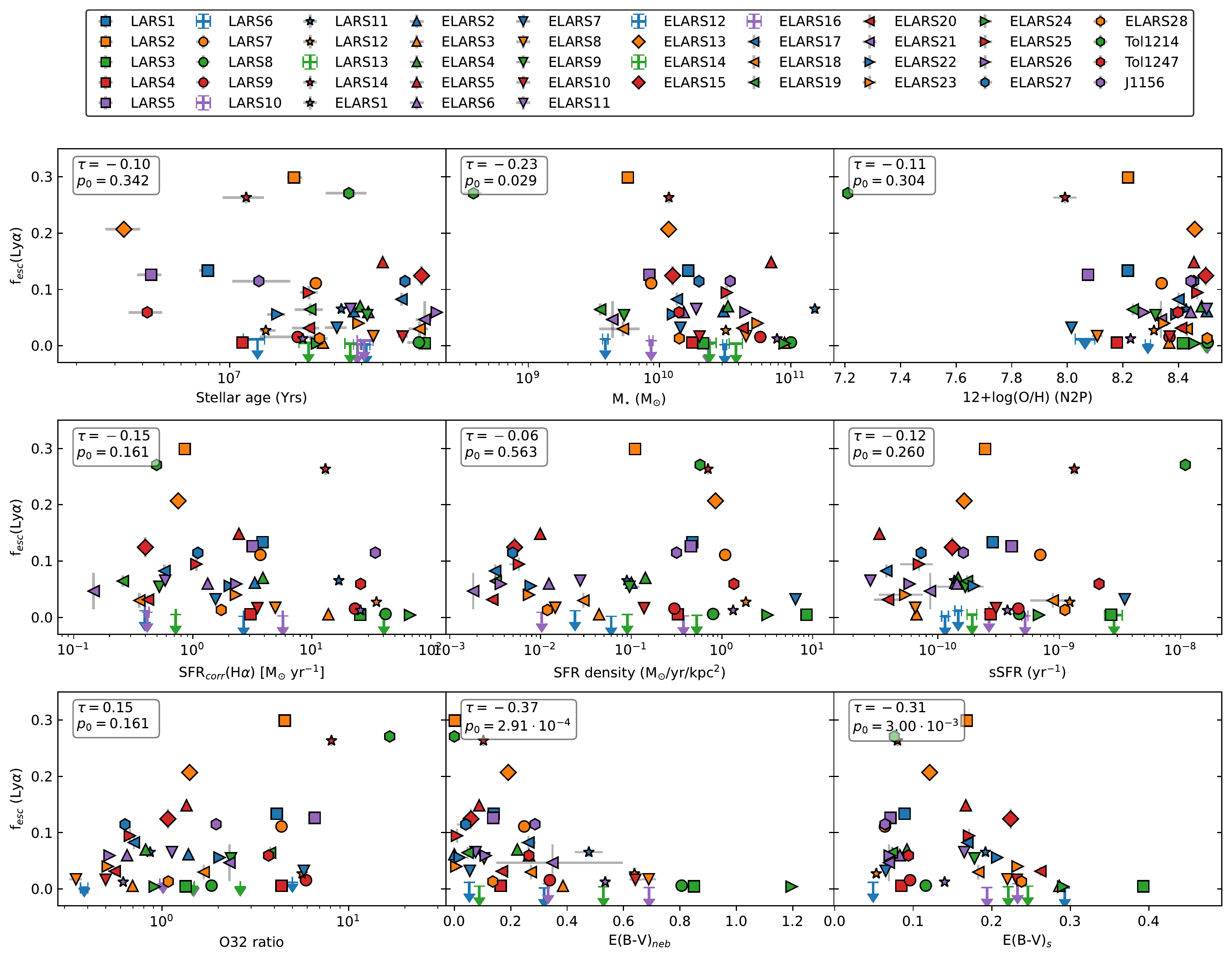}
\caption{Scatter plots of \fesc versus various properties of the galaxies.}
\label{fig:fesc_scatter}
\end{figure*}
In Figure~\ref{fig:fesc_scatter} we show the relation of \fesc to various
properties of the galaxies that have been suggested to affect the escape of
\lya photons. 

The stellar age and star formation quantities (global SFR, SFR density and
specific SFR) are all related to ISM feedback and will strongly affect the
kinematics of the ISM, both in terms of bulk motions and turbulence.
With high relative (to the \lya production sites) bulk motion of the neutral
ISM \lya photons will be able to escape in the wings of the ISM absorption
profile \citep[e.g., ][]{2017arXiv170403416D}. Feedback induced turbulence and shocks will cause
instabilities in the ISM that can create channels of low \hi column density
where \lya photons can travel relatively unimpeded, also causing galaxies to 
have dispersion-dominated kinematics \citep{2016A&A...587A..78H}. 
In the upper left
and three middle row panels of the figure we show scatter plots of \fesc versus
these quantities and find no strong correlations. For stellar age this may be
because we use a global average (FUV-weighted) value, which includes star
forming regions that are not directly involved in the \lya radiative transfer.
\citet{2019MNRAS.486.2215K} perform radiation-hydrodynamic simulations 
of \lya escape from star-forming regions and find that \fesc increases with stellar age up to 
$\sim 10$ Myr. Given that our total apertures contain multiple stellar populations with 
different ages and \lya output, that we do not see this effect in our sample is expected.
The overall lack of correlations with the SFR quantities could be
explained by high SFR being linked to other properties that makes escape
harder. For example, at low redshift the highest (extinction corrected) 
SFR is often found in the most dusty galaxies, and while similar systems 
are hard to find at high redshift they do exist \citep[e.g., ][]{2014PhR...541...45C}, and
many more are waiting to be detected by JWST. These dusty galaxies are also more likely to 
have higher global neutral gas masses. Hence, the increased mechanical energy injected
into the ISM from high SFR may be combined with higher dust and gas content which will limit
the escape of \lya photons.
Even though mechanical feedback from star formation
will increase the escape probability on scales of \hii regions, this is not seen 
in global measurements for the sample.

Both stellar mass and nebular metallicity are quantities that are linked to
each other and to other galaxy properties that may affect \lya escape, most
notably \hi mass (for the stellar mass), and dust extinction (for metallicity).
In the top middle and right panels of Figure~\ref{fig:fesc_scatter} we show
scatter plots of \fesc versus these two properties. For stellar mass we find a
weak anti-correlation ($\{\tau, p_0\}=\{-0.23, 0.029\}$) with substantial
scatter.  The anti-correlation is mainly driven by the lack of massive galaxies
with high \fesc.  From this data set it seems clear that \lya escape is low in
galaxies with the highest stellar mass, perhaps reflecting that these galaxies
have larger neutral gas reservoirs \citep{2014ApJ...794..101P}. The higher
neutral gas content will in turn cause longer scattering path lengths for the
\lya radiative transfer, and could also make it harder for the star formation
feedback to create clear channels for escape. For low stellar masses the
scatter in \fesc is large, this could be explained by variations in the neutral
gas fraction, e.g., some of these galaxies may have small stellar masses
compared to the neutral gas mass. In these galaxies \fesc will be low because
of high \hi column densities, while low mass galaxies that have used up a
greater portion of their neutral gas will have higher escape fractions. Lower
mass galaxies also have weaker gravitational potential which will make it
easier for feedback to create low column density channels for \lya escape
\citep{2009ApJ...704.1640L}. This effect would make \fesc higher in low-mass
galaxies on average, but would also show considerable scatter due to
geometrical effects.  A more detailed analysis of \hi column densities and
kinematics for this sample, and how they relate to \lya escape will be given in
a forthcoming paper (Le Reste et al., in prep.). The comparison with nebular
metallicity does not show anything conclusive, but we note that there is a lack
of data points in the low metallicity range.

The effect of ionisation state of the main star-forming regions in the galaxies
(as measured by the SDSS O32 ratio) on global \fesc is shown in the bottom row
left panel. We do not find any correlation of O32 with \fesc, but note that
there are very few galaxies with low O32 (indicating that the ISM is not
strongly ionized) that show strong \lya escape. A highly ionized ISM in the
regions where \lya is produced thus seems to be a necessary condition for high
($\gtrsim 0.1$) \fesc. This result is consistent with earlier studies that also
show a large scatter of \fesc at low O32 ratios \citep[e.g.,
][]{2015ApJ...809...19H, 2017ApJ...851L...9J, 2017ApJ...844..171Y,
2020MNRAS.491..468I}, but we do not find any strong correlation. However, it
should be noted that the sample selection is quite different for some of these
studies in that the galaxies are all \lya emitters and are selected to have
strong emission. Our sample contains a number of galaxies with moderate to
strong \lya absorption which will increase the scatter substantially which is
also seen in the study by \citet{2020MNRAS.491..468I}.

In the bottom row middle and right panels we show scatter plots of \fesc versus
nebular and stellar extinction. Both of these are strongly anticorrelated with
the escape fraction, $\{\tau, p_0\}=\{-0.37, 2.9\cdot 10^{-4}\}$ for
$E(B-V)_n$, and $\{\tau, p_0\}=\{-0.31, 3.00\cdot 10^{-3}\}$ for
$E(B-V)_s$. The correlation with nebular extinction was previously found by
\citet{2020ApJ...892...48R} in a rank correlation study, where \fesc was found
to have a strong dependence on $E(B-V)_n$. We will come back to the details of
nebular extinction and escape fraction in Section~\ref{sec:fescebv}. Given that
nebular and stellar extinction are strongly correlated it is not surprising to
find that the escape fraction is also anti-correlated to the stellar
extinction.

\begin{figure*}[pt!]
\centering
\includegraphics*[width=0.95\textwidth]{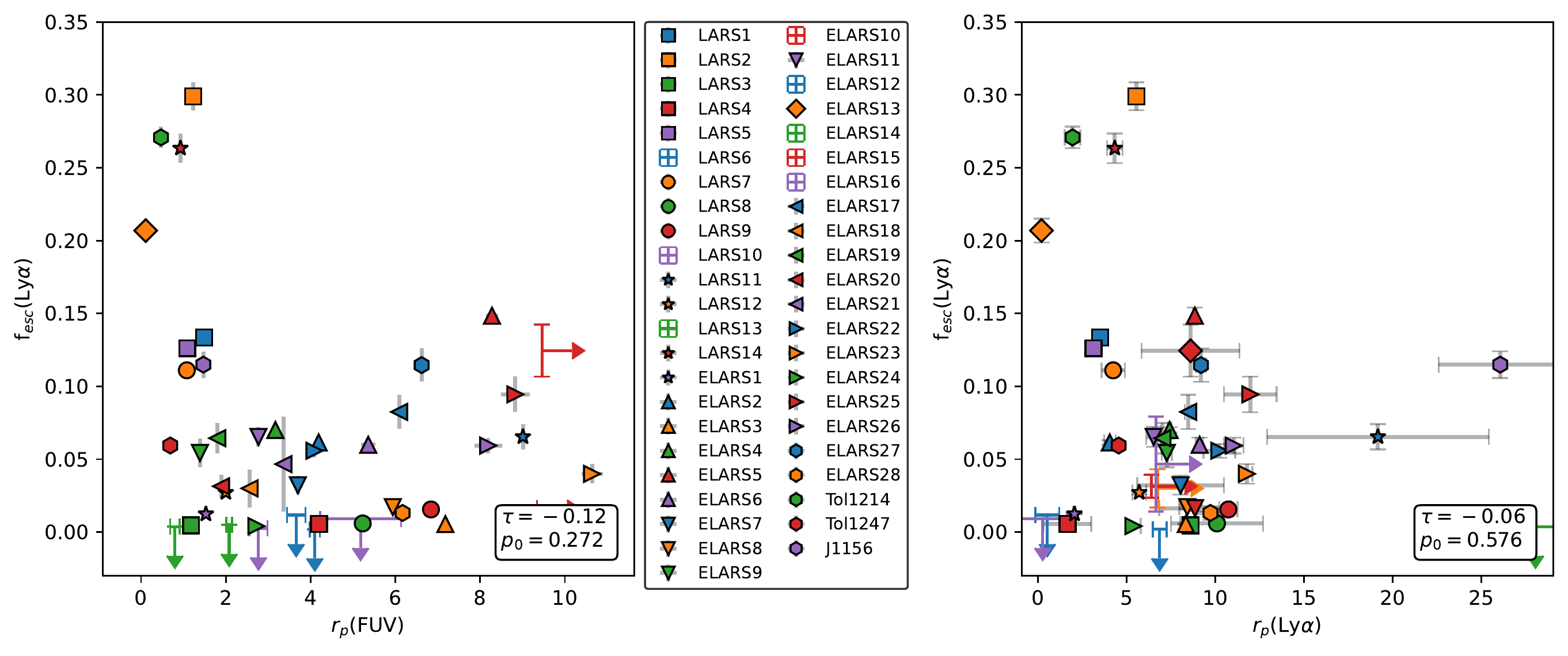}
\caption{Scatter plots of \fesc versus petrosian radii in FUV (left panel) and \lya (right panel). 
Note that the galaxy id labels refer to the left panel, therefore some of the galaxies with limits
in the right panel are not shown as such in the legend. Please refer to Table~\ref{tab:globprop} for individual galaxy limits.}
\label{fig:rpfig}
\end{figure*}
\begin{figure*}[pt!]
\centering
\includegraphics*[width=0.95\textwidth]{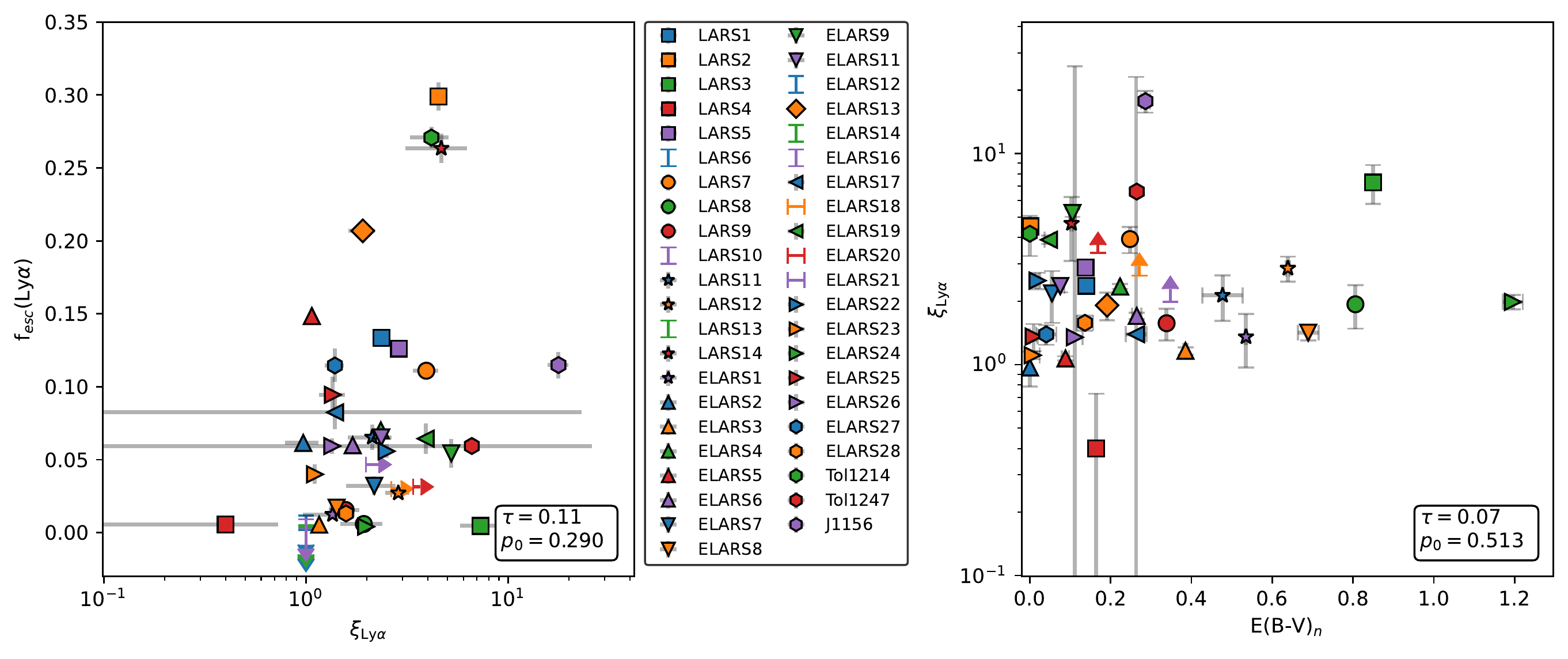}
\caption{Left panel: \fesc versus the $\xi_{\mathrm{Ly}\alpha}$ parameter. 
	Right panel: $\xi_{\mathrm{Ly}\alpha}$ versus nebular extinction.
	Galaxies with lower limits on $r_{\mathrm{p, FUV}}$ (ELARS15 and ELARS22) are
	excluded from these plots.}
\label{fig:xifig}
\end{figure*}
\subsection{The \lya spatial distribution and escape fractions}
\label{sec:lyaspatial}
As previously mentioned, a more thorough analysis and discussion of how the
spatial distribution of \lya relates to global observables for the sample in
this study is given in \citet{2022A&A...662A..64R}. In this paper we provide two simple 
observational diagnostics for the \lya spatial extent: $r_{\mathrm{p, Ly}\alpha}$, 
and $\xi_{\mathrm{Ly}\alpha}$ (defined below). We also investigate the effect 
of FUV compactness (as measured by $r_p(FUV)$) on the escape fraction. 

In Figure~\ref{fig:rpfig} we show the dependence of \fesc on the 
compactness (as measured by $r_{\mathrm{p}}$) of FUV and \lya emission. There are 
no correlations found for any of these properties, but we note that high escape fractions
are much less common in galaxies with large $r_{\mathrm{p, FUV}}$. Galaxies with very extended 
FUV emission thus seem to be less likely to have high \fesc. 

We define the \lya overextension parameter, $\xi_{\mathrm{Ly}\alpha}$, as the
ratio between the \lya and FUV continuum petrosian radii:  
\begin{equation}
\xi_{\mathrm{Ly}\alpha} = \frac{r_{\mathrm{p, Ly}\alpha}}{r_{\mathrm{p, FUV}}}.
\end{equation}
This ratio is high for galaxies which have extensive \lya halos caused by
radiative transfer of the \lya photons far away from where they were produced.
The parameter is also high for galaxies with substantial continuum absorption
in the central regions. Lower $\xi_{Ly\alpha}$ values are found in galaxies in 
which the Ly$\alpha$ escape happens along the line of sight or in which resonant 
scattering is less significant. 
In Figure~\ref{fig:xifig} we show how
$\xi_{\mathrm{Ly}\alpha}$ relates to \fesc and $E(B-V)_n$ for our sample. We do
not find any correlation between \fesc and $\xi_{\mathrm{Ly}\alpha}$.  Naively
one might expect higher \lya escape for galaxies where the photons escape
directly with little or no radiative transfer (giving low
$\xi_{\mathrm{Ly}\alpha}$), but there are likely a number of competing effects
which cause this prediction to fail. For example, in galaxies which have high
escape fractions due to low dust content photons tend to travel longer paths
before being destroyed, which naturally gives rise to larger \lya halos.
\citet{2014ApJ...782....6H} found a anti-correlation between $\xi_{\mathrm{Ly}\alpha}$ and
$E(B-V)_n$ for the original LARS galaxies, however, this result is not
confirmed with the updated calibration and when including the 31 additional galaxies in this sample.

\section{Discussion}
\label{sec:disc}

\begin{figure}[pt!]
\centering
\epsscale{1.2}
\plotone{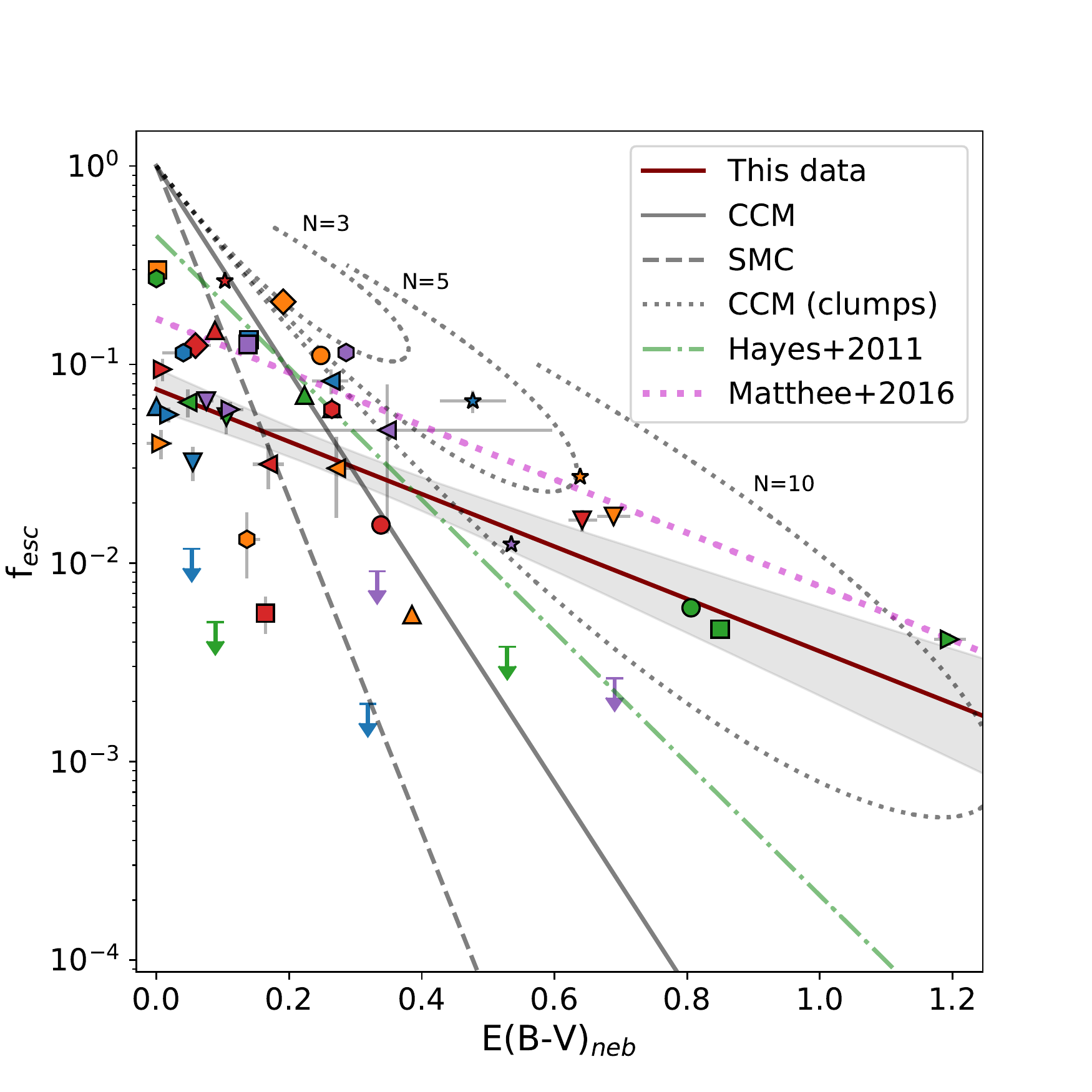}
\caption{Global escape fraction of \lya versus nebular extinction. The gray lines show 
	where the points would fall for different ISM extinction curves. The curve labeled CCM is
	the galactic extinction curve from \citet{1989ApJ...345..245C} and SMC is from \citet{1984AA...132..389P}.
	Coloured lines show curves derived from fitting a line to observation from this and other \lya studies. Note that that \citet{2011ApJ...730....8H} and 
	\citet{2016MNRAS.458..449M} use stellar extinction, for this comparison 
	we multiply their $E(B-V)$ values by a factor 2.27 \citep{2000ApJ...533..682C}. The dotted gray lines (labeled CCM (clumps)) show clumpy dust models with varying numbers of clumps along the line of sight.}
\label{fig:fescebvn}
\end{figure}

\subsection{\lya escape and dust extinction}
\label{sec:fescebv}
As shown in Section~\ref{sec:fesc_scatter} we find a strong anti-correlation
between \fesc and $E(B-V)_n$.  In Figure~\ref{fig:fescebvn} we show this
relation together with a number of extinction curves from the literature. When
comparing to dust screen models (CCM and SMC) it is clear that 
\lya escape cannot be described by these simple models, both in terms of the 
average slope of the curve and in terms of scatter. Similar to \citet{2011ApJ...730....8H} we
make the assumption that \fesc at $E(B-V)_n \approx 0$ can be $<1$ (even very small amounts
of dust will affect \lya photons due to the scattering), and fit a function of the form:
\begin{equation}
	f_{\mathrm{esc}} = C \cdot 10^{-0.4 k_{\mathrm{Ly}\alpha} E(B-V)_n}.
\end{equation}
We then fit a linear function to $\log$ \fesc (using a censored
bayesian linear regression method, \texttt{LINMIX}
\footnote{\url{https://github.com/jmeyers314/linmix} and \citet{2007ApJ...665.1489K}})
which yields the parameter values $C = 0.075\pm 0.019$ and
$k_{\mathrm{Ly}\alpha} = 3.30\pm 0.73$ where the slope and intercept
uncertainties are derived from the posterior probabilities. The best fit slope
is significantly flatter than the CCM and SMC curves. 

There are multiple causes for the scatter in the relation that can make
the \lya escape fraction move both up and down \citep[see also ][]{2011ApJ...730....8H}.
\begin{itemize}
\item{Resonant scattering of the \lya photons in the neutral hydrogen gas cause
the effective path lengths (or optical depth) to increase. This means that more
photons are absorbed by dust and \fesc consequently goes down \citep[e.g.,
][]{2017arXiv170403416D}. For this data set the effect is clearly seen at low
nebular extinction. Even with minimal dust content the effective optical depth
at the \lya wavelength becomes large and the escape fraction is strongly
decreased compared to the prediction of the dust screen model.}
\item{In a clumpy ISM (where most of the dust is confined within the dense
clumps of \hi gas), the \lya photons can scatter on the surface of the clumps
and thus avoid entering the regions where the dust content is highest. In this
situation \lya photons may suffer less from extinction than both the stellar
continuum and the optical nebular line emission \citep[an effect found in the
theoretical work of ][]{1991ApJ...370L..85N}, causing the escape fraction to be
higher than the dust screen models would imply. However, if this effect was
present for a majority of the galaxies we would also find a positive
correlation of \wlya and $E(B-V)_n$ \citep[][]{2009ApJ...704L..98S}, which we
do not find (in fact, they are anti-correlated to the same degree that \fesc
and $E(B-V)_n$ is).  More recent radiative transfer simulations have also found
that this type of scattering does not have a large effect on the output \lya
emission \citep[e.g., ][]{2013ApJ...766..124L,2016A&A...587A..77D}.  
It is not excluded that \fesc
could be boosted by this effect in some of the sources, but it does not seem to
be in play for the sample as a whole.}
\item{Pure dust geometrical effects can also cause \fesc to be higher than for
the uniform dust screen models. In Figure~\ref{fig:fescebvn} we show
the effect of a clumpy dust screen \citep{1984ApJ...287..228N,
1994ApJ...429..582C, 2009ApJ...704L..98S} on the \lya escape
fraction. As the dust content inside the clumps increases the
\fesc-$E(B-V)_n$ relation will move along these locii (N is the
average number of clumps along the line of sight).  The
extinction will then become more and more gray --- as the
individual clump extinction becomes high the attenuation will
tend toward a constant $e^{-\mathrm{N}}$. Many of the galaxies
in our sample (in particular the ones with high extinction)
show escape fractions that are consistent with clumpy dust
geometry.}
\end{itemize}

The overall correlation of extinction and \fesc was found already in the
original LARS publications \citep{2014ApJ...782....6H}, and also by other
theoretical and observational studies at low and high redshift
\citep{2008A&A...491...89V, 2009A&A...506L...1A,2009ApJ...704L..98S,
2010ApJ...711..693K, 2011ApJ...730....8H, 2016MNRAS.458..449M}. In order to
make the comparison to the studies using stellar extinction we use the
conversion $E(B-V)_n = 2.27 E(B-V)_s$ \citep{2000ApJ...533..682C}, and convert
the stellar extinctions given in the other publications to nebular. This ratio
of stellar to nebular extinction is explained by nebular emission originating
from regions within galaxies with higher density ISM, but was derived for a
small sample of nearby galaxies. The number is thus not necessarily applicable
to the high redshift samples compared with here, but a similar ratio ($2.077\pm
0.088$) was found by \citet{2020ApJ...902..123R} for sample of $z\sim 2$
galaxies.

Similar to this study these authors find that if \lya
escape were to be explained by extinction alone, the nebular attenuation curve
needs to be considerably more shallow than the standard dust screen models.  In
this paper we include galaxies with substantial extinction, and find a
shallower slope than all but one of the earlier studies. The slope is
remarkably similar to the one found by \citet{2016MNRAS.458..449M}, who uses a sample of
488 \ha-selected galaxies at $z\sim 2$. The sample in their study includes
galaxies with $E(B-V)_s \lesssim 0.5$ which, when applying the stellar to
nebular extinction conversion, gives a similar range of extinctions as in this
data set.

The shallow effective attenuation curve for \lya thus seems to be caused by a
combination of radiative scatter and dust geometry effects. At low extinction,
resonant scattering causes the optical depth for \lya photons to become longer
and \fesc consequently is lowered. The effect of a clumpy dust geometry on the
effective attenuation is small at low dust content, and the average \fesc thus
becomes lower than a simple dust screen would predict ($0.075$ at $E(B-V)_n = 0$
for the fit above).  At higher dust content resonant scattering seems to become
less important, and the effects of a clumpy dust geometry results in higher
\fesc than predicted. One interpretation of this result is that the \lya
radiative transfer in sufficiently dusty galaxies starts to be less dominated
by resonant scattering effects. Instead it is the dust content and clump
distribution that has the largest effect on the escape fraction. Remarkably,
none of the galaxies with $E(B-V)_n >0.6$ sits below the dust screen models in
the figure, indicating that optical depth effects from resonant scattering are
less prominent in this regime.  The caveat to this conclusion is that resonant
scattering in a clumpy \hi+dust medium can also lead to boosted \lya escape
\citep{1991ApJ...370L..85N}. However, as discussed above, this
effect is likely not enough to cause the observed escape fractions.

While a general prediction of the \fesc evolution with redshift based on the
overall changes to extinction over cosmic time \citep{2011ApJ...730....8H} is outside the
scope of this paper we note that the significantly lower slope found in this
study means that \fesc would evolve less with redshift compared to the earlier
study. 

\subsection{\lya luminosity as a diagnostic of star formation rate}
\label{sec:lyasfr}
\begin{figure}[pt!]
\centering
\epsscale{1.2}
\plotone{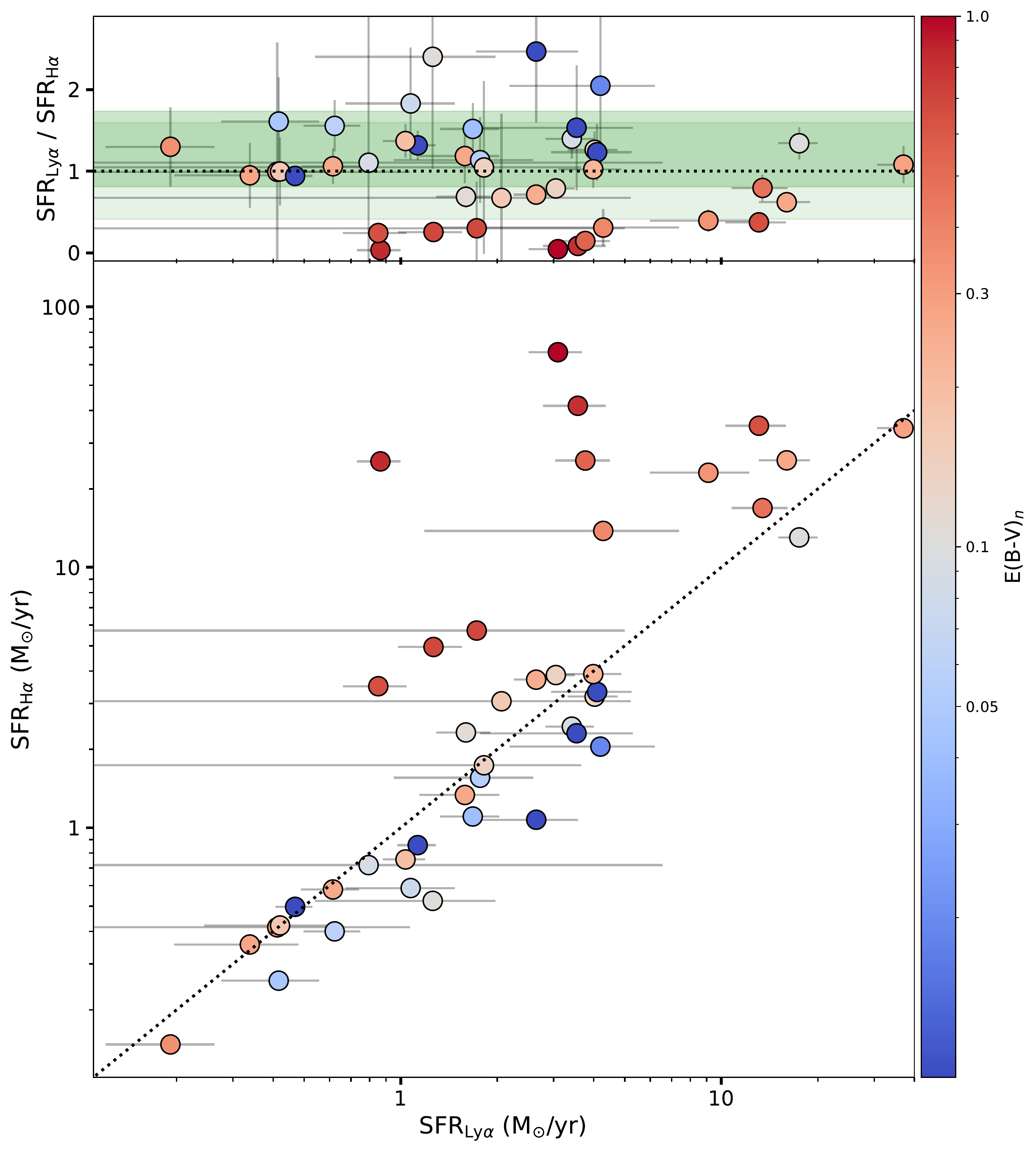}
\caption{Comparison of the \fesc-corrected \lya SFR diagnostic with SFR derived from \ha. The color scale of the 
	circles indicate the $E(B-V)_n$ of the galaxies. Galaxies with a zero or negative \lya luminosity are excluded from this plot. 
	The dotted line shows the 1-to-1 relation.
	The upper plot shows the ratio SFR$_{\mathrm{Ly}\alpha}/$SFR$_{\mathrm{H}\alpha}$. The light green band shows the
	standard deviation of the ratios for the full sample, while the darker green band shows the standard deviation for a sub-sample with $E(B-V)_n<0.3$.}
\label{fig:sfrcomp}
\end{figure}
The use of \lya luminosity to measure star formation rates in galaxies is
problematic because of the uncertainties in escape fraction. In contrast to
other nebular hydrogen lines (e.g., \ha or \hb) a simple dust correction is not
enough to get measurements of similar precision. Nevertheless earlier studies
showed  correlations of \lya luminosity with FUV \citep[e.g.,
][]{2016A&A...587A..98W} or \ha luminosity \citep[e.g.,
][]{2010Natur.464..562H}, albeit with large scatter.
\citet{2019A&A...623A.157S} used a sample of 218 \lya+\ha emitters at
$z\lesssim 0.3$ and $z\sim$2.2--2.6 to derive a SFR calibration valid for \lya luminosity.
They find a fairly tight linear relation betwen \fesc and \wlya and use this to
correct the observed \lya luminosities which can then be used to define a
\lya-based SFR calibration.  This calibration does not require any knowledge of
the extinction or escape fraction, but can be used with \wlya and luminosity
alone.

In Figure~\ref{fig:llya_fesc_comb} we show a linear fit of \fesc as a function of \wlya 
(again using \texttt{LINMIX} referenced above):
\begin{equation}
	f_{\mathrm{esc}} = \left(0.0040\pm 0.00050
	\cdot W_{\mathrm{Ly\alpha}} \right) \pm 0.013,
\end{equation}
where the error to the right is the uncertainty of the fitted intercept (the
best-fit intercept value itself is negligible compared to the uncertainty).
While the fit has a quite small statistical uncertainty ($\sim$5--10\% for the range 
of \wlya in this sample), there are also
systematic effects that come into play (see Section~\ref{sec:lumlya} for a discussion
of these), in particular the effect of extinction. The found relation is
significantly shallower than the one found by \citet{2019A&A...623A.157S}, even when
comparing to their low redshift fit. As discussed by these authors, the
relation of \fesc versus \wlya is affected strongly by the dust content of the
galaxies, and the difference in slope we find could be due to the inclusion of
more dusty galaxies in our sample (or other selection effects).

To find an expression for SFR as a function of \lya luminosity we rewrite 
Equation~\ref{eq:fesc} and use the \ha SFR calibration of \citet{2012ARA&A..50..531K}. 
We then replace \fesc with the relation found above and arrive at:
\begin{equation}
\mathrm{SFR}_{\mathrm{Ly}\alpha} = \frac{C_{\mathrm{H}\alpha}^{-1}}
	{8.7 f_{\mathrm{esc}}} L_{\mathrm{Ly}\alpha} =
	\frac{5.37\cdot 10^{-42}}{0.035 \cdot W_{\mathrm{Ly\alpha} }} 
	L_{\mathrm{Ly}\alpha},
\end{equation}
where $C_{\mathrm{H}\alpha}$ is the SFR calibration constant from 
\citet{2012ARA&A..50..531K}.
Note that this calibration is valid for extinction corrected \ha, and
SFR$_{\mathrm{Ly}\alpha}$ thus also provides a dust corrected estimate of the
star formation rate.  The uncertainty of the \fesc(\wlya) relation translates
into an uncertainty on the SFR of $\sim$5--10\%.  In Figure~\ref{fig:sfrcomp}
we show how the \lya-based SFR calibration compares to the \ha SFR for our
sample. Overall the \lya-based rate follows the \ha calibration well, but the
scatter is large. The average relative residual for the full sample is 5\%, but
with a scatter of 59\% (0.20 dex). The standard error of mean ($\sigma /
\sqrt{N}$) for the calibration is 9\%.  The high scatter seems to be driven by
the systemic uncertainties of the \wlya to \fesc conversion, and in particular
the high extinction galaxies that have severely underestimated SFR. If we
instead consider only galaxies with low extinction ($E(B-V)_n<0.3$) the average
relative residual is $+30$\%, the scatter is reduced to 46\% (0.16 dex), and
the standard error of mean stays the same. The positive offset found in this
case may reflect systematic effects on the \fesc relation resulting from
variations in the ionizing radiation production efficiency. Galaxies with 
less bursty star formation (older average stellar ages) will tend to 
have over-estimated escape fractions from the fit and will thus get an 
overestimated SFR from the diagnostic. The \fesc to \wlya relation 
stays the same (within the uncertainties) when excluding high extinction galaxies.

Another systematic uncertainty in the calibration is the assumption on the
value of $C_{\mathrm{H}\alpha}$, which itself depends on the age of the stellar
population generating the ionizing photons (and the ionized ISM temperature).
If the age is significantly lower than the assumption made by
\citet{2012ARA&A..50..531K}, with massive star formation already in
equilibrium, the SFR could be overestimated by a factor $\sim 2$ (for ionizing
clusters with age $\leq 3-4$ Myr). Note that if the \fesc--\wlya relation is
inverted and \wlya at 100\% escape calculated, the intrinsic equivalent widths
would fall around 200 -- 400 \AA. These values are typical of very young bursts
of massive star formation, which are the ones probably driving the observed
correlation.  In any case, we want to stress that this does not affect the
comparison shown in Figure~\ref{fig:sfrcomp}, given that the SFR indicators
there use the same constant. For simplicity and re-usability we choose to use
the standard calibration in this work.

The scatter found from comparing to SFR$_{\mathrm{H}\alpha}$ is comparabe to the
one reported by \citet{2019A&A...623A.157S}, who, comparing to dust corrected UV-based
SFR find a scatter of 0.2--0.3 dex. It is also similar to the typical scatter
found when comparing extinction corrected \ha SFR to FIR SFR, $\sim 0.3$ dex
\citep[e.g., ][]{2012MNRAS.426..330D}. By applying a cut on the extinction we
find a scatter which is somewhat lower. Variations in star formation histories of the 
galaxies (giving rise to variations in the ionizing radiation production efficiency) 
are likely responsible for part of the remaining scatter. The \lya
based SFR calibration presented here can thus provide estimates for SFR in
high redshift LAEs which do not have any optical line measurements. 
These estimates will be better if the galaxies have low extinction 
and similar star formation history as our sample. Without access to 
the optical lines it is hard to constrain the extinction, but a rough selection 
could be done using rest-frame UV stellar continuum slopes or SED fitting.

\subsection{The effect of aperture size on measurements of \fesc and \lya luminosity}
\label{sec:eff_aploss}
In Section~\ref{sec:aplosses} we show and describe how we use stacked growth curves to
estimate the aperture losses of \lya in the sample and how this translates to
some other observational cases. The quite severe losses estimated for our
sample do not change the results found for \fesc, but the SFR calibration
comparison discussed in the previous section could be affected. However, given the large
scatter for this relation the overall effect is too small to make a difference
to the comparison.

\lya emitters at $z\sim 0.3$ is an interesting case because it is possible 
to observe Lyman continuum (LyC) emission with HST/COS at this redshift. A number of 
studies targeting galaxies at this redshift have detected LyC emission, and also observe the
\lya spectrum with COS \citep[e.g., ][]{2018MNRAS.478.4851I, 2022ApJS..260....1F}. 
Escape of ionizing photons have been found to correlate with \fesc, but if
there is a significant amount of \lya flux missing from these measurement
these conclusions could be misleading. In this study we find that $27\pm 23$\%
of the \lya flux is outside of the COS aperture. An average offset of this
magnitude for the individual galaxies in these studies will not change the
found correlations, but if there are other dependencies on the amount of \lya
that is emitted from the outskirts of galaxies there may be systematics of the
relations that are poorly constrained. In addition, Rasekh et el. (in prep.) 
finds that the fraction of \lya emitted far outside the star-forming
regions for Green Peas seems to be considerably smaller for Lyman continuum
leaking galaxies \citep[see also][]{2015A&A...578A...7V}. The aperture losses found
for the sample in this paper may thus not be applicable to these galaxies and
are probably lower than the number reported here.

At high redshift ($z\sim 3$) the aperture losses are consistent with zero for
a typical slit width of 1\arcsec, and measurements using this setup should
provide good estimates of the total \lya flux. However, at extreme redshifts
and when using narrow slits (e.g., using the MOS mode with JWST/NIRSPEC and a
single slitlet width) the slit loss is very high. 
In addition, 
at high redshifts, much more of the circumgalactic medium and ISM inside 
galaxies is expected to be neutral. Radiative transfer simulations show that 
this can lead to more radiative transfer of \lya photons and thus larger 
spatial sizes of \lya halos \citep{2019A&A...627A..84L}. Our estimated 
aperture losses should thus be treated as lower limits in this case.  

When using the estimated average aperture losses 
to correct \fesc is should be noted that the values presented here require 
that all of the \ha emission is measured 
(i.e. the \ha aperture covers all of the star forming regions in the galaxy).
In a matched aperture COS (FUV) and SDSS (optical) setup observing nearby 
galaxies also \ha will be affected by aperture losses and the simple correction 
suggested here suffer from additional systematic uncertainty.
This could also
be true for the NIRSPEC setup pictured above where the very narrow slit width
will not cover the full star-forming extent of the galaxies. In the
intermediate redshift range all of the \ha emission will be easily measured and
our loss values should hold.

\section{Summary}
\label{sec:summary}
In this paper we present \lya imaging of a sample of 45 star-forming 
galaxies at low redshift. We describe the sample selection and data reduction, 
including specialized tools to match the FUV PSF to the optical data. 
We also provide a detailed description of the pixel SED-fitting code we use
to obtain secure \lya, \ha, and stellar population property 
maps, as well as error maps for these quantities, for the galaxies.

We then perform circular photometry on the resulting maps to obtain global
measurements of fluxes (\lya, \hb, \ha, and FUV stellar continuum) and stellar
properties (mass, luminosity-weighted stellar population age, and stellar dust
extinction). The measurements are also used to derive maps and global
measurements of \lya equivalent width, \lya escape fraction, star formation
rate, and nebular dust extinction. Different size estimates for the \lya, \ha,
and FUV emission are obtained using Petrosian radii and growth curve analysis.
The main results of the study are summarized below:
\begin{itemize}
\item{We detect \lya emission on a global scale from 39 out of 45 galaxies. The
equivalent widths and escape fractions range from $\sim 1$ \AA\ to $\sim 80$ \AA,
and 0.1\% to 30\%, respectively. The total FUV luminosities of the galaxies in
our sample are comparable (but slightly lower) to those in \lya emitter samples at high
redshift, but the \lya luminosities are considerably lower.}
\item{We find that \lya is more extended than the UV continuum (as measured by
Petrosian radius) in 35 out of the 39 galaxies with global \lya emission. The
distribution of \lya emission in the galaxies tends to be more clumpy and
irregular in the sources with low \fesc, but can be overall asymmetric also in
galaxies with high \fesc. Comparing the \lya maps to nebular extinction maps
(\hahb) we find that \lya escape tends to happen along directions in the
galaxies where there is less extinction. }
\item{We investigate how the escape of \lya depends on a number of galaxy
properties, both stellar (mass, age, dust, and SFR quantities) and ISM
(metallicity, O32, dust). We find that the escape fraction is strongly
anti-correlated with extinction, both for stars and nebular gas. We also find a weak
anti-correlation of \fesc with stellar mass and discuss various explanations
for this. Neither SFR, specific SFR, or SFR surface density is correlated to
\fesc in our study. The ionisation state of the ISM (as probed by the O32
ratio) in the galaxies seems not to be related to \fesc in a direct way, but we
note that most of the galaxies with O32$<1$ have \fesc$<10$\%.}
\item{The anti-correlation of \lya escape fraction with nebular extinction is
strong, but shows substantial scatter. Fitting a log-linear relation to the
measurements we find an intercept of $f_{\mathrm{esc}}(E(B-V)_n = 0) = 0.075\pm 0.019$ 
and a slope of $3.30\pm 0.73$. 
We compare the observations with the expected
relation for a number of attenuation curves and find that simple dust screen
models are too steep to explain the observations. The shallow slope we find is
comparable to those found by other \lya studies, in particular when the samples
include also weak and dusty \lya emitters. We conclude that the shallow slope
found --- with the observed \fesc being lower than expected at low $E(B-V)_n$,
and higher than expected at high $E(B-V)_n$ --- can be explained by a
combination of radiative transfer effects and a clumpy dust geometry.}
\item{The equivalent width of \lya is strongly correlated to \fesc as found by
multiple previous authors. We fit a linear function to the observed data points
and find a a relatively tight relation: $f_{\mathrm{esc}}
= \left(0.0040\pm 0.00050 \cdot W_{\mathrm{Ly\alpha}} \right) \pm 0.013$.
However, the scatter around the relation is large and we discuss two different
scenarios (high extinction and a non-standard ionisation production efficiency)
which could cause individual sources to deviate from the relation.  We then use
this relation to provide an approximate dust corrected SFR calibration for \lya
luminosity which uses only \wlya to correct the luminosity for the non-zero
escape fraction. When comparing the \lya SFR estimate to dust corrected \ha
estimates for our galaxies we find a good agreement, but with large scatter
(0.20 dex for the full sample, 0.16 dex for a sub-sample of galaxies with
$E(B-V)_n<0.3$). }
\item{We characterize the aperture losses of \lya in limited aperture spectroscopic 
observations by using stacked growth curves of our sample. We find that the losses 
can be quite severe for small slit widths or aperture sizes at certain redshift ranges, 
and that this needs to be taken into account when discussing results or planning new surveys.}
\end{itemize}

The fully reduced data and maps are available to the community along with our
measurements as a High Level Science Product at MAST
(\dataset[10.17909/pbe1-m743]{http://dx.doi.org/10.17909/pbe1-m743}). The
\lya images for our reference sample, and the supporting observations can thus
be used to interpret and compare to future \lya observations in both nearby and
very distant galaxies.


\begin{acknowledgments}
This work has been supported by the Swedish National Space Agency (SNSA).  M.M.
acknowledges the support of the Swedish Research Council,
Vetenskapsrådet (grant 2019-00502). JMMH is funded by Spanish
MCIN/AEI/10.13039/501100011033 grant PID2019-107061GB-C61.  The Cosmic
Dawn Center (DAWN) is funded by the Danish National Research Foundation
under grant No. 140. 

This research is based on observations made with
the NASA/ESA Hubble Space Telescope obtained from the Space Telescope
Science Institute, which is operated by the Association of Universities
for Research in Astronomy, Inc., under NASA contract NAS 5–26555. These
observations are associated with programs 12310, 13483, 13027, 14923, 
and 13656.  The research
has also made use of the NASA/IPAC Infrared Science Archive, which is funded
by the National Aeronautics and Space Administration and operated by
the California Institute of Technology.
\end{acknowledgments}
%

\vspace{5mm}
\facilities{Hubble Space Telescope (ACS, WFC3), MAST, IRSA}


\software{Astropy \citep{2013A&A...558A..33A},  
          Drizzlepac \citep{2012drzp.book.....G},
          Matplotlib \citep{Hunter:2007},
          SCOOP \citep{10.1145/2616498.2616565},
	  LACosmic \citep{2001PASP..113.1420V},
	  TinyTim \citep{2011SPIE.8127E..0JK},
	  LaXS \citep{2014ApJ...782....6H,2014ApJ...797...11O},  
	  Photutils \citep{2022zndo...6825092B}
          }

\clearpage
\appendix
\section{HST filters used for each galaxy}
\onecolumngrid
\begin{deluxetable*}{lccccc}[b!]
\tabletypesize{\scriptsize}
\tablecaption{HST imaging for individual galaxies\label{tab:obs}}
\tablehead{
    \colhead{ID} &
        \colhead{\lya (SBC)} & 
        \colhead{\ha (UVIS/WFC)} & 
        \colhead{\hb (UVIS/WFC)} & 
        \colhead{FUV cont. (SBC)} & 
        \colhead{UV/optical cont. (UVIS/WFC)} \\ 
    \colhead{(1)} & 
	    \colhead{(2)} & 
	    \colhead{(3)} & 
	    \colhead{(4)} & 
	    \colhead{(5)} &  
	    \colhead{(6)}   
}
\startdata
LARS01    & F125LP & UVIS/F673N & UVIS/F502N  & F140LP, F150LP & UVIS/F336W, UVIS/F438W, UVIS/F775W \\ 
LARS02    & F125LP & UVIS/F673N & UVIS/F502N  & F140LP, F150LP & UVIS/F336W, UVIS/F438W, UVIS/F775W \\ 
LARS03    & F125LP & UVIS/F673N & UVIS/F502N  & F140LP, F150LP & UVIS/F336W, WFC/F435W,  WFC/F814W  \\ 
LARS04    & F125LP & UVIS/F673N & UVIS/F502N  & F140LP, F150LP & UVIS/F336W, UVIS/F438W, UVIS/F775W \\ 
LARS05    & F125LP & UVIS/F673N & UVIS/F502N  & F140LP, F150LP & UVIS/F336W, UVIS/F438W, UVIS/F775W \\ 
LARS06    & F125LP & UVIS/F673N & UVIS/F502N  & F140LP, F150LP & UVIS/F336W, UVIS/F438W, UVIS/F775W \\ 
LARS07    & F125LP & WFC/FR656N & WFC/F502N   & F140LP, F150LP & UVIS/F336W, UVIS/F438W, UVIS/F775W \\ 
LARS08    & F125LP & WFC/FR656N & WFC/F502N   & F140LP, F150LP & UVIS/F336W, UVIS/F438W, UVIS/F775W \\ 
LARS09    & F125LP & WFC/FR716N & WFC/FR505N  & F140LP, F150LP & UVIS/F336W, UVIS/F438W, UVIS/F775W \\ 
LARS10   & F125LP & WFC/FR716N & WFC/FR505N  & F140LP, F150LP & UVIS/F336W, UVIS/F438W, UVIS/F775W \\ 
LARS11   & F125LP & WFC/FR716N & WFC/FR551N  & F140LP, F150LP & UVIS/F336W, UVIS/F438W, UVIS/F775W \\ 
LARS12   & F125LP & WFC/FR716N & WFC/FR551N  & F140LP, F150LP & UVIS/F336W, UVIS/F438W, UVIS/F775W \\ 
LARS13   & F140LP & WFC/FR782N & WFC/FR551N  & F150LP         & UVIS/F390W, UVIS/F475W, UVIS/F850LP\\ 
LARS14   & F140LP & WFC/FR782N & WFC/FR551N  & F150LP         & UVIS/F390W, UVIS/F475W, WFC/F850LP \\ 
ELARS01   & F125LP & UVIS/F673N & UVIS/F502N  & F140LP, F150LP & UVIS/F336W, WFC/F435W,  WFC/F814W  \\ 
ELARS02   & F125LP & UVIS/F680N & UVIS/FQ508N & F140LP, F150LP & UVIS/F336W, UVIS/F438W, UVIS/F775W \\ 
ELARS03   & F125LP & UVIS/F673N & UVIS/F502N  & F140LP, F150LP & UVIS/F336W, UVIS/F438W, UVIS/F775W \\ 
ELARS04   & F125LP & UVIS/F673N & UVIS/F502N  & F140LP, F150LP & UVIS/F336W, UVIS/F438W, UVIS/F775W \\ 
ELARS05   & F125LP & UVIS/F673N & UVIS/F502N  & F140LP, F150LP & UVIS/F336W, UVIS/F438W, UVIS/F775W \\ 
ELARS06   & F125LP & UVIS/F673N & UVIS/F502N  & F140LP, F150LP & UVIS/F336W, UVIS/F438W, UVIS/F775W \\ 
ELARS07   & F125LP & UVIS/F673N & UVIS/F502N  & F140LP, F150LP & UVIS/F336W, UVIS/F438W, UVIS/F775W \\ 
ELARS08   & F125LP & UVIS/F673N & UVIS/F502N  & F140LP, F150LP & UVIS/F336W, UVIS/F438W, UVIS/F775W \\ 
ELARS09   & F125LP & UVIS/F673N & UVIS/F502N  & F140LP, F150LP & UVIS/F336W, UVIS/F438W, UVIS/F775W \\ 
ELARS10 & F125LP & UVIS/F673N & UVIS/F502N  & F140LP, F150LP & UVIS/F336W, UVIS/F438W, UVIS/F775W  \\ 
ELARS11 & F125LP & UVIS/F673N & UVIS/F502N  & F140LP, F150LP & UVIS/F336W, UVIS/F438W, UVIS/F775W  \\ 
ELARS12 & F125LP & UVIS/F673N & UVIS/F502N  & F140LP, F150LP & UVIS/F336W, UVIS/F438W, UVIS/F775W  \\ 
ELARS13 & F125LP & UVIS/F673N & UVIS/F502N  & F140LP, F150LP & UVIS/F336W, UVIS/F438W, UVIS/F775W  \\ 
ELARS14 & F125LP & UVIS/F673N & UVIS/F502N  & F140LP, F150LP & UVIS/F336W, UVIS/F438W, UVIS/F775W  \\ 
ELARS15 & F125LP & UVIS/F673N & UVIS/F502N  & F140LP, F150LP & UVIS/F336W, UVIS/F438W, UVIS/F775W  \\ 
ELARS16 & F125LP & UVIS/F673N & UVIS/F502N  & F140LP, F150LP & UVIS/F336W, UVIS/F438W, UVIS/F775W  \\ 
ELARS17 & F125LP & UVIS/F673N & UVIS/F502N  & F140LP, F150LP & UVIS/F336W, UVIS/F438W, UVIS/F775W  \\ 
ELARS18 & F125LP & UVIS/F673N & UVIS/F502N  & F140LP, F150LP & UVIS/F336W, UVIS/F438W, UVIS/F775W  \\ 
ELARS19 & F125LP & UVIS/F673N & UVIS/F502N  & F140LP, F150LP & UVIS/F336W, UVIS/F438W, UVIS/F775W  \\ 
ELARS20 & F125LP & UVIS/F673N & UVIS/F502N  & F140LP, F150LP & UVIS/F336W, UVIS/F438W, UVIS/F775W  \\ 
ELARS21 & F125LP & UVIS/F673N & UVIS/F502N  & F140LP, F150LP & UVIS/F336W, UVIS/F438W, UVIS/F775W  \\ 
ELARS22 & F125LP & UVIS/F680N & UVIS/FQ508N & F140LP, F150LP & UVIS/F336W, UVIS/F438W, UVIS/F775W  \\ 
ELARS23 & F125LP & UVIS/F680N & UVIS/FQ508N & F140LP, F150LP & UVIS/F336W, UVIS/F438W, UVIS/F775W  \\ 
ELARS24 & F125LP & UVIS/F680N & UVIS/FQ508N & F140LP, F150LP & UVIS/F336W, UVIS/F438W, UVIS/F775W  \\ 
ELARS25 & F125LP & UVIS/F680N & UVIS/FQ508N & F140LP, F150LP & UVIS/F336W, UVIS/F438W, UVIS/F775W  \\ 
ELARS26 & F125LP & UVIS/F680N & UVIS/FQ508N & F140LP, F150LP & UVIS/F336W, UVIS/F438W, UVIS/F775W  \\ 
ELARS27 & F125LP & UVIS/F680N & UVIS/FQ508N & F140LP, F150LP & UVIS/F336W, UVIS/F438W, UVIS/F775W  \\ 
ELARS28 & F125LP & UVIS/F680N & UVIS/FQ508N & F140LP, F150LP & UVIS/F336W, UVIS/F438W, UVIS/F775W  \\ 
Tol1214 & F125LP & UVIS/F680N & UVIS/F502N  & F140LP, F150LP & UVIS/F336W, UVIS/F438W, UVIS/F775W \\ 
Tol1247 & F125LP & UVIS/FQ674N& WFC/FR505N  & F140LP         & UVIS/F336W, UVIS/F438W, UVIS/F775W \\ 
J1156    & F140LP & WFC/FR782N & WFC/FR601N & F150LP, F165LP & UVIS/F390W, UVIS/F475W, UVIS/F850LP\\ 
\enddata
\tablecomments{The filter columns shows the instruments and filters used.
}
\end{deluxetable*}
\twocolumngrid

\clearpage
\section{Point spread function matching}
\label{app:psfmatch}
\subsection{Building PSF models}
\begin{figure}[b!]
\centering
\epsscale{1.3}
\plotone{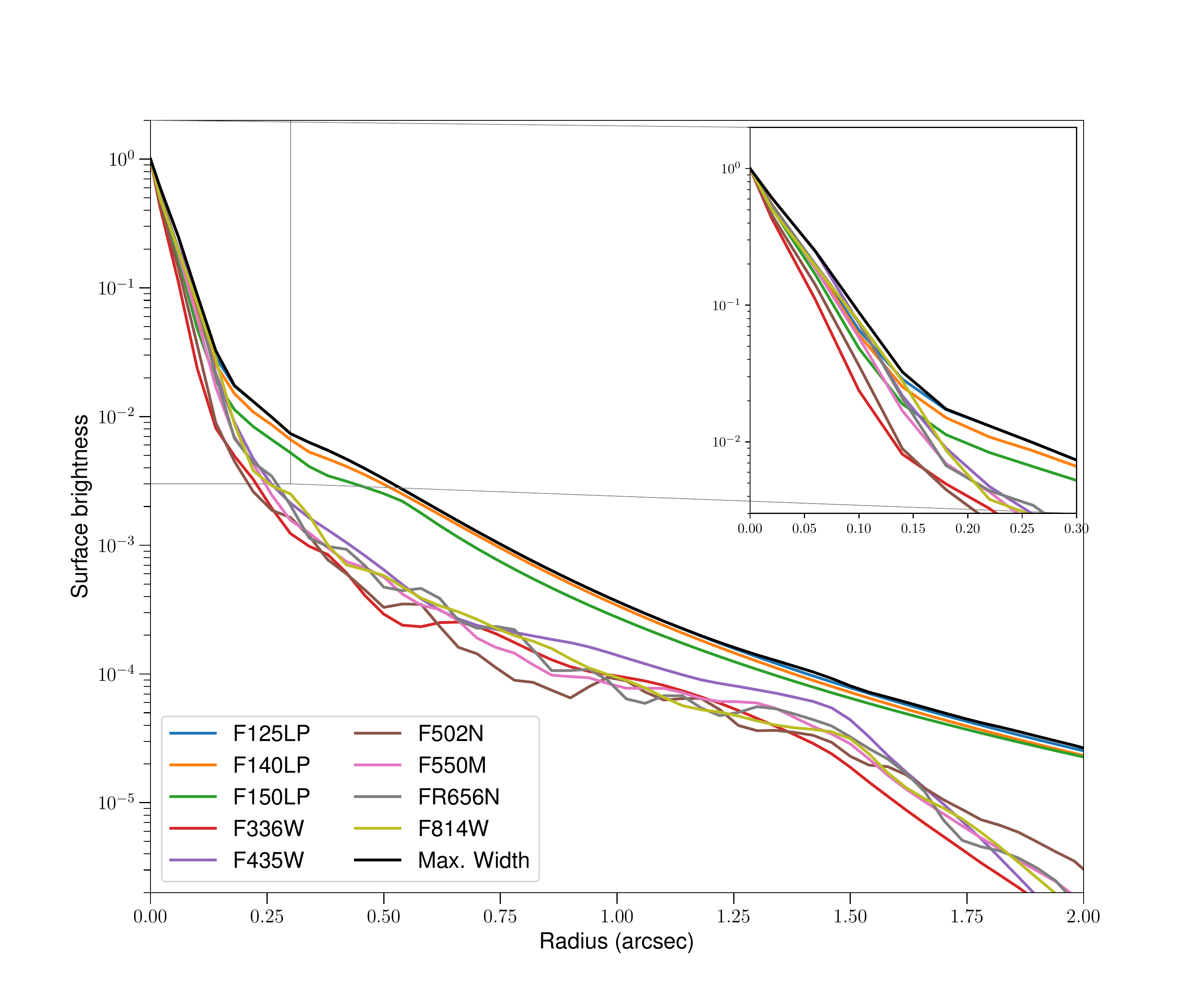}
\caption{PSF models for a selection of filters.}
\label{fig:psfmod1d}
\end{figure}

Accurate PSF models are available for most of the filters in HST instruments,
but some of these only cover the inner parts of the PSF. In order to correctly
estimate low surface brightness emission of \lya using SBC we need match the
PSF over a larger area than what is usually done. For the study reported in
this paper we thus make custom models that reach a radius of at least 1 \arcsec (25
pixels for the resampled pixel scale in use). In Figure~\ref{fig:psfmod1d} we 
show models for a selection of filters used.

Models for the optical (ACS/WFC and WFC3/UVIS) filters were generated by adding
a synthetic point source to individual reduced data frames, drizzling these
using the same procedure as for the science frames, and then cutting out a
thumbnail image of the PSF which was then properly background subtracted and
normalized.  The synthetic point source used for this comes from the
\texttt{TinyTim} software, which is able to produce models out to the needed
radius, and each individual frame has the correct defocus parameter and
position on chip (we use the center of the galaxy) set for the model. We thus
obtain one PSF model per filter and galaxy which has undergone the same
treatment (drizzling and distortion correction) as the real data. 

\begin{figure}
\centering
\epsscale{1.2}
\plotone{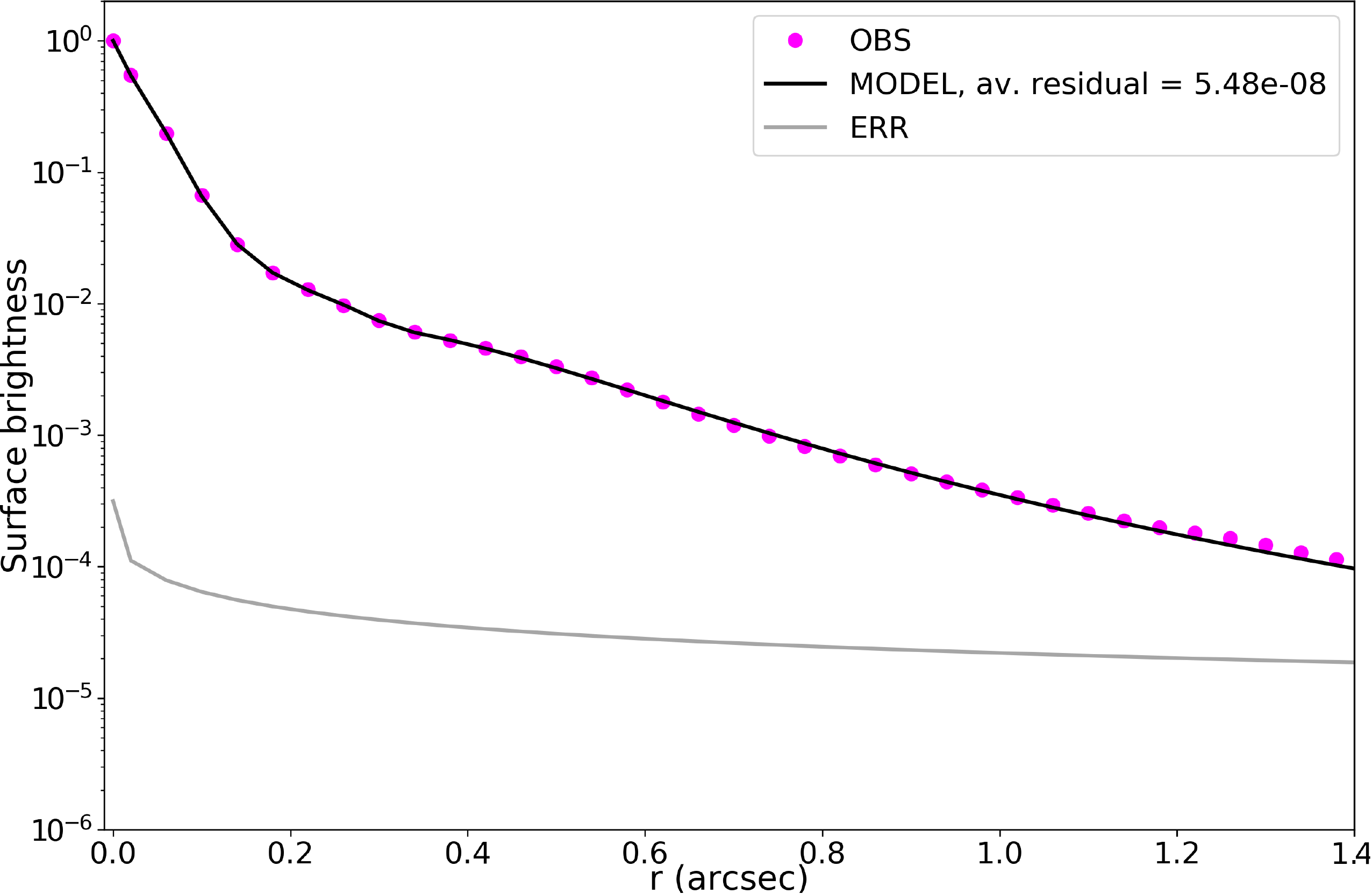}
\caption{Best fit PSF model for the SBC F125LP filter. Observed data points 
come from a stack of stars in NGC~6681. The average residual for the observed 
data points is given in the legend.}
\label{fig:sbcpsfmod}
\end{figure}
For ACS/SBC the models from \texttt{TinyTim} are inaccurate in the wings and
are therefore unusable for the matching required in this study. Instead we use
HST calibration data targeting the stellar cluster NGC~6681. We drizzle this
data to the same pixel scale as the one used in our data set (0.04 \arcsec per
pixel), find sufficiently isolated stars (roughly 50 stars) in the frame and
fit a 2D model to them, making sure to subtract background. This is done
iteratively and nearby stars that may affect the fit are removed from the image
before the next pass is done. The model that provides the best fit to the SBC 
PSF consists of three components: (i) a Moffat model with seven
free parameters, including the center coordinates and ellipticity, (ii) a
Gaussian model with three free parameters, and (iii) a radially symmetric 4th
order polynomial function. The Moffat function dominates in the center and the
polynomial function in the outer parts. The Gaussian function is needed to
correctly characterize the observed bump in the SBC PSF wings at a radius of
$\sim 0.5$ \arcsec.  An example of a SBC PSF model fit is shown in
Figure~\ref{fig:sbcpsfmod}.  

A number of different functions have been tried to
find a suitable model for the SBC filters, and this is by far the best option we have
found. Another option would be to use a thumbnail of stacked stars directly,
but the resulting PSF model is then too noisy to be used in the PSF matching
code. It should be noted that the PSF models for SBC are not dependent on
defocus, nor location on chip, they are instead an average model per filter
that we use for all galaxies. The uncertainty resulting from not taking these
second order effects into account is negligible compared to the inaccuracy
found for the SBC \texttt{TinyTim} models. Note that, somewhat unintuitively, the bluest SBC
filter (F125LP for most of the galaxies) has the widest PSF in terms of power
contained in the wings.

From the set of PSF models for a given galaxy (one per filter) we construct a
maximum width 2D PSF model by finding the maximum value per pixel in the set of
normalized models. This is then the PSF model that all of the filters are later
on matched to. We have found that this step is necessary to limit the
appearance of convolution artifacts in the matching kernel. The maximum width
PSF model is generally dominated by the reddest filter in the center regions,
and by the bluest SBC filter in the wings.

\subsection{The PSF matching technique}
\begin{figure*}
\centering
\plotone{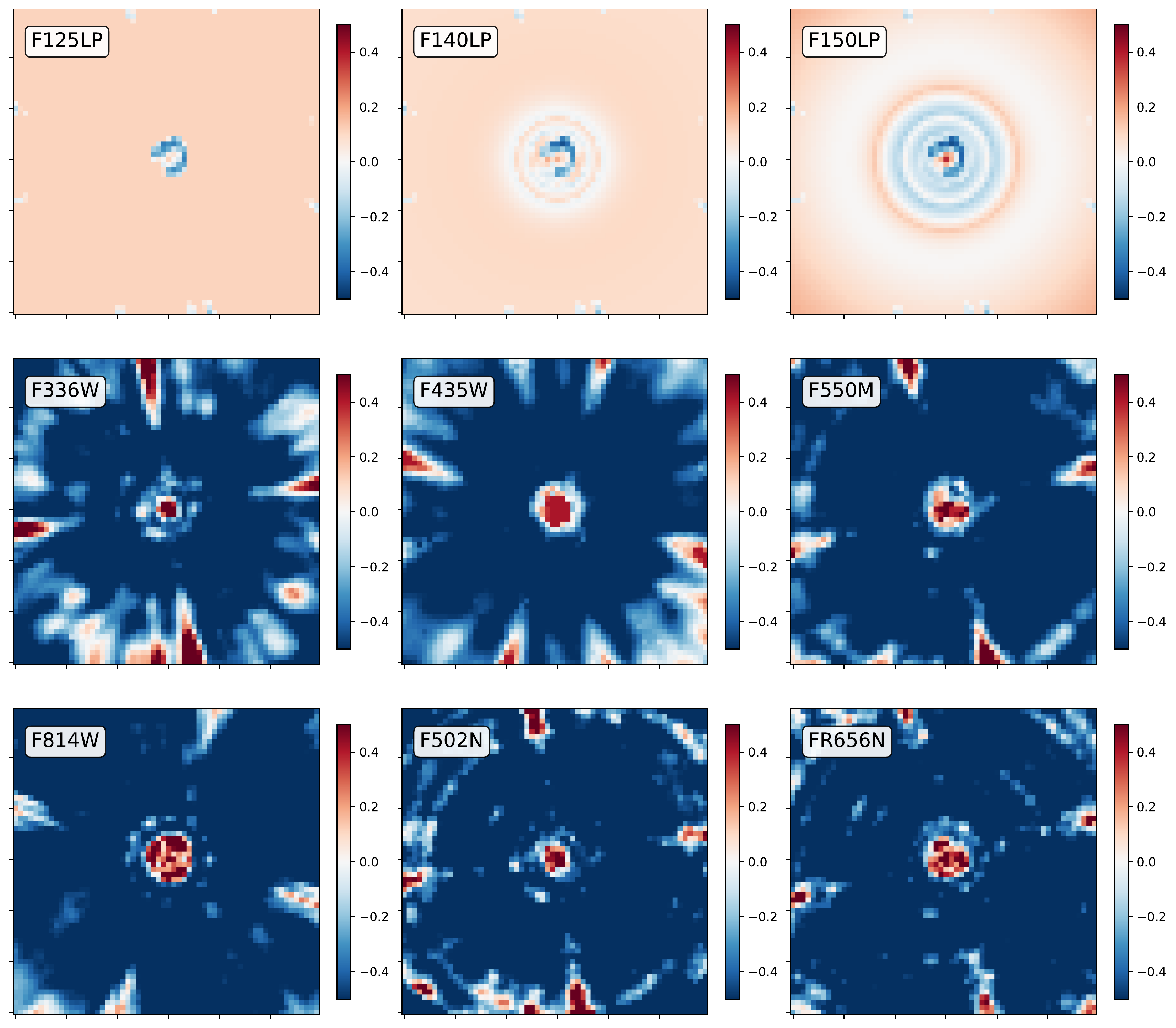}
\caption{Relative residual of PSF models compared to a maximum width PSF 
(filter setup and individual models from the ELARS01 data set). This figure shows
residuals without performing any matching.}
\label{fig:compraw}
\end{figure*}

\begin{figure*}
\centering
\plotone{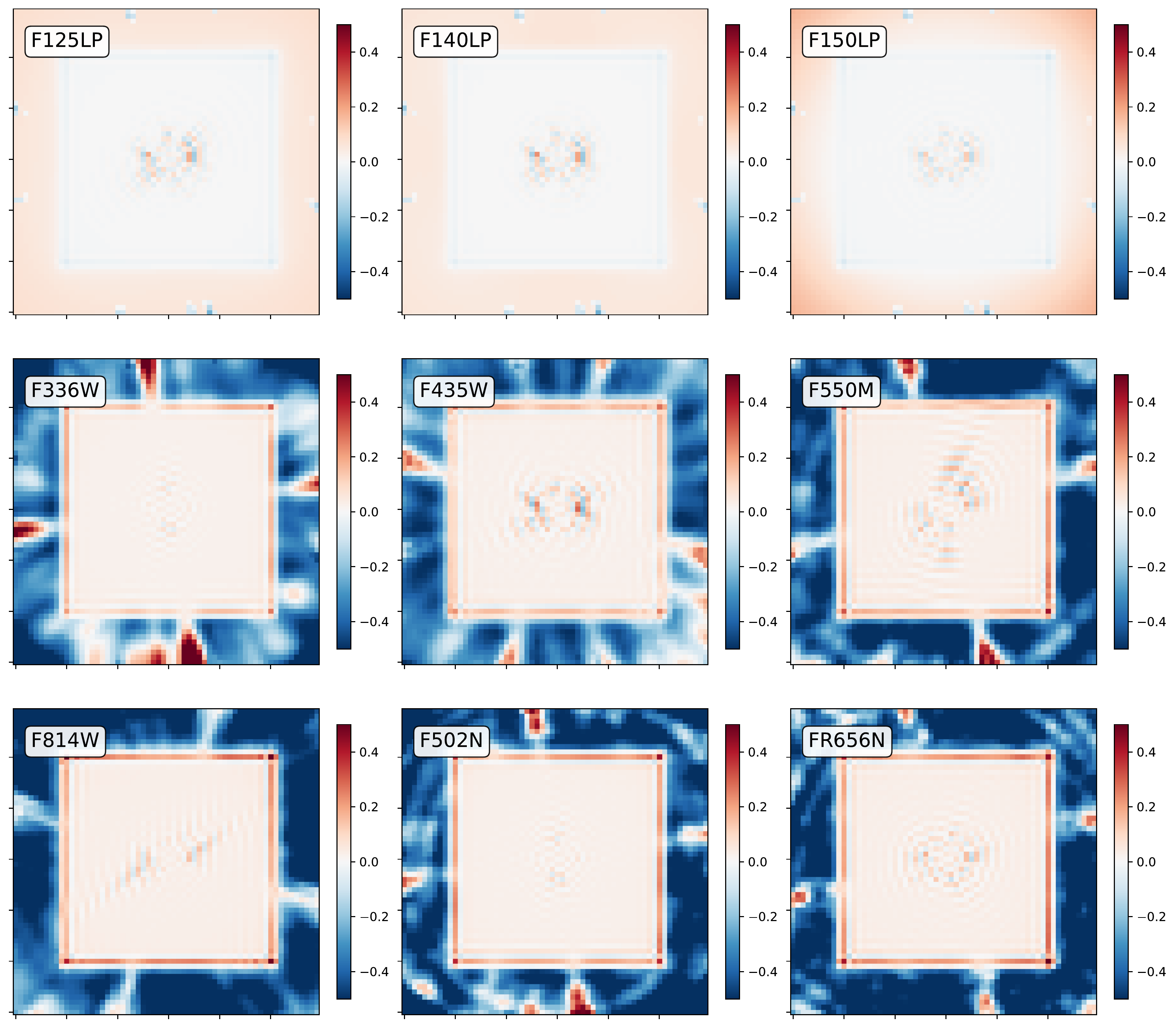}
\caption{Relative residual of PSF models after matching them to the maximum width PSF using
the technique described in this paper.}
\label{fig:compconv}
\end{figure*}
Initial attempts to match the PSFs from FUV to red optical using a single
Gaussian convolution kernel showed very poor results, in particular in the
wings. The reason for this is the non-standard shape of the SBC PSF, which is
very narrow in the center but has strong extended wings. Motivated by this we
have developed a matching tool with a more general kernel base, which has the
freedom to correctly find the optimum matching solution also for complicated
PSFs.  We base our algorithm on the work of \citet{2012MNRAS.425.1341B}, who
use a convolution kernel built from a linear combination of multiple delta
functions (one per pixel in the kernel). A kernel of this type is shown to be
very powerful in finding an optimal matching solution, but is very sensitive to
noise (both in the data and numerical), due to the high number of degrees of
freedom in finding the matching kernel. The authors address
this issue by adding regularisation to the kernel solving. They show that
moderate regularisation makes it possible to find robust matching kernels
without significant loss of accuracy.

Following \citet{2012MNRAS.425.1341B} the expression for the kernel we use is:
\begin{equation}
\label{eq:kernmod}
K_{i,j}(u,v) = \delta(u-i)\delta(v-j),
\end{equation}
which represents one delta function per kernel pixel. The final convolution kernel is then modelled as a
linear combination of these basis functions, $K(u,v) = \Sigma_{i,j} a_{i,j} K_{i,j}(u,v)$.
The convolution of the 
input PSF model ($\mathbf{R}$) with the kernel basis funtions can then be written as 
$\mathbf{C} = \mathbf{R} \bigotimes K_{i,j}$.
We then want to find the optimum matching kernel (i.e., solve for $a_{i,j}$) 
that best transforms the input model into
the maximum width filter PSF ($\mathbf{S}$). This is done by solving the following matrix equation:
\begin{equation}
\mathbf{b} = \mathbf{M}_{\lambda}\, a = (\mathbf{M} + \lambda_r \mathbf{H})\, a,
\end{equation}
where $\mathbf{b} = \mathbf{C}^T \mathbf{\Sigma}^{-1} \mathbf{S}$, 
 $\mathbf{M} = \mathbf{C}^T \mathbf{\Sigma}^{-1} \mathbf{C}$
and $\lambda_r \mathbf{H}$ is the regularisation term. For details of the
matrix equation and definitions refer to \citet{2012MNRAS.425.1341B}.  
The best value to use for
$\lambda_r$ depends on the input models and we run tests to find a good number
that limits the amount of over-fitting artifacts in the matched PSF models
while still keeping the overall matching residuals low. For the models used in
this project we have found that $\lambda_r \sim 10^{-5}$ is usually good, but
for some of the galaxies a higher regularisation strength is needed (up to
$10^{-4}$).  The algorithm is implemented as a parallelised python code which
can find the optimum matching kernel for PSF models using a given kernel size.
The number of operations performed by the code scale with kernel size to the
power of three, which means that this size should be kept as low as possible.
For our data set we have found that a kernel size of 40 pixels (1.6\arcsec,
corresponding to a radius of $\sim$ 0.8\arcsec) is adequate to get accurate
matching results while keeping the computing time reasonable.

In Figure~\ref{fig:compraw} we show example relative residual maps
($(\mathbf{R} - \mathbf{S})/\mathbf{S}$) of comparing individual filter models
(using the same filters as the PSF models displayed in
Figure~\ref{fig:psfmod1d}) to the maximum width model without performing any
PSF matching. As expected the residuals are very strong in the wings of the
optical filter PSFs, and quite substantial also in the core. The SBC filters
look slightly better, but the core regions are problematic also there. In
Figure~\ref{fig:compconv} we show the result of applying the optimum PSF
matching with a kernel size of 40 pixels. This changes the residual maps
drastically, decreasing the average residual by a factor $\sim$10--20 (depending on the
individual filter).
\clearpage

\clearpage
\section{Surface brightness, \wlya, and \fesc profiles}
This section contains tables of surface brightness, \wlya, and \fesc 
profiles for the galaxies. Note that the we have elected to use the same number
of significant digits for all the entries, even if the errors are sometimes large. 
The dynamic range of the measurements is very large, and the errors depend both 
on the measurements and the fitting uncertainties. This means that the errors are often 
very small far away from the bright galaxy centres, and the number of significant digits
have been chosen to be able to show these errors.
\label{app:sbtables}
\begin{deluxetable*}{lcccccc}[b]
\tabletypesize{\scriptsize}
\renewcommand{\arraystretch}{1.3}
\tablecaption{Surface brightness, \wlya, and \fesc profiles\label{tab:sbprofs}}
\tablehead{
    \colhead{Galaxy ID} &
    \colhead{$r$} &
	\colhead{SB$_{FUV}$} &
	\colhead{SB$_{\mathrm{H}\alpha}$} &
	\colhead{SB$_{\mathrm{Ly}\alpha}$} &
    \colhead{\wlya} &
	\colhead{\fesc} \\
	\colhead{} &
	\colhead{kpc} &
	\colhead{$10^{39}$erg/s/\AA/kpc$^2$} &
	\colhead{$10^{40}$erg/s/kpc$^2$} &
	\colhead{$10^{40}$erg/s/kpc$^2$} &
    \colhead{\AA} &
	\colhead{} \\
	\colhead{(1)} & 
	\colhead{(2)} & 
	\colhead{(3)} & 
	\colhead{(4)} & 
	\colhead{(5)} &  
	\colhead{(6)} &   
	\colhead{(7)}    
}
\startdata
LARS01  &   0.10 & 70.99200$_{0.29892}^{0.21275}$ & 38.51040$_{0.19535}^{0.32427}$ & 158.01700    $_{6.68443}^{8.49713}$ & 22.25840     $_{0.98618}^{1.10507}$ & 0.42986       $_{0.01997}^{0.01997}$\\  
LARS01  &   0.33 & 13.05120$_{0.04493}^{0.03799}$ & 28.19240$_{0.06222}^{0.06018}$ & 36.39420     $_{1.09275}^{0.82643}$ & 25.86760     $_{0.47232}^{0.54082}$ & 0.16459       $_{0.00309}^{0.00309}$\\
LARS01  &   0.56 & 3.83610 $_{0.01470}^{0.01055}$ & 10.98370$_{0.02848}^{0.02638}$ & 12.20720     $_{0.29510}^{0.24303}$ & 27.53570     $_{0.42431}^{0.45634}$ & 0.13933       $_{0.00251}^{0.00251}$\\
LARS01  &   0.78 & 1.99101 $_{0.00719}^{0.00827}$ & 6.78822 $_{0.02206}^{0.01633}$ & 8.20741      $_{0.18433}^{0.18546}$ & 30.00710     $_{0.41020}^{0.37692}$ & 0.13022       $_{0.00187}^{0.00187}$\\
LARS01  &   1.01 & 1.31142 $_{0.00568}^{0.00459}$ & 5.01932 $_{0.01414}^{0.01185}$ & 5.55087      $_{0.11409}^{0.13061}$ & 31.70100     $_{0.34369}^{0.38338}$ & 0.12243       $_{0.00139}^{0.00139}$\\
LARS01  &   1.24 & 0.70184 $_{0.00446}^{0.00375}$ & 3.16691 $_{0.01451}^{0.00720}$ & 3.01222      $_{0.07796}^{0.07880}$ & 32.64860     $_{0.31905}^{0.37942}$ & 0.11510       $_{0.00124}^{0.00124}$\\
LARS01  &   1.47 & 0.48885 $_{0.00377}^{0.00262}$ & 1.29346 $_{0.00808}^{0.00739}$ & 2.02582      $_{0.07115}^{0.06760}$ & 33.22940     $_{0.35021}^{0.31417}$ & 0.11564       $_{0.00136}^{0.00136}$\\
LARS01  &   1.69 & 0.38911 $_{0.00263}^{0.00254}$ & 0.76669 $_{0.00601}^{0.00729}$ & 1.47740      $_{0.05702}^{0.06151}$ & 33.50380     $_{0.36378}^{0.35916}$ & 0.11658       $_{0.00131}^{0.00131}$\\
LARS01  &   1.92 & 0.24877 $_{0.00191}^{0.00193}$ & 0.56295 $_{0.00470}^{0.00503}$ & 1.08880      $_{0.03840}^{0.03709}$ & 33.92070     $_{0.35802}^{0.37290}$ & 0.11715       $_{0.00125}^{0.00125}$\\
LARS01  &   2.15 & 0.16905 $_{0.00192}^{0.00162}$ & 0.42472 $_{0.00481}^{0.00466}$ & 0.95145      $_{0.03103}^{0.04196}$ & 34.59480     $_{0.31824}^{0.34963}$ & 0.11843       $_{0.00125}^{0.00125}$\\
LARS01  &   2.38 & 0.13984 $_{0.00167}^{0.00142}$ & 0.56876 $_{0.00374}^{0.00442}$ & 0.85177      $_{0.03095}^{0.04317}$ & 35.30470     $_{0.38651}^{0.29758}$ & 0.11819       $_{0.00136}^{0.00136}$\\
LARS01  &   2.60 & 0.10606 $_{0.00127}^{0.00140}$ & 0.43856 $_{0.00408}^{0.00355}$ & 0.65158      $_{0.03461}^{0.02706}$ & 35.88010     $_{0.36342}^{0.29408}$ & 0.11771       $_{0.00130}^{0.00130}$\\
LARS01  &   2.83 & 0.10035 $_{0.00142}^{0.00143}$ & 0.26576 $_{0.00252}^{0.00410}$ & 0.57028      $_{0.02338}^{0.03272}$ & 36.34590     $_{0.35660}^{0.34410}$ & 0.11874       $_{0.00129}^{0.00129}$\\
LARS01  &   3.06 & 0.10276 $_{0.00162}^{0.00120}$ & 0.28880 $_{0.00203}^{0.00460}$ & 0.46350      $_{0.02990}^{0.03008}$ & 36.55690     $_{0.33116}^{0.36894}$ & 0.11906       $_{0.00129}^{0.00129}$\\
LARS01  &   3.29 & 0.06105 $_{0.00107}^{0.00114}$ & 0.16322 $_{0.00314}^{0.00358}$ & 0.39626      $_{0.01937}^{0.02541}$ & 36.98720     $_{0.33296}^{0.33533}$ & 0.12001       $_{0.00126}^{0.00126}$\\
LARS01  &   3.51 & 0.04505 $_{0.00081}^{0.00081}$ & 0.11332 $_{0.00277}^{0.00277}$ & 0.36863      $_{0.02067}^{0.02556}$ & 37.51910     $_{0.33064}^{0.35437}$ & 0.12177       $_{0.00135}^{0.00135}$\\
LARS01  &   3.74 & 0.03694 $_{0.00103}^{0.00091}$ & 0.09549 $_{0.00297}^{0.00212}$ & 0.34906      $_{0.02205}^{0.02507}$ & 38.10440     $_{0.42559}^{0.35960}$ & 0.12314       $_{0.00137}^{0.00137}$\\
LARS01  &   3.97 & 0.02005 $_{0.00082}^{0.00094}$ & 0.05743 $_{0.00261}^{0.00252}$ & 0.28614      $_{0.01302}^{0.02067}$ & 38.72050     $_{0.41239}^{0.36458}$ & 0.12436       $_{0.00132}^{0.00132}$\\
LARS01  &   4.20 & 0.01615 $_{0.00079}^{0.00077}$ & 0.04557 $_{0.00232}^{0.00257}$ & 0.24280      $_{0.01352}^{0.01844}$ & 39.27850     $_{0.40120}^{0.37909}$ & 0.12549       $_{0.00137}^{0.00137}$\\
LARS01  &   4.42 & 0.01513 $_{0.00073}^{0.00082}$ & 0.03664 $_{0.00231}^{0.00274}$ & 0.22110      $_{0.01102}^{0.01709}$ & 39.80430     $_{0.39713}^{0.40556}$ & 0.12631       $_{0.00137}^{0.00137}$\\
LARS01  &   4.65 & 0.02175 $_{0.00071}^{0.00060}$ & 0.03915 $_{0.00200}^{0.00217}$ & 0.18782      $_{0.01542}^{0.01392}$ & 40.14830     $_{0.40845}^{0.47300}$ & 0.12664       $_{0.00144}^{0.00144}$\\
LARS01  &   4.88 & 0.02118 $_{0.00061}^{0.00066}$ & 0.03577 $_{0.00249}^{0.00185}$ & 0.16818      $_{0.01297}^{0.01260}$ & 40.44280     $_{0.40367}^{0.47217}$ & 0.12682       $_{0.00144}^{0.00144}$\\
LARS01  &   5.11 & 0.01575 $_{0.00068}^{0.00069}$ & 0.03014 $_{0.00237}^{0.00234}$ & 0.14806      $_{0.01511}^{0.01234}$ & 40.75410     $_{0.39275}^{0.44259}$ & 0.12685       $_{0.00137}^{0.00137}$\\
LARS01  &   5.33 & 0.01151 $_{0.00055}^{0.00068}$ & 0.01777 $_{0.00199}^{0.00249}$ & 0.11454      $_{0.01322}^{0.01025}$ & 41.01380     $_{0.37782}^{0.41764}$ & 0.12789       $_{0.00127}^{0.00127}$\\
LARS01  &   5.56 & 0.01064 $_{0.00053}^{0.00062}$ & 0.01521 $_{0.00190}^{0.00209}$ & 0.09811      $_{0.01009}^{0.01347}$ & 41.23110     $_{0.37352}^{0.42272}$ & 0.12881       $_{0.00127}^{0.00127}$\\
LARS01  &   5.79 & 0.00773 $_{0.00052}^{0.00043}$ & 0.01026 $_{0.00229}^{0.00176}$ & 0.09727      $_{0.01083}^{0.01223}$ & 41.50220     $_{0.38337}^{0.45148}$ & 0.12986       $_{0.00135}^{0.00135}$\\
LARS01  &   6.02 & 0.00674 $_{0.00046}^{0.00049}$ & 0.00840 $_{0.00166}^{0.00181}$ & 0.08448      $_{0.01706}^{0.00951}$ & 41.74500     $_{0.40574}^{0.42879}$ & 0.13082       $_{0.00140}^{0.00140}$\\
LARS01  &   6.24 & 0.00600 $_{0.00048}^{0.00042}$ & 0.00742 $_{0.00157}^{0.00189}$ & 0.06429      $_{0.00996}^{0.01176}$ & 41.91980     $_{0.41548}^{0.41685}$ & 0.13155       $_{0.00141}^{0.00141}$\\
LARS01  &   6.47 & 0.00514 $_{0.00047}^{0.00043}$ & 0.00801 $_{0.00184}^{0.00200}$ & 0.05201      $_{0.01377}^{0.01151}$ & 42.06010     $_{0.42472}^{0.42333}$ & 0.13212       $_{0.00148}^{0.00148}$\\
LARS01  &   6.70 & 0.00485 $_{0.00037}^{0.00046}$ & 0.00783 $_{0.00182}^{0.00208}$ & 0.04873      $_{0.01241}^{0.01181}$ & 42.19510     $_{0.45770}^{0.39695}$ & 0.13266       $_{0.00156}^{0.00156}$\\
LARS02  &   0.10 & 17.01260$_{0.12730}^{0.09033}$ & 25.70370$_{0.17739}^{0.22320}$ &$<2.83704$                           &$<1.63848$                           &$<0.00958$                           \\  
LARS02  &   0.38 & 1.83723 $_{0.01057}^{0.01042}$ & 5.91272 $_{0.03036}^{0.02139}$ & 7.51897      $_{0.17952}^{0.28318}$ & 22.34760     $_{1.15808}^{1.01958}$ & 0.09356       $_{0.00451}^{0.00451}$\\
LARS02  &   0.65 & 0.96809 $_{0.00461}^{0.00567}$ & 1.97144 $_{0.01155}^{0.01040}$ & 2.94720      $_{0.11656}^{0.11996}$ & 25.59460     $_{1.09964}^{0.82924}$ & 0.10887       $_{0.00431}^{0.00431}$\\
LARS02  &   0.93 & 0.52246 $_{0.00492}^{0.00399}$ & 0.86781 $_{0.00516}^{0.00474}$ & 2.23548      $_{0.08832}^{0.09785}$ & 29.88720     $_{0.88144}^{0.83957}$ & 0.13106       $_{0.00382}^{0.00382}$\\
LARS02  &   1.21 & 0.41289 $_{0.00350}^{0.00235}$ & 0.81836 $_{0.00491}^{0.00759}$ & 2.07652      $_{0.07493}^{0.05451}$ & 34.17820     $_{0.97234}^{0.70548}$ & 0.15393       $_{0.00411}^{0.00411}$\\
\enddata
\tablecomments{Table~\ref{tab:sbprofs} is published in its entirety in machine-readable format.
      A portion (profiles for LARS01 and part of LARS02) is shown here for guidance regarding its form and content.}
\renewcommand{\arraystretch}{1}
\end{deluxetable*}
\twocolumngrid 
\clearpage

\bibliography{larscat_subm}{}

\begin{thebibliography}{}
\expandafter\ifx\csname natexlab\endcsname\relax\def\natexlab#1{#1}\fi
\providecommand{\url}[1]{\href{#1}{#1}}
\providecommand{\dodoi}[1]{doi:~\href{http://doi.org/#1}{\nolinkurl{#1}}}
\providecommand{\doeprint}[1]{\href{http://ascl.net/#1}{\nolinkurl{http://ascl.net/#1}}}
\providecommand{\doarXiv}[1]{\href{https://arxiv.org/abs/#1}{\nolinkurl{https://arxiv.org/abs/#1}}}

\bibitem[{{Adams} {et~al.}(2011){Adams}, {Blanc}, {Hill}, {Gebhardt}, {Drory},
  {Hao}, {Bender}, {Byun}, {Ciardullo}, {Cornell}, {Finkelstein}, {Fry},
  {Gawiser}, {Gronwall}, {Hopp}, {Jeong}, {Kelz}, {Kelzenberg}, {Komatsu},
  {MacQueen}, {Murphy}, {Odoms}, {Roth}, {Schneider}, {Tufts}, \&
  {Wilkinson}}]{2011ApJS..192....5A}
{Adams}, J.~J., {Blanc}, G.~A., {Hill}, G.~J., {et~al.} 2011, \apjs, 192, 5,
  \dodoi{10.1088/0067-0049/192/1/5}

\bibitem[{Akritas {et~al.}(1995)Akritas, Murphy, \& LaValley}]{10.2307/2291140}
Akritas, M.~G., Murphy, S.~A., \& LaValley, M.~P. 1995, Journal of the American
  Statistical Association, 90, 170.
\newblock \url{http://www.jstor.org/stable/2291140}

\bibitem[{{Aller}(1984)}]{1984ASSL..112.....A}
{Aller}, L.~H. 1984, {Physics of thermal gaseous nebulae},
  \dodoi{10.1007/978-94-010-9639-3}

\bibitem[{{Anders} \& {Fritze-v. Alvensleben}(2003)}]{2003A&A...401.1063A}
{Anders}, P., \& {Fritze-v. Alvensleben}, U. 2003, \aap, 401, 1063,
  \dodoi{10.1051/0004-6361:20030151}

\bibitem[{{Astropy Collaboration} {et~al.}(2013){Astropy Collaboration},
  {Robitaille}, {Tollerud}, {Greenfield}, {Droettboom}, {Bray}, {Aldcroft},
  {Davis}, {Ginsburg}, {Price-Whelan}, {Kerzendorf}, {Conley}, {Crighton},
  {Barbary}, {Muna}, {Ferguson}, {Grollier}, {Parikh}, {Nair}, {Unther},
  {Deil}, {Woillez}, {Conseil}, {Kramer}, {Turner}, {Singer}, {Fox}, {Weaver},
  {Zabalza}, {Edwards}, {Azalee Bostroem}, {Burke}, {Casey}, {Crawford},
  {Dencheva}, {Ely}, {Jenness}, {Labrie}, {Lim}, {Pierfederici}, {Pontzen},
  {Ptak}, {Refsdal}, {Servillat}, \& {Streicher}}]{2013A&A...558A..33A}
{Astropy Collaboration}, {Robitaille}, T.~P., {Tollerud}, E.~J., {et~al.} 2013,
  \aap, 558, A33, \dodoi{10.1051/0004-6361/201322068}

\bibitem[{{Atek} {et~al.}(2009){Atek}, {Kunth}, {Schaerer}, {Hayes},
  {Deharveng}, {{\"O}stlin}, \& {Mas-Hesse}}]{2009A&A...506L...1A}
{Atek}, H., {Kunth}, D., {Schaerer}, D., {et~al.} 2009, \aap, 506, L1,
  \dodoi{10.1051/0004-6361/200912787}

\bibitem[{{Atek} {et~al.}(2014){Atek}, {Kunth}, {Schaerer}, {Mas-Hesse},
  {Hayes}, {{\"O}stlin}, \& {Kneib}}]{2014A&A...561A..89A}
---. 2014, \aap, 561, A89, \dodoi{10.1051/0004-6361/201321519}

\bibitem[{{Baldwin} {et~al.}(1981){Baldwin}, {Phillips}, \&
  {Terlevich}}]{1981PASP...93....5B}
{Baldwin}, J.~A., {Phillips}, M.~M., \& {Terlevich}, R. 1981, \pasp, 93, 5,
  \dodoi{10.1086/130766}

\bibitem[{{Becker} {et~al.}(2012){Becker}, {Homrighausen}, {Connolly},
  {Genovese}, {Owen}, {Bickerton}, \& {Lupton}}]{2012MNRAS.425.1341B}
{Becker}, A.~C., {Homrighausen}, D., {Connolly}, A.~J., {et~al.} 2012, \mnras,
  425, 1341, \dodoi{10.1111/j.1365-2966.2012.21542.x}

\bibitem[{{Blanton} {et~al.}(2001){Blanton}, {Dalcanton}, {Eisenstein},
  {Loveday}, {Strauss}, {SubbaRao}, {Weinberg}, {Anderson}, {Annis}, {Bahcall},
  {Bernardi}, {Brinkmann}, {Brunner}, {Burles}, {Carey}, {Castander},
  {Connolly}, {Csabai}, {Doi}, {Finkbeiner}, {Friedman}, {Frieman}, {Fukugita},
  {Gunn}, {Hennessy}, {Hindsley}, {Hogg}, {Ichikawa}, {Ivezi{\'c}}, {Kent},
  {Knapp}, {Lamb}, {Leger}, {Long}, {Lupton}, {McKay}, {Meiksin}, {Merelli},
  {Munn}, {Narayanan}, {Newcomb}, {Nichol}, {Okamura}, {Owen}, {Pier}, {Pope},
  {Postman}, {Quinn}, {Rockosi}, {Schlegel}, {Schneider}, {Shimasaku},
  {Siegmund}, {Smee}, {Snir}, {Stoughton}, {Stubbs}, {Szalay}, {Szokoly},
  {Thakar}, {Tremonti}, {Tucker}, {Uomoto}, {Vanden Berk}, {Vogeley},
  {Waddell}, {Yanny}, {Yasuda}, \& {York}}]{2001AJ....121.2358B}
{Blanton}, M.~R., {Dalcanton}, J., {Eisenstein}, D., {et~al.} 2001, \aj, 121,
  2358, \dodoi{10.1086/320405}

\bibitem[{{Bradley} {et~al.}(2022){Bradley}, {Sip{\H{o}}cz}, {Robitaille},
  {Tollerud}, {Vin{\'\i}cius}, {Deil}, {Barbary}, {Wilson}, {Busko}, {Donath},
  {G{\"u}nther}, {Cara}, {Lim}, {Me{\ss}linger}, {Conseil}, {Bostroem},
  {Droettboom}, {Bray}, {Andersen Bratholm}, {Barentsen}, {Craig}, {Rathi},
  {Pascual}, {Perren}, {Georgiev}, {De Val-Borro}, {Kerzendorf}, {Bach},
  {Quint}, \& {Souchereau}}]{2022zndo...6825092B}
{Bradley}, L., {Sip{\H{o}}cz}, B., {Robitaille}, T., {et~al.} 2022,
  {astropy/photutils: 1.5.0}, 1.5.0, Zenodo,  Zenodo,
  \dodoi{10.5281/zenodo.6825092}

\bibitem[{{Bridge} {et~al.}(2018){Bridge}, {Hayes}, {Melinder}, {{\"O}stlin},
  {Gronwall}, {Ciardullo}, {Atek}, {Cannon}, {Gronke}, {Guaita}, {Hagen},
  {Herenz}, {Kunth}, {Laursen}, {Mas-Hesse}, \& {Pardy}}]{2018ApJ...852....9B}
{Bridge}, J.~S., {Hayes}, M., {Melinder}, J., {et~al.} 2018, \apj, 852, 9,
  \dodoi{10.3847/1538-4357/aa9932}

\bibitem[{{Brinchmann} {et~al.}(2004){Brinchmann}, {Charlot}, {White},
  {Tremonti}, {Kauffmann}, {Heckman}, \& {Brinkmann}}]{2004MNRAS.351.1151B}
{Brinchmann}, J., {Charlot}, S., {White}, S.~D.~M., {et~al.} 2004, \mnras, 351,
  1151, \dodoi{10.1111/j.1365-2966.2004.07881.x}

\bibitem[{{Calzetti} {et~al.}(2000){Calzetti}, {Armus}, {Bohlin}, {Kinney},
  {Koornneef}, \& {Storchi-Bergmann}}]{2000ApJ...533..682C}
{Calzetti}, D., {Armus}, L., {Bohlin}, R.~C., {et~al.} 2000, \apj, 533, 682,
  \dodoi{10.1086/308692}

\bibitem[{{Calzetti} {et~al.}(1994){Calzetti}, {Kinney}, \&
  {Storchi-Bergmann}}]{1994ApJ...429..582C}
{Calzetti}, D., {Kinney}, A.~L., \& {Storchi-Bergmann}, T. 1994, \apj, 429,
  582, \dodoi{10.1086/174346}

\bibitem[{{Cardelli} {et~al.}(1989){Cardelli}, {Clayton}, \&
  {Mathis}}]{1989ApJ...345..245C}
{Cardelli}, J.~A., {Clayton}, G.~C., \& {Mathis}, J.~S. 1989, \apj, 345, 245,
  \dodoi{10.1086/167900}

\bibitem[{{Casey} {et~al.}(2014){Casey}, {Narayanan}, \&
  {Cooray}}]{2014PhR...541...45C}
{Casey}, C.~M., {Narayanan}, D., \& {Cooray}, A. 2014, \physrep, 541, 45,
  \dodoi{10.1016/j.physrep.2014.02.009}

\bibitem[{{Cowie} {et~al.}(2011){Cowie}, {Barger}, \&
  {Hu}}]{2011ApJ...738..136C}
{Cowie}, L.~L., {Barger}, A.~J., \& {Hu}, E.~M. 2011, \apj, 738, 136,
  \dodoi{10.1088/0004-637X/738/2/136}

\bibitem[{{Cowie} \& {Hu}(1998)}]{1998AJ....115.1319C}
{Cowie}, L.~L., \& {Hu}, E.~M. 1998, \aj, 115, 1319, \dodoi{10.1086/300309}

\bibitem[{{Davidsen} {et~al.}(1977){Davidsen}, {Hartig}, \&
  {Fastie}}]{1977Natur.269..203D}
{Davidsen}, A.~F., {Hartig}, G.~F., \& {Fastie}, W.~G. 1977, \nat, 269, 203,
  \dodoi{10.1038/269203a0}

\bibitem[{{Dijkstra}(2014)}]{2014PASA...31...40D}
{Dijkstra}, M. 2014, \pasa, 31, e040, \dodoi{10.1017/pasa.2014.33}

\bibitem[{{Dijkstra}(2017)}]{2017arXiv170403416D}
---. 2017, arXiv e-prints, arXiv:1704.03416.
\newblock \doarXiv{1704.03416}

\bibitem[{{Dom{\'\i}nguez S{\'a}nchez} {et~al.}(2012){Dom{\'\i}nguez
  S{\'a}nchez}, {Mignoli}, {Pozzi}, {Calura}, {Cimatti}, {Gruppioni}, {Cepa},
  {S{\'a}nchez Portal}, {Zamorani}, {Berta}, {Elbaz}, {Le Floc'h}, {Granato},
  {Lutz}, {Maiolino}, {Matteucci}, {Nair}, {Nordon}, {Pozzetti}, {Silva},
  {Silverman}, {Wuyts}, {Carollo}, {Contini}, {Kneib}, {Le F{\`e}vre}, {Lilly},
  {Mainieri}, {Renzini}, {Scodeggio}, {Bardelli}, {Bolzonella}, {Bongiorno},
  {Caputi}, {Coppa}, {Cucciati}, {de la Torre}, {de Ravel}, {Franzetti},
  {Garilli}, {Iovino}, {Kampczyk}, {Knobel}, {Kova{\v{c}}}, {Lamareille}, {Le
  Borgne}, {Le Brun}, {Maier}, {Magnelli}, {Pell{\'o}}, {Peng},
  {Perez-Montero}, {Ricciardelli}, {Riguccini}, {Tanaka}, {Tasca}, {Tresse},
  {Vergani}, \& {Zucca}}]{2012MNRAS.426..330D}
{Dom{\'\i}nguez S{\'a}nchez}, H., {Mignoli}, M., {Pozzi}, F., {et~al.} 2012,
  \mnras, 426, 330, \dodoi{10.1111/j.1365-2966.2012.21710.x}

\bibitem[{{Duval} {et~al.}(2016){Duval}, {{\"O}stlin}, {Hayes}, {Zackrisson},
  {Verhamme}, {Orlitova}, {Adamo}, {Guaita}, {Melinder}, {Cannon}, {Laursen},
  {Rivera-Thorsen}, {Herenz}, {Gruyters}, {Mas-Hesse}, {Kunth}, {Sandberg},
  {Schaerer}, \& {M{\aa}nsson}}]{2016A&A...587A..77D}
{Duval}, F., {{\"O}stlin}, G., {Hayes}, M., {et~al.} 2016, \aap, 587, A77,
  \dodoi{10.1051/0004-6361/201526876}

\bibitem[{{Erb} {et~al.}(2014){Erb}, {Steidel}, {Trainor}, {Bogosavljevi{\'c}},
  {Shapley}, {Nestor}, {Kulas}, {Law}, {Strom}, {Rudie}, {Reddy}, {Pettini},
  {Konidaris}, {Mace}, {Matthews}, \& {McLean}}]{2014ApJ...795...33E}
{Erb}, D.~K., {Steidel}, C.~C., {Trainor}, R.~F., {et~al.} 2014, \apj, 795, 33,
  \dodoi{10.1088/0004-637X/795/1/33}

\bibitem[{{Flury} {et~al.}(2022){Flury}, {Jaskot}, {Ferguson}, {Worseck},
  {Makan}, {Chisholm}, {Saldana-Lopez}, {Schaerer}, {McCandliss}, {Wang},
  {Ford}, {Heckman}, {Ji}, {Giavalisco}, {Amorin}, {Atek}, {Blaizot},
  {Borthakur}, {Carr}, {Castellano}, {Cristiani}, {De Barros}, {Dickinson},
  {Finkelstein}, {Fleming}, {Fontanot}, {Garel}, {Grazian}, {Hayes}, {Henry},
  {Mauerhofer}, {Micheva}, {Oey}, {Ostlin}, {Papovich}, {Pentericci},
  {Ravindranath}, {Rosdahl}, {Rutkowski}, {Santini}, {Scarlata}, {Teplitz},
  {Thuan}, {Trebitsch}, {Vanzella}, {Verhamme}, \& {Xu}}]{2022ApJS..260....1F}
{Flury}, S.~R., {Jaskot}, A.~E., {Ferguson}, H.~C., {et~al.} 2022, \apjs, 260,
  1, \dodoi{10.3847/1538-4365/ac5331}

\bibitem[{{Fricke} {et~al.}(2001){Fricke}, {Izotov}, {Papaderos}, {Guseva}, \&
  {Thuan}}]{2001AJ....121..169F}
{Fricke}, K.~J., {Izotov}, Y.~I., {Papaderos}, P., {Guseva}, N.~G., \& {Thuan},
  T.~X. 2001, \aj, 121, 169, \dodoi{10.1086/318016}

\bibitem[{{Fruchter} \& {Hook}(2002)}]{2002PASP..114..144F}
{Fruchter}, A.~S., \& {Hook}, R.~N. 2002, \pasp, 114, 144,
  \dodoi{10.1086/338393}

\bibitem[{{Giavalisco} {et~al.}(1996){Giavalisco}, {Koratkar}, \&
  {Calzetti}}]{1996ApJ...466..831G}
{Giavalisco}, M., {Koratkar}, A., \& {Calzetti}, D. 1996, \apj, 466, 831,
  \dodoi{10.1086/177557}

\bibitem[{{Gonzaga} {et~al.}(2012){Gonzaga}, {Hack}, {Fruchter}, \&
  {Mack}}]{2012drzp.book.....G}
{Gonzaga}, S., {Hack}, W., {Fruchter}, A., \& {Mack}, J. 2012, {The DrizzlePac
  Handbook}

\bibitem[{{Gronke} \& {Dijkstra}(2016)}]{2016ApJ...826...14G}
{Gronke}, M., \& {Dijkstra}, M. 2016, \apj, 826, 14,
  \dodoi{10.3847/0004-637X/826/1/14}

\bibitem[{{Guaita} {et~al.}(2015){Guaita}, {Melinder}, {Hayes}, {{\"O}stlin},
  {Gonzalez}, {Micheva}, {Adamo}, {Mas-Hesse}, {Sandberg},
  {Ot{\'{\i}}-Floranes}, {Schaerer}, {Verhamme}, {Freeland}, {Orlitov{\'a}},
  {Laursen}, {Cannon}, {Duval}, {Rivera-Thorsen}, {Herenz}, {Kunth}, {Atek},
  {Puschnig}, {Gruyters}, \& {Pardy}}]{2015A&A...576A..51G}
{Guaita}, L., {Melinder}, J., {Hayes}, M., {et~al.} 2015, \aap, 576, A51,
  \dodoi{10.1051/0004-6361/201425053}

\bibitem[{{Guseva} {et~al.}(2011){Guseva}, {Izotov}, {Stasi{\'n}ska}, {Fricke},
  {Henkel}, \& {Papaderos}}]{2011AA...529A.149G}
{Guseva}, N.~G., {Izotov}, Y.~I., {Stasi{\'n}ska}, G., {et~al.} 2011, \aap,
  529, A149, \dodoi{10.1051/0004-6361/201016291}

\bibitem[{{Hayes}(2015)}]{2015PASA...32...27H}
{Hayes}, M. 2015, \pasa, 32, e027, \dodoi{10.1017/pasa.2015.25}

\bibitem[{{Hayes} {et~al.}(2016){Hayes}, {Melinder}, {{\"O}stlin}, {Scarlata},
  {Lehnert}, \& {Mannerstr{\"o}m-Jansson}}]{2016ApJ...828...49H}
{Hayes}, M., {Melinder}, J., {{\"O}stlin}, G., {et~al.} 2016, \apj, 828, 49,
  \dodoi{10.3847/0004-637X/828/1/49}

\bibitem[{{Hayes} {et~al.}(2007){Hayes}, {{\"O}stlin}, {Atek}, {Kunth},
  {Mas-Hesse}, {Leitherer}, {Jim{\'e}nez-Bail{\'o}n}, \&
  {Adamo}}]{2007MNRAS.382.1465H}
{Hayes}, M., {{\"O}stlin}, G., {Atek}, H., {et~al.} 2007, \mnras, 382, 1465,
  \dodoi{10.1111/j.1365-2966.2007.12482.x}

\bibitem[{{Hayes} {et~al.}(2009){Hayes}, {{\"O}stlin}, {Mas-Hesse}, \&
  {Kunth}}]{2009AJ....138..911H}
{Hayes}, M., {{\"O}stlin}, G., {Mas-Hesse}, J.~M., \& {Kunth}, D. 2009, \aj,
  138, 911, \dodoi{10.1088/0004-6256/138/3/911}

\bibitem[{{Hayes} {et~al.}(2005){Hayes}, {{\"O}stlin}, {Mas-Hesse}, {Kunth},
  {Leitherer}, \& {Petrosian}}]{2005A&A...438...71H}
{Hayes}, M., {{\"O}stlin}, G., {Mas-Hesse}, J.~M., {et~al.} 2005, \aap, 438,
  71, \dodoi{10.1051/0004-6361:20052702}

\bibitem[{{Hayes} {et~al.}(2011){Hayes}, {Schaerer}, {{\"O}stlin}, {Mas-Hesse},
  {Atek}, \& {Kunth}}]{2011ApJ...730....8H}
{Hayes}, M., {Schaerer}, D., {{\"O}stlin}, G., {et~al.} 2011, \apj, 730, 8,
  \dodoi{10.1088/0004-637X/730/1/8}

\bibitem[{{Hayes} {et~al.}(2010){Hayes}, {{\"O}stlin}, {Schaerer}, {Mas-Hesse},
  {Leitherer}, {Atek}, {Kunth}, {Verhamme}, {de Barros}, \&
  {Melinder}}]{2010Natur.464..562H}
{Hayes}, M., {{\"O}stlin}, G., {Schaerer}, D., {et~al.} 2010, \nat, 464, 562,
  \dodoi{10.1038/nature08881}

\bibitem[{{Hayes} {et~al.}(2013){Hayes}, {{\"O}stlin}, {Schaerer}, {Verhamme},
  {Mas-Hesse}, {Adamo}, {Atek}, {Cannon}, {Duval}, {Guaita}, {Herenz}, {Kunth},
  {Laursen}, {Melinder}, {Orlitov{\'a}}, {Ot{\'{\i}}-Floranes}, \&
  {Sandberg}}]{2013ApJ...765L..27H}
---. 2013, \apjl, 765, L27, \dodoi{10.1088/2041-8205/765/2/L27}

\bibitem[{{Hayes} {et~al.}(2014){Hayes}, {{\"O}stlin}, {Duval}, {Sandberg},
  {Guaita}, {Melinder}, {Adamo}, {Schaerer}, {Verhamme}, {Orlitov{\'a}},
  {Mas-Hesse}, {Cannon}, {Atek}, {Kunth}, {Laursen}, {Ot{\'{\i}}-Floranes},
  {Pardy}, {Rivera-Thorsen}, \& {Herenz}}]{2014ApJ...782....6H}
{Hayes}, M., {{\"O}stlin}, G., {Duval}, F., {et~al.} 2014, \apj, 782, 6,
  \dodoi{10.1088/0004-637X/782/1/6}

\bibitem[{{Heckman} {et~al.}(2011){Heckman}, {Borthakur}, {Overzier},
  {Kauffmann}, {Basu-Zych}, {Leitherer}, {Sembach}, {Martin}, {Rich},
  {Schiminovich}, \& {Seibert}}]{2011ApJ...730....5H}
{Heckman}, T.~M., {Borthakur}, S., {Overzier}, R., {et~al.} 2011, \apj, 730, 5,
  \dodoi{10.1088/0004-637X/730/1/5}

\bibitem[{{Henry} {et~al.}(2015){Henry}, {Scarlata}, {Martin}, \&
  {Erb}}]{2015ApJ...809...19H}
{Henry}, A., {Scarlata}, C., {Martin}, C.~L., \& {Erb}, D. 2015, \apj, 809, 19,
  \dodoi{10.1088/0004-637X/809/1/19}

\bibitem[{{Herenz} {et~al.}(2016){Herenz}, {Gruyters}, {Orlitova}, {Hayes},
  {{\"O}stlin}, {Cannon}, {Roth}, {Bik}, {Pardy}, {Ot{\'{\i}}-Floranes},
  {Mas-Hesse}, {Adamo}, {Atek}, {Duval}, {Guaita}, {Kunth}, {Laursen},
  {Melinder}, {Puschnig}, {Rivera-Thorsen}, {Schaerer}, \&
  {Verhamme}}]{2016A&A...587A..78H}
{Herenz}, E.~C., {Gruyters}, P., {Orlitova}, I., {et~al.} 2016, \aap, 587, A78,
  \dodoi{10.1051/0004-6361/201527373}

\bibitem[{{Herenz} {et~al.}(2017){Herenz}, {Urrutia}, {Wisotzki}, {Kerutt},
  {Saust}, {Werhahn}, {Schmidt}, {Caruana}, {Diener}, {Bacon}, {Brinchmann},
  {Schaye}, {Maseda}, \& {Weilbacher}}]{2017A&A...606A..12H}
{Herenz}, E.~C., {Urrutia}, T., {Wisotzki}, L., {et~al.} 2017, \aap, 606, A12,
  \dodoi{10.1051/0004-6361/201731055}

\bibitem[{Hold-Geoffroy {et~al.}(2014)Hold-Geoffroy, Gagnon, \&
  Parizeau}]{10.1145/2616498.2616565}
Hold-Geoffroy, Y., Gagnon, O., \& Parizeau, M. 2014, in Proceedings of the 2014
  Annual Conference on Extreme Science and Engineering Discovery Environment,
  XSEDE '14 (New York, NY, USA: Association for Computing Machinery),
  \dodoi{10.1145/2616498.2616565}

\bibitem[{Hunter(2007)}]{Hunter:2007}
Hunter, J.~D. 2007, Computing In Science \& Engineering, 9, 90,
  \dodoi{10.1109/MCSE.2007.55}

\bibitem[{{Izotov} {et~al.}(2020){Izotov}, {Schaerer}, {Worseck}, {Verhamme},
  {Guseva}, {Thuan}, {Orlitov{\'a}}, \& {Fricke}}]{2020MNRAS.491..468I}
{Izotov}, Y.~I., {Schaerer}, D., {Worseck}, G., {et~al.} 2020, \mnras, 491,
  468, \dodoi{10.1093/mnras/stz3041}

\bibitem[{{Izotov} {et~al.}(2018){Izotov}, {Worseck}, {Schaerer}, {Guseva},
  {Thuan}, {Fricke}, \& {Orlitov{\'a}}}]{2018MNRAS.478.4851I}
{Izotov}, Y.~I., {Worseck}, G., {Schaerer}, D., {et~al.} 2018, \mnras, 478,
  4851, \dodoi{10.1093/mnras/sty1378}

\bibitem[{{Jaskot} {et~al.}(2017){Jaskot}, {Oey}, {Scarlata}, \&
  {Dowd}}]{2017ApJ...851L...9J}
{Jaskot}, A.~E., {Oey}, M.~S., {Scarlata}, C., \& {Dowd}, T. 2017, \apjl, 851,
  L9, \dodoi{10.3847/2041-8213/aa9d83}

\bibitem[{{Jensen} {et~al.}(2013){Jensen}, {Laursen}, {Mellema}, {Iliev},
  {Sommer-Larsen}, \& {Shapiro}}]{2013MNRAS.428.1366J}
{Jensen}, H., {Laursen}, P., {Mellema}, G., {et~al.} 2013, \mnras, 428, 1366,
  \dodoi{10.1093/mnras/sts116}

\bibitem[{{Kelly}(2007)}]{2007ApJ...665.1489K}
{Kelly}, B.~C. 2007, \apj, 665, 1489, \dodoi{10.1086/519947}

\bibitem[{{Kennicutt} \& {Evans}(2012)}]{2012ARA&A..50..531K}
{Kennicutt}, R.~C., \& {Evans}, N.~J. 2012, \araa, 50, 531,
  \dodoi{10.1146/annurev-astro-081811-125610}

\bibitem[{{Kewley} \& {Dopita}(2002)}]{2002ApJS..142...35K}
{Kewley}, L.~J., \& {Dopita}, M.~A. 2002, \apjs, 142, 35,
  \dodoi{10.1086/341326}

\bibitem[{{Kimm} {et~al.}(2019){Kimm}, {Blaizot}, {Garel}, {Michel-Dansac},
  {Katz}, {Rosdahl}, {Verhamme}, \& {Haehnelt}}]{2019MNRAS.486.2215K}
{Kimm}, T., {Blaizot}, J., {Garel}, T., {et~al.} 2019, \mnras, 486, 2215,
  \dodoi{10.1093/mnras/stz989}

\bibitem[{{Kornei} {et~al.}(2010){Kornei}, {Shapley}, {Erb}, {Steidel},
  {Reddy}, {Pettini}, \& {Bogosavljevi{\'c}}}]{2010ApJ...711..693K}
{Kornei}, K.~A., {Shapley}, A.~E., {Erb}, D.~K., {et~al.} 2010, \apj, 711, 693,
  \dodoi{10.1088/0004-637X/711/2/693}

\bibitem[{{Krist} {et~al.}(2011){Krist}, {Hook}, \&
  {Stoehr}}]{2011SPIE.8127E..0JK}
{Krist}, J.~E., {Hook}, R.~N., \& {Stoehr}, F. 2011, in Society of
  Photo-Optical Instrumentation Engineers (SPIE) Conference Series, Vol. 8127,
  Optical Modeling and Performance Predictions V, ed. M.~A. {Kahan}, 81270J,
  \dodoi{10.1117/12.892762}

\bibitem[{{Kroupa} {et~al.}(1993){Kroupa}, {Tout}, \&
  {Gilmore}}]{1993MNRAS.262..545K}
{Kroupa}, P., {Tout}, C.~A., \& {Gilmore}, G. 1993, \mnras, 262, 545,
  \dodoi{10.1093/mnras/262.3.545}

\bibitem[{{Kunth} {et~al.}(2003){Kunth}, {Leitherer}, {Mas-Hesse},
  {{\"O}stlin}, \& {Petrosian}}]{2003ApJ...597..263K}
{Kunth}, D., {Leitherer}, C., {Mas-Hesse}, J.~M., {{\"O}stlin}, G., \&
  {Petrosian}, A. 2003, \apj, 597, 263, \dodoi{10.1086/378396}

\bibitem[{{Kunth} {et~al.}(1998){Kunth}, {Mas-Hesse}, {Terlevich}, {Terlevich},
  {Lequeux}, \& {Fall}}]{1998A&A...334...11K}
{Kunth}, D., {Mas-Hesse}, J.~M., {Terlevich}, E., {et~al.} 1998, \aap, 334, 11.
\newblock \doarXiv{astro-ph/9802253}

\bibitem[{{Laursen} {et~al.}(2013){Laursen}, {Duval}, \&
  {{\"O}stlin}}]{2013ApJ...766..124L}
{Laursen}, P., {Duval}, F., \& {{\"O}stlin}, G. 2013, \apj, 766, 124,
  \dodoi{10.1088/0004-637X/766/2/124}

\bibitem[{{Laursen} {et~al.}(2009){Laursen}, {Sommer-Larsen}, \&
  {Andersen}}]{2009ApJ...704.1640L}
{Laursen}, P., {Sommer-Larsen}, J., \& {Andersen}, A.~C. 2009, \apj, 704, 1640,
  \dodoi{10.1088/0004-637X/704/2/1640}

\bibitem[{{Laursen} {et~al.}(2019){Laursen}, {Sommer-Larsen}, {Milvang-Jensen},
  {Fynbo}, \& {Razoumov}}]{2019A&A...627A..84L}
{Laursen}, P., {Sommer-Larsen}, J., {Milvang-Jensen}, B., {Fynbo}, J. P.~U., \&
  {Razoumov}, A.~O. 2019, \aap, 627, A84, \dodoi{10.1051/0004-6361/201833645}

\bibitem[{{Le Reste} {et~al.}(2022){Le Reste}, {Hayes}, {Cannon}, {Herenz},
  {Melinder}, {Menacho}, {{\"O}stlin}, {Puschnig}, {Rivera-Thorsen}, {Kunth},
  \& {Velikonja}}]{2022ApJ...934...69L}
{Le Reste}, A., {Hayes}, M., {Cannon}, J.~M., {et~al.} 2022, \apj, 934, 69,
  \dodoi{10.3847/1538-4357/ac77ed}

\bibitem[{{Leclercq} {et~al.}(2017){Leclercq}, {Bacon}, {Wisotzki}, {Mitchell},
  {Garel}, {Verhamme}, {Blaizot}, {Hashimoto}, {Herenz}, {Conseil},
  {Cantalupo}, {Inami}, {Contini}, {Richard}, {Maseda}, {Schaye}, {Marino},
  {Akhlaghi}, {Brinchmann}, \& {Carollo}}]{2017A&A...608A...8L}
{Leclercq}, F., {Bacon}, R., {Wisotzki}, L., {et~al.} 2017, \aap, 608, A8,
  \dodoi{10.1051/0004-6361/201731480}

\bibitem[{{Leitherer} {et~al.}(1999){Leitherer}, {Schaerer}, {Goldader},
  {Delgado}, {Robert}, {Kune}, {de Mello}, {Devost}, \&
  {Heckman}}]{1999ApJS..123....3L}
{Leitherer}, C., {Schaerer}, D., {Goldader}, J.~D., {et~al.} 1999, \apjs, 123,
  3, \dodoi{10.1086/313233}

\bibitem[{{Malhotra} \& {Rhoads}(2004)}]{2004ApJ...617L...5M}
{Malhotra}, S., \& {Rhoads}, J.~E. 2004, \apjl, 617, L5, \dodoi{10.1086/427182}

\bibitem[{{Matthee} {et~al.}(2016){Matthee}, {Sobral}, {Oteo}, {Best}, {Smail},
  {R{\"o}ttgering}, \& {Paulino-Afonso}}]{2016MNRAS.458..449M}
{Matthee}, J., {Sobral}, D., {Oteo}, I., {et~al.} 2016, \mnras, 458, 449,
  \dodoi{10.1093/mnras/stw322}

\bibitem[{{Meier} \& {Terlevich}(1981)}]{1981ApJ...246L.109M}
{Meier}, D.~L., \& {Terlevich}, R. 1981, \apjl, 246, L109,
  \dodoi{10.1086/183565}

\bibitem[{{Momose} {et~al.}(2014){Momose}, {Ouchi}, {Nakajima}, {Ono},
  {Shibuya}, {Shimasaku}, {Yuma}, {Mori}, \& {Umemura}}]{2014MNRAS.442..110M}
{Momose}, R., {Ouchi}, M., {Nakajima}, K., {et~al.} 2014, \mnras, 442, 110,
  \dodoi{10.1093/mnras/stu825}

\bibitem[{{Montero-Dorta} {et~al.}(2016){Montero-Dorta}, {Bolton},
  {Brownstein}, {Swanson}, {Dawson}, {Prada}, {Eisenstein}, {Maraston},
  {Thomas}, {Comparat}, {Chuang}, {McBride}, {Favole}, {Guo},
  {Rodr{\'{\i}}guez-Torres}, \& {Schneider}}]{2016MNRAS.461.1131M}
{Montero-Dorta}, A.~D., {Bolton}, A.~S., {Brownstein}, J.~R., {et~al.} 2016,
  \mnras, 461, 1131, \dodoi{10.1093/mnras/stw1352}

\bibitem[{{Nakajima} {et~al.}(2018){Nakajima}, {Fletcher}, {Ellis},
  {Robertson}, \& {Iwata}}]{2018MNRAS.477.2098N}
{Nakajima}, K., {Fletcher}, T., {Ellis}, R.~S., {Robertson}, B.~E., \& {Iwata},
  I. 2018, \mnras, 477, 2098, \dodoi{10.1093/mnras/sty750}

\bibitem[{{Natta} \& {Panagia}(1984)}]{1984ApJ...287..228N}
{Natta}, A., \& {Panagia}, N. 1984, \apj, 287, 228, \dodoi{10.1086/162681}

\bibitem[{{Neufeld}(1991)}]{1991ApJ...370L..85N}
{Neufeld}, D.~A. 1991, \apjl, 370, L85, \dodoi{10.1086/185983}

\bibitem[{{{\"O}stlin} {et~al.}(2009){{\"O}stlin}, {Hayes}, {Kunth},
  {Mas-Hesse}, {Leitherer}, {Petrosian}, \& {Atek}}]{2009AJ....138..923O}
{{\"O}stlin}, G., {Hayes}, M., {Kunth}, D., {et~al.} 2009, \aj, 138, 923,
  \dodoi{10.1088/0004-6256/138/3/923}

\bibitem[{{{\"O}stlin} {et~al.}(2014){{\"O}stlin}, {Hayes}, {Duval},
  {Sandberg}, {Rivera-Thorsen}, {Marquart}, {Orlitov{\'a}}, {Adamo},
  {Melinder}, {Guaita}, {Atek}, {Cannon}, {Gruyters}, {Herenz}, {Kunth},
  {Laursen}, {Mas-Hesse}, {Micheva}, {Ot{\'{\i}}-Floranes}, {Pardy}, {Roth},
  {Schaerer}, \& {Verhamme}}]{2014ApJ...797...11O}
{{\"O}stlin}, G., {Hayes}, M., {Duval}, F., {et~al.} 2014, \apj, 797, 11,
  \dodoi{10.1088/0004-637X/797/1/11}

\bibitem[{{Ouchi} {et~al.}(2018){Ouchi}, {Harikane}, {Shibuya}, {Shimasaku},
  {Taniguchi}, {Konno}, {Kobayashi}, {Kajisawa}, {Nagao}, {Ono}, {Inoue},
  {Umemura}, {Mori}, {Hasegawa}, {Higuchi}, {Komiyama}, {Matsuda}, {Nakajima},
  {Saito}, \& {Wang}}]{2018PASJ...70S..13O}
{Ouchi}, M., {Harikane}, Y., {Shibuya}, T., {et~al.} 2018, \pasj, 70, S13,
  \dodoi{10.1093/pasj/psx074}

\bibitem[{{Pardy} {et~al.}(2014){Pardy}, {Cannon}, {{\"O}stlin}, {Hayes},
  {Rivera-Thorsen}, {Sandberg}, {Adamo}, {Freeland}, {Herenz}, {Guaita},
  {Kunth}, {Laursen}, {Mas-Hesse}, {Melinder}, {Orlitov{\'a}},
  {Ot{\'{\i}}-Floranes}, {Puschnig}, {Schaerer}, \&
  {Verhamme}}]{2014ApJ...794..101P}
{Pardy}, S.~A., {Cannon}, J.~M., {{\"O}stlin}, G., {et~al.} 2014, \apj, 794,
  101, \dodoi{10.1088/0004-637X/794/2/101}

\bibitem[{{Partridge} \& {Peebles}(1967)}]{1967ApJ...147..868P}
{Partridge}, R.~B., \& {Peebles}, P.~J.~E. 1967, \apj, 147, 868,
  \dodoi{10.1086/149079}

\bibitem[{{Petrosian}(1976)}]{1976ApJ...209L...1P}
{Petrosian}, V. 1976, \apjl, 210, L53, \dodoi{10.1086/182301}

\bibitem[{{Prevot} {et~al.}(1984){Prevot}, {Lequeux}, {Maurice}, {Prevot}, \&
  {Rocca-Volmerange}}]{1984AA...132..389P}
{Prevot}, M.~L., {Lequeux}, J., {Maurice}, E., {Prevot}, L., \&
  {Rocca-Volmerange}, B. 1984, \aap, 132, 389

\bibitem[{{Puschnig} {et~al.}(2017){Puschnig}, {Hayes}, {{\"O}stlin},
  {Rivera-Thorsen}, {Melinder}, {Cannon}, {Menacho}, {Zackrisson}, {Bergvall},
  \& {Leitet}}]{2017MNRAS.469.3252P}
{Puschnig}, J., {Hayes}, M., {{\"O}stlin}, G., {et~al.} 2017, \mnras, 469,
  3252, \dodoi{10.1093/mnras/stx951}

\bibitem[{{Rasekh} {et~al.}(2022){Rasekh}, {Melinder}, {{\"O}stlin}, {Hayes},
  {Herenz}, {Runnholm}, {Kunth}, {Mas Hesse}, {Verhamme}, \&
  {Cannon}}]{2022A&A...662A..64R}
{Rasekh}, A., {Melinder}, J., {{\"O}stlin}, G., {et~al.} 2022, \aap, 662, A64,
  \dodoi{10.1051/0004-6361/202140734}

\bibitem[{{Reddy} {et~al.}(2020){Reddy}, {Shapley}, {Kriek}, {Steidel},
  {Shivaei}, {Sanders}, {Mobasher}, {Coil}, {Siana}, {Freeman}, {Azadi},
  {Fetherolf}, {Leung}, {Price}, \& {Zick}}]{2020ApJ...902..123R}
{Reddy}, N.~A., {Shapley}, A.~E., {Kriek}, M., {et~al.} 2020, \apj, 902, 123,
  \dodoi{10.3847/1538-4357/abb674}

\bibitem[{{Rivera-Thorsen} {et~al.}(2015){Rivera-Thorsen}, {Hayes},
  {{\"O}stlin}, {Duval}, {Orlitov{\'a}}, {Verhamme}, {Mas-Hesse}, {Schaerer},
  {Cannon}, {Ot{\'{\i}}-Floranes}, {Sandberg}, {Guaita}, {Adamo}, {Atek},
  {Herenz}, {Kunth}, {Laursen}, \& {Melinder}}]{2015ApJ...805...14R}
{Rivera-Thorsen}, T.~E., {Hayes}, M., {{\"O}stlin}, G., {et~al.} 2015, \apj,
  805, 14, \dodoi{10.1088/0004-637X/805/1/14}

\bibitem[{{Rodighiero} {et~al.}(2011){Rodighiero}, {Daddi}, {Baronchelli},
  {Cimatti}, {Renzini}, {Aussel}, {Popesso}, {Lutz}, {Andreani}, {Berta},
  {Cava}, {Elbaz}, {Feltre}, {Fontana}, {F{\"o}rster Schreiber},
  {Franceschini}, {Genzel}, {Grazian}, {Gruppioni}, {Ilbert}, {Le Floch},
  {Magdis}, {Magliocchetti}, {Magnelli}, {Maiolino}, {McCracken}, {Nordon},
  {Poglitsch}, {Santini}, {Pozzi}, {Riguccini}, {Tacconi}, {Wuyts}, \&
  {Zamorani}}]{2011ApJ...739L..40R}
{Rodighiero}, G., {Daddi}, E., {Baronchelli}, I., {et~al.} 2011, \apjl, 739,
  L40, \dodoi{10.1088/2041-8205/739/2/L40}

\bibitem[{{Runnholm} {et~al.}(2020){Runnholm}, {Hayes}, {Melinder},
  {Rivera-Thorsen}, {{\"O}stlin}, {Cannon}, \& {Kunth}}]{2020ApJ...892...48R}
{Runnholm}, A., {Hayes}, M., {Melinder}, J., {et~al.} 2020, \apj, 892, 48,
  \dodoi{10.3847/1538-4357/ab7a91}

\bibitem[{{Salim} {et~al.}(2007){Salim}, {Rich}, {Charlot}, {Brinchmann},
  {Johnson}, {Schiminovich}, {Seibert}, {Mallery}, {Heckman}, {Forster},
  {Friedman}, {Martin}, {Morrissey}, {Neff}, {Small}, {Wyder}, {Bianchi},
  {Donas}, {Lee}, {Madore}, {Milliard}, {Szalay}, {Welsh}, \&
  {Yi}}]{2007ApJS..173..267S}
{Salim}, S., {Rich}, R.~M., {Charlot}, S., {et~al.} 2007, \apjs, 173, 267,
  \dodoi{10.1086/519218}

\bibitem[{{Scarlata} {et~al.}(2009){Scarlata}, {Colbert}, {Teplitz}, {Panagia},
  {Hayes}, {Siana}, {Rau}, {Francis}, {Caon}, {Pizzella}, \&
  {Bridge}}]{2009ApJ...704L..98S}
{Scarlata}, C., {Colbert}, J., {Teplitz}, H.~I., {et~al.} 2009, \apjl, 704,
  L98, \dodoi{10.1088/0004-637X/704/2/L98}

\bibitem[{{Schlafly} \& {Finkbeiner}(2011)}]{2011ApJ...737..103S}
{Schlafly}, E.~F., \& {Finkbeiner}, D.~P. 2011, \apj, 737, 103,
  \dodoi{10.1088/0004-637X/737/2/103}

\bibitem[{{Smith} {et~al.}(2015){Smith}, {Safranek-Shrader}, {Bromm}, \&
  {Milosavljevi{\'c}}}]{2015MNRAS.449.4336S}
{Smith}, A., {Safranek-Shrader}, C., {Bromm}, V., \& {Milosavljevi{\'c}}, M.
  2015, \mnras, 449, 4336, \dodoi{10.1093/mnras/stv565}

\bibitem[{{Sobral} \& {Matthee}(2019)}]{2019A&A...623A.157S}
{Sobral}, D., \& {Matthee}, J. 2019, \aap, 623, A157,
  \dodoi{10.1051/0004-6361/201833075}

\bibitem[{{Sobral} {et~al.}(2018){Sobral}, {Matthee}, {Darvish}, {Smail},
  {Best}, {Alegre}, {R{\"o}ttgering}, {Mobasher}, {Paulino-Afonso}, {Stroe}, \&
  {Oteo}}]{2018MNRAS.477.2817S}
{Sobral}, D., {Matthee}, J., {Darvish}, B., {et~al.} 2018, \mnras, 477, 2817,
  \dodoi{10.1093/mnras/sty782}

\bibitem[{{Steidel} {et~al.}(2011){Steidel}, {Bogosavljevi{\'c}}, {Shapley},
  {Kollmeier}, {Reddy}, {Erb}, \& {Pettini}}]{2011ApJ...736..160S}
{Steidel}, C.~C., {Bogosavljevi{\'c}}, M., {Shapley}, A.~E., {et~al.} 2011,
  \apj, 736, 160, \dodoi{10.1088/0004-637X/736/2/160}

\bibitem[{{Terlevich} {et~al.}(1993){Terlevich}, {Diaz}, {Terlevich}, \&
  {Vargas}}]{1993MNRAS.260....3T}
{Terlevich}, E., {Diaz}, A.~I., {Terlevich}, R., \& {Vargas}, M.~L.~G. 1993,
  \mnras, 260, 3, \dodoi{10.1093/mnras/260.1.3}

\bibitem[{{Terlevich} {et~al.}(1991){Terlevich}, {Melnick}, {Masegosa},
  {Moles}, \& {Copetti}}]{1991A&AS...91..285T}
{Terlevich}, R., {Melnick}, J., {Masegosa}, J., {Moles}, M., \& {Copetti},
  M.~V.~F. 1991, \aaps, 91, 285

\bibitem[{{Trainor} {et~al.}(2019){Trainor}, {Strom}, {Steidel}, {Rudie},
  {Chen}, \& {Theios}}]{2019ApJ...887...85T}
{Trainor}, R.~F., {Strom}, A.~L., {Steidel}, C.~C., {et~al.} 2019, \apj, 887,
  85, \dodoi{10.3847/1538-4357/ab4993}

\bibitem[{van~der Walt {et~al.}(2014)van~der Walt, {S}ch\"onberger,
  {Nunez-Iglesias}, {B}oulogne, {W}arner, {Y}ager, {G}ouillart, {Y}u, \& the
  scikit-image contributors}]{scikit-image}
van~der Walt, S., {S}ch\"onberger, J.~L., {Nunez-Iglesias}, J., {et~al.} 2014,
  PeerJ, 2, e453, \dodoi{10.7717/peerj.453}

\bibitem[{{van Dokkum}(2001)}]{2001PASP..113.1420V}
{van Dokkum}, P.~G. 2001, \pasp, 113, 1420,
  \dodoi{10.1086/32389410.48550/arXiv.astro-ph/0108003}

\bibitem[{{Verhamme} {et~al.}(2012){Verhamme}, {Dubois}, {Blaizot}, {Garel},
  {Bacon}, {Devriendt}, {Guiderdoni}, \& {Slyz}}]{2012A&A...546A.111V}
{Verhamme}, A., {Dubois}, Y., {Blaizot}, J., {et~al.} 2012, \aap, 546, A111,
  \dodoi{10.1051/0004-6361/201218783}

\bibitem[{{Verhamme} {et~al.}(2015){Verhamme}, {Orlitov{\'a}}, {Schaerer}, \&
  {Hayes}}]{2015A&A...578A...7V}
{Verhamme}, A., {Orlitov{\'a}}, I., {Schaerer}, D., \& {Hayes}, M. 2015, \aap,
  578, A7, \dodoi{10.1051/0004-6361/201423978}

\bibitem[{{Verhamme} {et~al.}(2008){Verhamme}, {Schaerer}, {Atek}, \&
  {Tapken}}]{2008A&A...491...89V}
{Verhamme}, A., {Schaerer}, D., {Atek}, H., \& {Tapken}, C. 2008, \aap, 491,
  89, \dodoi{10.1051/0004-6361:200809648}

\bibitem[{{Wisotzki} {et~al.}(2016){Wisotzki}, {Bacon}, {Blaizot},
  {Brinchmann}, {Herenz}, {Schaye}, {Bouch{\'e}}, {Cantalupo}, {Contini},
  {Carollo}, {Caruana}, {Courbot}, {Emsellem}, {Kamann}, {Kerutt}, {Leclercq},
  {Lilly}, {Patr{\'\i}cio}, {Sandin}, {Steinmetz}, {Straka}, {Urrutia},
  {Verhamme}, {Weilbacher}, \& {Wendt}}]{2016A&A...587A..98W}
{Wisotzki}, L., {Bacon}, R., {Blaizot}, J., {et~al.} 2016, \aap, 587, A98,
  \dodoi{10.1051/0004-6361/201527384}

\bibitem[{{Yang} {et~al.}(2017{\natexlab{a}}){Yang}, {Malhotra}, {Rhoads},
  {Leitherer}, {Wofford}, {Jiang}, \& {Wang}}]{2017ApJ...838....4Y}
{Yang}, H., {Malhotra}, S., {Rhoads}, J.~E., {et~al.} 2017{\natexlab{a}}, \apj,
  838, 4, \dodoi{10.3847/1538-4357/aa6337}

\bibitem[{{Yang} {et~al.}(2017{\natexlab{b}}){Yang}, {Malhotra}, {Gronke},
  {Rhoads}, {Leitherer}, {Wofford}, {Jiang}, {Dijkstra}, {Tilvi}, \&
  {Wang}}]{2017ApJ...844..171Y}
{Yang}, H., {Malhotra}, S., {Gronke}, M., {et~al.} 2017{\natexlab{b}}, \apj,
  844, 171, \dodoi{10.3847/1538-4357/aa7d4d}

\bibitem[{{Yin} {et~al.}(2007){Yin}, {Liang}, {Hammer}, {Brinchmann}, {Zhang},
  {Deng}, \& {Flores}}]{2007AA...462..535Y}
{Yin}, S.~Y., {Liang}, Y.~C., {Hammer}, F., {et~al.} 2007, \aap, 462, 535,
  \dodoi{10.1051/0004-6361:20065798}

\end{thebibliography}
\bibliographystyle{aasjournal}



\end{document}